\documentclass[twocolumn]{aastex701}

\usepackage{rotating}
\usepackage{placeins}
\usepackage{amsmath}

\received{2024-12-27}
\revised{2025-07-29}
\accepted{2025-07-29}
\published{2025-09-17}
\submitjournal{Astrophysical Journal}

\reportnum{arXiv:2505.11374}

\begin{document}

\title{A magnetic field detection in the massive O-type bright giant 63 Oph}

\author[0000-0003-2088-0706]{James A. Barron}
\affiliation{Department of Physics, Engineering Physics \& Astronomy, Queen’s University, 64 Bader Lane, Kingston, ON K7L 3N6, Canada}
\email{NA}
\affiliation{Department of Physics \& Space Science, Royal Military College of Canada, P.O. Box 17000, Station Forces, Kingston, ON K7K 7B4, Canada}
\author[0000-0002-1854-0131]{Gregg A. Wade}
\affiliation{Department of Physics \& Space Science, Royal Military College of Canada, PO Box 17000, Station Forces, Kingston, ON K7K 7B4, Canada}
\email{NA}
\author[0000-0002-9296-8259]{Gonzalo Holgado}
\affiliation{Instituto de Astrof\'isica de Canarias, E-38200 La Laguna, Tenerife, Spain}
\email{NA}
\author[0000-0003-1168-3524]{Sergio Simón-Díaz}
\affiliation{Instituto de Astrof\'isica de Canarias, E-38200 La Laguna, Tenerife, Spain}
\affiliation{Departamento de Astrof\'isica, Universidad de La Laguna, E-38205 La Laguna, Tenerife, Spain}
\email{NA}

\begin{abstract}
Surface magnetic fields are detected in less than $10\%$ of the massive O-type star population and even less frequently among `old' massive stars approaching the terminal-age main sequence (TAMS). It is unclear to what extent the rarity of magnetic detections in massive stars near the TAMS is due to magnetic field decay or observational biases. We report the detection of a weak surface magnetic field in the O-type giant 63~Oph ($T_{\mathrm{eff}}=35.0\pm0.3$\,kK, $\log g=3.51\pm0.03$) from new ESPaDOnS circularly polarized spectra. The mean longitudinal field strength associated with the magnetic detection is $\langle B_{z}\rangle=84\pm14$\,G, which we use to set a lower limit on the dipolar field strength of $B_{\mathrm{p}}\geq300\pm50$\,G. We report Balmer line-core equivalent widths (EWs) and radial velocity measurements from the analysis of spectra primarily obtained by the IACOB project with the FEROS, FIES and HERMES spectrographs. We identify a dominant period of $\sim19.8$\,d in the EWs, which we attribute to the effects of a rotating magnetosphere under the oblique rotator model. We do not identify any coherent signals in a time series analysis of archival Hipparcos, ASAS-SN and K2 photometry. Our findings show that 63~Oph may be a rare link between strongly magnetic massive stars detected on or near the zero-age main sequence and weakly magnetic O-type supergiants. Additional observations are needed to fully constrain 63~Oph's magnetic field geometry and magnetospheric properties. 
\end{abstract}

\keywords{Massive stars (732) -- Magnetic stars (995) -- O giant stars (1136) -- Stellar magnetic fields (1610) -- Spectropolarimetry (1973)}

\section{Introduction}\label{sec:intro}
Detectable surface magnetic fields are relatively rare in massive O-type stars, occurring in less than 10\% of the Galactic population \citep{grunhut_2017, Scholler_2017, Petit_2019}. Magnetic fields in O stars are typically strong with dipolar field strengths on the order of $1-10$\,kG (\citealt{wade_2015} and references therein). Radiatively driven stellar winds are channelled by the magnetic fields into corotating magnetospheres \citep{ud-Doula_2002, petit_2013, ud-Doula_2013, Subramanian_2022}, which significantly impact massive star evolutionary paths by reducing mass loss and slowing rotation through magnetic braking \citep{ud-Doula_2009, Townsend_2010, Keszthelyi_2022}. Massive star magnetic fields are generally considered to be `fossil fields' generated at an earlier stellar evolutionary stage and no longer sustained by a contemporaneous dynamo. Several evolutionary channels for the generation of fossil fields have been proposed, including the amplification of molecular cloud seed fields during pre-main-sequence (MS) stages \citep{Keszthelyi_2023}, or stellar mergers \citep{Ferrario_2009, Schneider_2019, Frost_2024}.

It is not well understood how fossil fields change during a star's MS lifetime as the stellar radius increases and the star evolves toward post-MS stages. \cite{Fossati_2016} found evidence of magnetic field decay in massive magnetic stars toward the end of the MS using the BONNSAI Bayesian tool \citep{Schneider_2014} and \cite{Brott_2011} single-star evolutionary tracks. Fossil-field decay may be due to ohmic diffusion, which is poorly constrained in the most massive stars \citep{Keszthelyi_2023}, or from tidal dissipation due to a close companion \citep{Vidal_2019}. Alternatively, the dearth of old detectably magnetic massive stars could be attributed to the conservation of magnetic flux, where the surface field strength decreases with increasing radius. \cite{Petit_2019} argued that detection biases cannot be ruled out due to systematic uncertainties in using stellar evolutionary tracks that do not fully account for the effects of strong fossil magnetic fields (such as those employed by \cite{Fossati_2016}). As there are only about a dozen confirmed Galactic magnetic O-type stars (\citealt{grunhut_2017} and references therein), our understanding of evolved magnetic massive stars is hindered by the small sample size. 

Fossil-field decay has important implications for understanding a variety of astrophysical objects and processes. Massive magnetic stars have been suggested as the progenitors of exotic stellar end products and transients, including heavy stellar mass black holes \citep{Petit_2017}, pair-instability supernovae \citep{Georgy_2017}, magnetars \citep{Ferrario_2006} and fast radio bursts \citep{Kritti_2024}. Recently, \cite{Manzari_2024} proposed that the core-collapse supernovae of high-mass, evolved magnetic stars can be used to constrain axion particle models. 

The Magnetism in Massive Stars Survey (MiMeS) was a large, high-resolution, spectropolarimetric survey designed to characterize the magnetic properties of O and early B-type stars \citep{wade_2016}. Three magnetic O-star candidates, including the O-type giant 63~Oph, were identified by \cite{grunhut_2017} from analysis of the MiMeS O star sample. These magnetic candidates presented tentative evidence of a Zeeman signature but lacked a sufficient number of observations to confirm. 

\begin{figure}
    \centering
    \includegraphics[width=\linewidth]{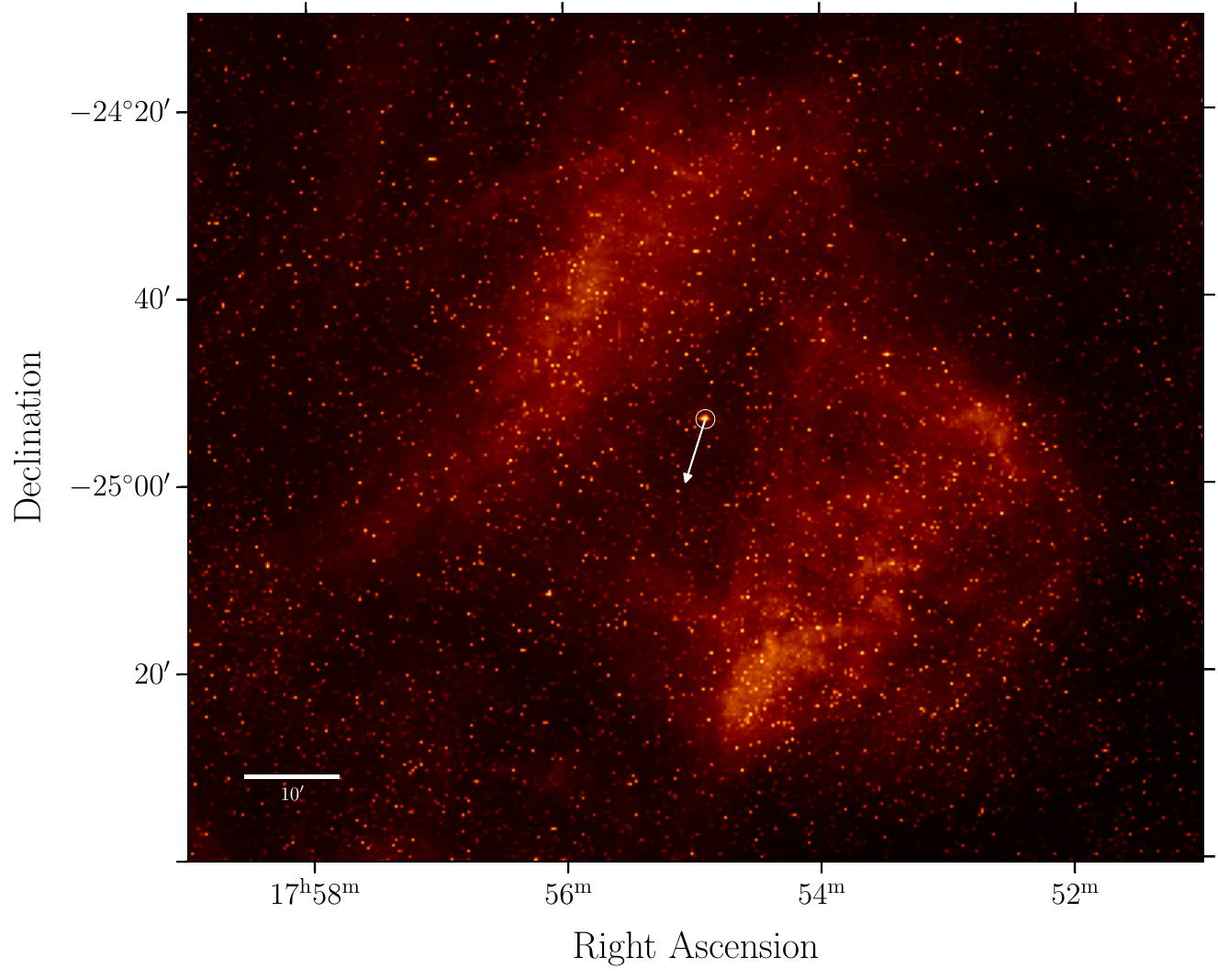}
    \caption{SuperCOSMOS H$\alpha$ image of the H\,\textsc{ii} nebula  Sh 2-22 surrounding 63~Oph \citep{Parker_2005}. 63~Oph is enclosed by the white circle, and the white arrow denotes the proper-motion vector. The scale bar corresponds to $10^{\prime}$.}
    \label{fig:63Oph_nebula}
\end{figure}

\object{63 Oph} (\object{HD 162978}) is an O8~II((f)) bright giant \citep{sota_2014, Martins_2018} that shows clear variability in the wind diagnostic Balmer and helium lines \citep{holgado_2018}. The star appears to be the principal ionizing source of \object{Sh 2-22} (\object{RCW 144}), a poorly studied H\,\textsc{ii} ring nebula \citep{lozinskaya_1983} located in the Galactic plane at the edge of the historical Sgr~OB1 association \citep{Blaha_1989, Melnik_2017}. Figure~\ref{fig:63Oph_nebula} shows a SuperCOSMOS H$\alpha$ image \citep{Parker_2005} of 63~Oph and the surrounding nebula. 

63~Oph has not been previously identified as a spectroscopic binary \citep{garmany_1980, chini_2012}, and several imaging and interferometric studies have not observed any bright, close companions \citep{Mason_1998, turner_2008, mason_2009, sana_2014}. A single, faint ($\Delta H = 6.76\pm0.25$~mag) visual companion has been identified at a wide angular separation of $4.46\pm0.04$~arcsec \citep{sana_2014}. Using Hipparcos and Gaia~DR2 data \cite{Kervella_2019} calculated the proper-motion anomaly (PMa) for nearby stars. The authors did not detect a significant PMa for 63~Oph and found no clear evidence of a binary companion. Previous studies have reported normal He and N surface abundances, providing no obvious indication of previous binary interaction \citep{martins_2015, holgado_2018, Martinez_2025}.

Table~\ref{tab:params_63Oph} provides stellar parameters for 63~Oph from the literature and derived in this work.

\begin{table}
\centering
 \caption{Stellar Parameters of 63~Oph.}
\footnotesize
\begin{tabular}{lrcrc}
\hline\hline
Parameter&Quantity&Ref.&Quantity&Ref.\\
\hline
$m_{V}$ [mag] &6.20&D02& & \\
$M_{V}$ [mag] &$-5.14\pm0.22$&$-$& & \\
$\log L$ [$L_{\odot}$] &$5.29\pm0.09$&$-$& & \\
$T_{\textrm{eff}}$ [kK]&$35.0\pm0.3$&$-$&$33.5\pm2.0$&M17\\
$\log g_{\textrm{corr}}$&$3.51\pm0.03$&$-$&$3.41\pm0.20$&M17\\
$\log \mathcal{L}$ [$\mathcal{L_{\odot}}$]&$4.06\pm0.03$&$-$\\
$v_{\mathrm{eq}}\sin i$ [km\,s$^{-1}$]&$48\pm5$&$-$&$53\pm10$&M17\\
$v_{\textrm{mac}}$ [km\,s$^{-1}$]&$101\pm5$&$-$&$93\pm20$&M17\\
$R_{\textrm{spec}}$ [$R_{\odot}$]&$12.0\pm1.2$&$-$\\
$M_{\textrm{spec}}$ [$M_{\odot}$]&$17\pm4$&$-$\\
$R_{\textrm{evol}}$ [$R_{\odot}$]&$\sim20$&$-$&$13.41$&P19\\
$M_{\textrm{evol}}$ [$M_{\odot}$]&$\sim48$&$-$&30&P19\\
Age [Myr]&3.7&P19\\
MS Frac. Age & 0.72&P19\\
He/H&$0.10\pm0.3$&H18&&\\
N/H [$10^{-4}$]&$3.9^{+1.4}_{-0.8}$&M15&&\\
$\varpi$ [mas]&$0.91\pm0.09$&DR2&$1.04\pm0.12$&DR3\\
d [pc]&$1082.3^{+119.6\dagger}_{-98.2}$&BJ18&$997.9_{-109.1}^{+135.9}$&BJ21\\
\hline
\end{tabular}
\tablecomments{\footnotesize\textbf{References}: BJ18 \citep{Bailer_Jones_2018}, BJ21 \citep{bailer_jones_2021}, D02 \citep{Ducati_2002}, DR2 \citep{Gaia_2018}, DR3 \citep{Gaia_2023}, H18 \citep{holgado_2018}, M15 \citep{martins_2015}, M17 \citep{martins_2017}, P19 \citep{Petit_2019}}
\tablecomments{\footnotesize Values without a provided reference were derived in this work.}
\tablecomments{\footnotesize$^{\dagger}$We adopted the BJ18 distance in our analysis (see Section~\ref{sec:distance}).}
\label{tab:params_63Oph}
\end{table}

Early magnetic measurements of 63~Oph were obtained by \cite{hubrig_2008} using the low-resolution FORS1 spectropolarimeter at the Very Large Telescope. No magnetic field was detected in these FORS1 observations by the authors or in a subsequent analysis by \cite{bagnulo_2015}. During the MiMeS survey, 63~Oph was observed twice by ESPaDOnS at the Canada-France-Hawaii Telescope (CFHT), once in 2008 and once in 2012. \cite{grunhut_2017} obtained a marginal magnetic detection in the Stokes~$V$ profile of the 2008 observation, with a $\sim5\sigma$ detection of the longitudinal field ($\langle B_{z}\rangle=111\pm23$~G). However, an analysis of the 2012 observation yielded a nondetection ($\langle B_{z}\rangle=-13\pm14$~G) despite a signal-to-noise ratio (S/N) almost twice as large. \cite{grunhut_2017} noted that while 63~Oph shows uncharacteristically weak emission in key magnetospheric lines, the first marginally detected observation shows higher levels of emission than the second nondetected observation. As \cite{grunhut_2017} discuss, this line profile variability can be explained under the oblique rotator model (ORM; \citealt{Stibbs_1950}) if the magnetosphere was observed at different orientations during the star's rotation (e.g. \citealt{sundqvist_2012}). Due to the single marginal magnetic detection, \cite{grunhut_2017} classified 63~Oph as a magnetic candidate. This classification was supported by a subsequent Bayesian analysis of the Stokes~$V$ signatures \citep{Petit_2019}. 

In this paper, we confirm 63~Oph as a magnetic O-type star from analysis of new ESPaDOnS spectropolarimetric observations. In Section~\ref{sec:obs} we describe the spectroscopic and photometric observations. In Section~\ref{sec:stellar_parameters} we derive spectroscopic stellar parameters. Section~\ref{sec:specpol_analysis} presents the magnetic analysis of the ESPaDOnS spectropolarimetric observations. Section~\ref{sec:var_analysis} presents a variability analysis of the radial velocity (RV) measurements, Balmer lines and time series photometry. We provide additional discussion in Section~\ref{sec:discussion} and summarize conclusions in Section~\ref{sec:Conclusions}.

\section{Observations}\label{sec:obs}
\subsection{Spectropolarimetry}
We obtained three spectropolarimetric observations of 63~Oph in July 2022 using the ESPaDOnS high-resolution spectropolarimeter \citep{donati_2006} at CFHT under observing program 22AC31 (PI: Barron). We also reanalyzed the two archival ESPaDOnS observations obtained by the MiMeS survey. The ESPaDOnS spectropolarimeter is comprised of a bench-mounted, cross-dispersed echelle spectrograph, which is fibre-fed from a Cassegrain-mounted polarimeter. The instrument has a high resolving power of $R\sim65,000$ and a spectral range of $370-1050$\,nm over 40 overlapping spectral orders. The ESPaDOnS data were reduced using the Upena pipeline, which is based on the Libre-ESpRIT reduction package \citep{donati_1997}. 

A single spectropolarimetric sequence consists of four subexposures, each corresponding to a different angle of the Fresnel rhombs. These subexposures are combined to give one unpolarized Stokes~$I$ spectrum, one circularly polarized Stokes~$V$ spectrum and two diagnostic null $N$ spectra. Circularly polarized Stokes~$V$ spectra are sensitive to the line-of-sight or longitudinal component of a stellar magnetic field (see review by \citealt{donati_2009}). The diagnostic null spectra are used to identify spurious instrumental signals in Stokes~$V$ \citep{donati_1997}. Successive spectropolarimetric sequences were coadded and normalized using the interactive IDL program \texttt{norm\_gui} described by \cite{Grunhut_2012_norm}.

For each coadded spectrum, we calculated the S/N per 1.8\,km\,s$^{-1}$ spectral bin at 500\,nm from the flux uncertainty. The S/N of the new observations range from $\sim1100$ to 2000, comparable to the archival observations. Table~\ref{tab:mag_log} provides a log of the spectropolarimetric observations.

\begin{table*}
\centering
\caption{ESPaDOnS spectropolarimetric observations and magnetic results for 63~Oph.}
\begin{tabular}{cccccccccc}
\hline
\hline
Date&HJD\,$-$&Exp. Time&S/N&Mask&$\langle B_{z} \rangle\pm\sigma_{B}$&$|\langle B_{z} \rangle|/\sigma_{B}$&$\textrm{FAP}_{V}$&$\textrm{FAP}_{N_{1}}$&Detection\\
&2,450,000&[s]&(at 500\,nm)&&[G]&&&&Flag\\
\hline
2008-07-29&4676.7997&$1\times4\times700$&1120&Optimal&$115\pm 19$&6.0&$3.34\times 10^{-5}$&$9.99\times 10^{-1}$&$\textbf{MD}$\\
&&&&$d>0.05$&$116\pm 20$&6.0&$6.52\times 10^{-5}$&$10.0\times 10^{-1}$&$\textbf{MD}$\\
&&&&Metal&$33\pm 72$&0.5&$4.77\times 10^{-2}$&$9.91\times 10^{-1}$&ND\\
&&&&He&$127\pm 20$&6.3&$4.34\times 10^{-5}$&$9.98\times 10^{-1}$&$\textbf{MD}$\\
&&&&Of?p&$78\pm 44$&1.8&$2.8\times 10^{-1}$&$9.2\times 10^{-1}$&ND\\
2012-06-22&6100.9523&$4\times4\times720$&1973 &Optimal&$-1\pm 11$&0.1&$5.23\times 10^{-1}$&$8.14\times 10^{-1}$&ND\\
&&&&$d>0.05$&$-1\pm 11$&0.1&$6.43\times 10^{-1}$&$8.21\times 10^{-1}$&ND\\
&&&&Metal&$-21\pm 41$&0.5&$9.29\times 10^{-1}$&$8.6\times 10^{-1}$&ND\\
&&&&He&$1\pm 11$&0.1&$7.17\times 10^{-1}$&$8.73\times 10^{-1}$&ND\\
&&&&Of?p&$-18\pm 25$&0.7&$9.98\times 10^{-1}$&$9.94\times 10^{-1}$&ND\\
2022-07-07&9767.9648&$3\times4\times530$&1051&Optimal&$45\pm 24$&1.9&$7.76\times 10^{-1}$&$8.59\times 10^{-1}$&ND\\
&&&&$d>0.05$&$46\pm 25$&1.8&$8.74\times 10^{-1}$&$9.03\times 10^{-1}$&ND\\
&&&&Metal&$-29\pm 85$&0.3&$9.67\times 10^{-1}$&$3.04\times 10^{-1}$&ND\\
&&&&He&$52\pm 25$&2.1&$5.66\times 10^{-1}$&$9.34\times 10^{-1}$&ND\\
&&&&Of?p&$-6\pm 50$&0.1&$9.58\times 10^{-1}$&$7.41\times 10^{-1}$&ND\\
2022-07-08&9768.9499&$3\times4\times530$&1937 &Optimal&$84\pm 14$&6.2&$1.06\times 10^{-10}$&$9.99\times 10^{-1}$&$\textbf{DD}$\\
&&&&$d>0.05$&$85\pm 15$&5.6&$5.85\times 10^{-8}$&$10.0\times 10^{-1}$&$\textbf{DD}$\\
&&&&Metal&$187\pm 42$&4.4&$5.68\times 10^{-6}$&$9.32\times 10^{-1}$&$\textbf{DD}$\\
&&&&He&$76\pm 14$&5.4&$1.13\times 10^{-5}$&$9.91\times 10^{-1}$&$\textbf{MD}$\\
&&&&Of?p&$105\pm 26$&4.1&$2.12\times 10^{-5}$&$8.09\times 10^{-1}$&$\textbf{MD}$\\
2022-07-11&9771.9016&$3\times4\times530$&1671 &Optimal&$34\pm 12$&2.7&$9.13\times 10^{-3}$&$7.14\times 10^{-1}$&ND\\
&&&&$d>0.05$&$36\pm 12$&2.9&$1.02\times 10^{-2}$&$8.08\times 10^{-1}$&ND\\
&&&&Metal&$42\pm 43$&1.0&$2.42\times 10^{-2}$&$9.36\times 10^{-1}$&ND\\
&&&&He&$34\pm 13$&2.7&$9.33\times 10^{-2}$&$4.39\times 10^{-1}$&ND\\
&&&&Of?p&$65\pm 28$&2.3&$2.29\times 10^{-3}$&$9.84\times 10^{-1}$&ND\\
\hline
\end{tabular}
\tablecomments{The reported S/N is per 1.8~km\,s$^{-1}$ pixel at 500 nm of the co-added ESPaDOnS spectrum.}
\tablecomments{Marginal (MD) and definite (DD) detection flags are in bold.}
\label{tab:mag_log}
\end{table*}

\subsection{Optical Spectra}
We analyzed 30 Stokes~$I$ spectra in addition to the five ESPaDOnS observations. The observations were obtained between April 2007 and August 2022. Twenty-nine Stokes~$I$ optical spectra were obtained and archived by the IACOB project \citep{simon-diaz_2015, simon-diaz_2020} using the FIES, HERMES and FEROS spectrographs. One additional HERMES spectrum was retrieved from the MELCHIORS database \citep{Royer_2024}. The FIES cross-dispersed echelle spectrograph is mounted on the 2.5~m Nordic Optical Telescope at Observatorio del Roque de los Muchachos, La Palma \citep{telting_2014}. We analyzed FIES spectra obtained in medium-resolution mode ($R=46,000$), which cover a spectral range of $370-830$~nm. The high-resolution ($R=85,000$) HERMES spectrograph is attached to the 1.2~m  Mercator telescope on La Palma and covers a spectral range of $380-900$~nm \citep{raskin_2011}. The FEROS spectrograph is mounted on the MPG/ESO 2.2~m telescope at La Silla Observatory in Chile and has a spectral resolution of $R=48,000$ and a spectral range of $360-920$~nm \citep{kaufer_1999}. The peak S/N of the FIES, HERMES and FEROS spectra range from approximately 150 to 400. Table~\ref{tab:obs_log} provides a log of the analyzed Stokes~$I$ spectra including the S/N at a uniform wavelength of 500\,nm.

\subsection{Time Series Photometry}\label{sec:obs_photometry}
We retrieved archival Hipparcos, All-Sky Automated Survey for Supernovae (ASAS-SN) and K2 time series photometry of 63~Oph in an effort to help constrain the rotation period and look for evidence of a binary companion. $Gaia$ epoch photometry has not yet been released for 63~Oph. The full $Gaia$ photometric catalogue is expected to be released in $Gaia$ DR4 in mid 2026 or later. No \textit{TESS} \citep{ricker_2015} observations of 63~Oph were released at the time of writing. The first \textit{TESS} observations of 63~Oph were obtained in April and May 2025 (sectors 91 and 92).

\subsubsection{Hipparcos}
Photometric observations of 63~Oph were obtained by the Hipparcos satellite \citep{vanLeeuwen_1997_hipparocs_mission} between 1990 and 1993 \citep{ESA_1997, vanLeeuwen_1997}. The Hipparcos passband $H_{p}$ has an effective wavelength $\lambda_{\mathrm{eff}}$ of approximately 520\,nm and a full width at half-maximum (FWHM) of 230\,nm \citep{Bessell_2000, Rodrigo_2020}. Following \cite{Koen_2002}, we discard all observations with a Hipparcos flag greater than 7, leaving 100 observations over a time span of $\sim1000$\,d.

\subsubsection{ASAS-SN}
ASAS-SN \citep{Shappee_2014, Hart_2023} employs a network of camera stations to image the entire sky on a nightly basis in Johnson $V$-band and Sloan $g$-band. Targets begin saturating at $g\sim12$\,mag, leading to worse photometric precision. \cite{Winecki_2024} designed a neural network pipeline to extract light curves, improving the dispersion over the standard pipeline. We employ this ``Saturated Stars (Machine
Learning)" method as implemented in ASAS-SN Sky Patrol v1.0\footnote{https://asas-sn.osu.edu} to extract $V$ and $g$-band light curves for 63~Oph \citep{Konachek_2017}. As the machine learning method does not provide magnitude uncertainties, the pipeline reports uncertainties of 37\,mmag for 63~Oph's light curves based on typical light-curve dispersions. We removed obvious outliers from both data sets, leaving 115 observations over $\sim550$\,d in the $V$-band and 536 observations over $\sim2200$\,d in the $g$-band light curve.

\subsubsection{Kepler/K2}
63~Oph (\object{EPIC 200069370}) was observed by the Kepler space telescope \citep{Borucki_2010} at 30 min cadence during the K2 mission. The K2 mission was a repurposing of Kepler after the loss of two reaction wheels and observed near-ecliptic fields in campaigns $\sim80$\,d in length \citep{Howell_2014}. 63~Oph was observed under Guest Observer Program GO9923 during Campaign 9 (C9; April - July, 2016), and was included in the K2 Bright Star Survey (HALO) \citep{Pope_2019}. C9 was primarily dedicated to the study of gravitational microlensing events toward the Galactic bulge and had a $\sim3$\,d break near the middle of observations, which split the campaign into C9a and C9b \citep{Barensten_2020}.

The Kepler passband $K_{p}$ spans a wide wavelength range between approximately 420 and 900 nm with $\lambda_{\mathrm{eff}}=598$\,nm and $\mathrm{FWHM}=424$\,nm \citep{Rodrigo_2020}. As Kepler was designed primarily to detect exoplanets around faint stars, the detector saturates at $K_{p}\sim11.3$\,mag \citep{Gilliland_2010}, causing electrons to `bleed' into adjacent pixels along the column. It is challenging to recover accurate light curves of saturated K2 targets due to size limitations of the pixel `postage stamps' and instrumental systematics due to the roll motion of the spacecraft \citep{van_cleve_2016}. \cite{Pope_2019} successfully extracted K2 light curves of bright ($V<7$\,mag) saturated targets using the \texttt{halophot} halo photometry software package \citep{White_2017}. The halo photometry method applies a total variation based regularization to the halo of scattered light surrounding a saturated target. After applying the halo procedure \cite{Pope_2019} applied the \textsc{k2sc} Gaussian process regression code \citep{Aigrain_2016} to correct for small residual systematic errors producing final light curves which are hosted on the Mikulski Archive for Space  Telescopes (MAST) \citep{10.17909/t9-6wj4-eb32}. However, the C9a and C9b target pixel files do not have \texttt{POS\_CORR} information and so \cite{Pope_2019} did not produce \textsc{k2sc}-corrected light curves for C9.

We performed a custom extraction of the 63~Oph's K2 photometry using \texttt{halophot}. We retrieved the C9a and C9b target pixels from MAST and generated a custom mask that excludes the saturated columns and contaminating stars within the halo region. The contaminating stars ($G<18$\,mag) were visually identified using the \texttt{interact\_sky} method in the \texttt{lightkurve} software package \citep{lightkurve_2018}. After reducing the data with \texttt{halophot} we removed the remaining long-term trend using a cubic polynomial. We chose to use a low-order polynomial in an effort to preserve possible long periods. Figure~\ref{fig:K2_reduction} illustrates the data reduction process for the C9a light curve.

\section{Stellar Parameters}\label{sec:stellar_parameters}
\subsection{Distance}\label{sec:distance}
 The \textit{Gaia} DR3 parallax for 63~Oph is $\varpi=1.0376\pm0.1187$\,mas with a high associated renormalised unit weight error (RUWE) of 3.52 \citep{Lindegren_2021_astrometric, Gaia_2023}. The RUWE is a goodness-of-fit (GOF) statistic for the astrometric solution, and values greater than 1.4 indicate poor-quality fits. \cite{Jesus_2023} reported a corrected DR3 parallax of $\varpi_{\mathrm{c}}=1.09632\pm0.48196$\,mas, which includes a correction factor in the uncertainty due to the high RUWE \citep{Jesus_2022_parallax, Jesus_2021_parallax}. The corresponding distance of $912\pm401$\,pc has a large uncertainty, which results in a poorly constrained spectroscopic radius (see Section~\ref{sec:parameters}). In comparison, \textit{Gaia} DR2 reports a parallax of $\varpi=0.9071\pm0.0882$\,mas with a lower RUWE of 0.95 \citep{Gaia_2018, Lindegren_2018}. Accordingly, we adopt the \cite{Bailer_Jones_2018} Bayesian DR2 distance estimate of $1082.3^{+119.6}_{-98.2}$\,pc in our analysis.

We note that the Hipparcos satellite provided early astrometric measurements for 63~Oph. However, a Hipparcos-based distance cannot be inferred as the reported parallax is negative \citep{van_Leeuwen_2007}.

\subsection{Spectroscopic Modelling}\label{sec:parameters}
We derived spectroscopic stellar parameters ($T_{\mathrm{eff}}$, $\log g$, $\log L$, $M_{\mathrm{spec}}$, $R_{\mathrm{spec}}$ and $M_{V}$) for 63~Oph by modelling the 2011-08-28 FIES spectrum. We used the IACOB-GBAT software tool \citep{simon_diaz_2011}, which is based on a grid of \texttt{FASTWIND} stellar atmosphere models \citep{Santolaya-Rey_1997, puls_2005}. The methodology is discussed in detail by \cite{holgado_2018} and \cite{holgado_2019}. The derived parameters are reported in Table~\ref{tab:params_63Oph}. Our $T_{\mathrm{eff}}$ and $\log g$ measurements are consistent with a previous spectroscopic analysis using \texttt{CMFGEN} atmosphere models \citep{martins_2017}. Our determination of $M_{\mathrm{spec}}=17\pm4$\,$M_{\odot}$ is in tension with the evolutionary mass of $M_{\mathrm{evol}}=30$\,$M_{\odot}$ determined by \cite{Petit_2019} and the mass of $\sim48\,M_{\odot}$ we infer in Section~\ref{sec:sHRD} from \cite{Keszthelyi_2022} evolutionary tracks. This result is not unexpected as the evolutionary/spectroscopic mass discrepancy problem is a long-standing unresolved issue in massive star studies (e.g. \citealt{Herrero_1992, Markova_2018}). 

\subsection{Rotation and Macroturbulence}\label{sec:rotation_macro}
We measured the projected rotational velocity $v_{\mathrm{eq}}\sin i$ and macroturbulence $v_{\mathrm{mac}}$ using the GOF method with the IACOB-BROAD tool \citep{Simon_2014}. We assumed a radial-tangential macroturbulent broadening profile and adopted a suitable limb-darkening coefficient of $u=0.3$ \citep{Claret_2000}. We fit the O~\textsc{iii} $\lambda5592$ line for all 35 observations listed in Table~\ref{tab:obs_log} and report the weighted mean values of $v_{\mathrm{eq}}\sin i=48\pm5$\,km\,s$^{-1}$ and $v_{\mathrm{mac}}=101\pm5$\,km\,s$^{-1}$.

From $v_{\mathrm{eq}}\sin i$ and $R_{\mathrm{spec}}$ we estimate a maximum surface equatorial rotation period of $P_{\mathrm{rot}}^{\mathrm{max}}=13\pm2$\,d. The critical rotation velocity $v_{\mathrm{crit}}$ is given by (e.g. \citealt{Ekstrom_2008})
\begin{equation}\label{eq:crit_vel}
    v_{\mathrm{crit}}=\sqrt{\frac{2}{3}\frac{GM_{\star}}{R_{\mathrm{p, crit}}}}
\end{equation}
 where $R_{\mathrm{p, crit}}$ is equal to the polar radius at critical velocity. Assuming that $R_{\mathrm{spec}}$ equals $R_{\mathrm{p, crit}}$ and adopting $M_{\mathrm{spec}}$ for the mass give $v_{\mathrm{crit}}=430\pm50$\,km\,s$^{-1}$. This implies a minimum surface rotation period of $P_{\mathrm{rot}}^{\mathrm{min}}=1.4\pm0.3$\,d.

 In Section~\ref{sec:Balmer_line_analysis} we identify a period of $\sim19.8$\,d as the likely stellar rotation period from an analysis of the H$\alpha$ and H$\beta$ line profiles. The equatorial rotation velocity is given by
 \begin{equation}
     v_{\mathrm{eq}}=\frac{2\pi R_{\mathrm{\star}}}{P_{\mathrm{rot}}}.
 \end{equation}
Adopting $R_{\star}=R_{\mathrm{spec}}$ and $P_{\mathrm{rot}}=19.8$\,d implies $v_{\mathrm{eq}}=31\pm3$\,km\,s$^{-1}$, which is in tension with the measured $v_{\mathrm{eq}}\sin i$. This may imply that $v_{\mathrm{eq}}\sin i$ is overestimated due to limitations in disentangling rotational broadening from other broadening mechanisms when $v_{\mathrm{eq}}\sin i\lesssim 50$\,kms$^{-1}$ \citep{Sundqvist_2013, Simon_2014, Simon_2017}. 

Alternatively, a larger stellar radius of $\sim20R_{\odot}$ can be inferred from magnetic stellar evolution models (see Section~\ref{sec:sHRD}). Assuming $R_{\star}=20\pm2\,R_{\odot}$ gives $P_{\mathrm{rot}}^{\mathrm{max}}=21\pm3\,$d and $v_{\mathrm{eq}} = 54\pm8$\,km\,s$^{-1}$ which are consistent with the inferred $P_{\mathrm{rot}}$ and $v_{\mathrm{eq}}\sin i$.

We further discuss the discrepancy between $R_{\mathrm{spec}}$ and $P_{\mathrm{rot}}$ in Section~\ref{sec:interp_ew_var}.

\subsection{Spectroscopic Hertzsprung-Russell Diagram}\label{sec:sHRD}
\begin{figure}
\includegraphics[width=\linewidth]{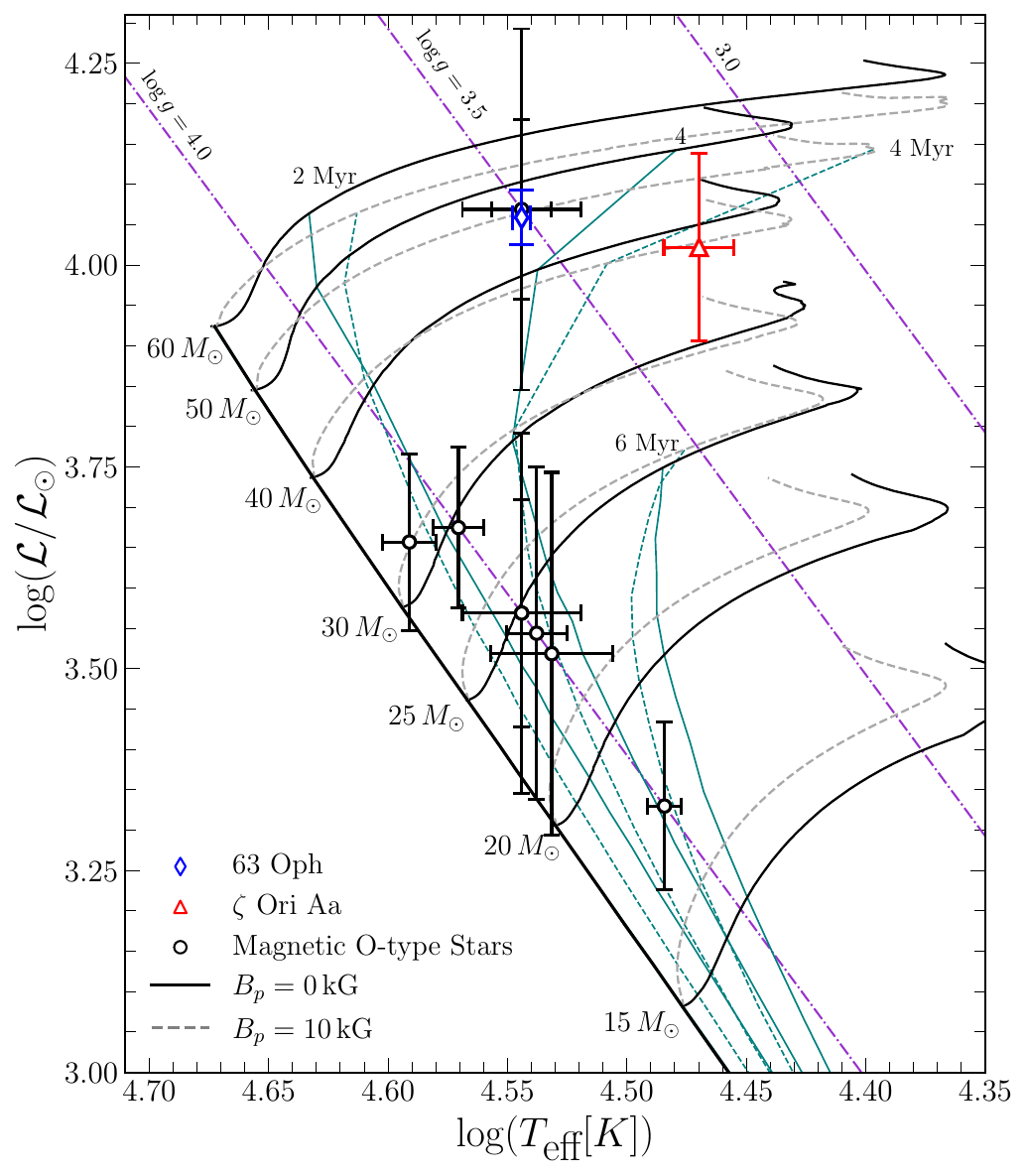}
    \caption{Location of 63~Oph (blue square) in the sHRD ($\mathcal{L}=T_{\mathrm{eff}}^{4}/g$) with illustrative MESA evolutionary tracks computed by \cite{Keszthelyi_2022}. Nonmagnetic ($B_{\mathrm{p}}=0\,\mathrm{kG}$) models are denoted by the solid black lines, and models with an initial dipolar field strength of $B_{\mathrm{p}}=10\,\mathrm{kG}$ are denoted by the dashed gray lines. Solid and dashed teal lines show $B_{\mathrm{p}}=0\,\mathrm{kG}$ and $B_{\mathrm{p}}=10\,\mathrm{kG}$ isochrones respectively. All models were computed at solar metallicity with an initial rotation rate of $\Omega/\Omega_{\mathrm{crit}}=0.5$. The magnetic O supergiant $\zeta$~Ori Aa (HD 37742) and other confirmed magnetic O-type stars are plotted for reference. Note that the location of 63~Oph overlaps with HD~108 and HD~191612, and CPD-28$^{\circ}$~2561 overlaps with NGC 1624-2 in the sHRD. The 1$\sigma$ error bars for 63~Oph are shown in blue. The dash-dotted purple lines denote constant $\log g$ of 4.0, 3.5 and 3.0.}
    \label{fig:sHRD}
\end{figure}
In Figure~\ref{fig:sHRD} we show the location of 63~Oph in the spectroscopic Hertzsprung-Russell Diagram (sHRD; \citealt{Langer_2014}) with illustrative magnetic evolutionary tracks and isochrones\footnote{\url{https://zenodo.org/records/7069766}} computed by \cite{Keszthelyi_2022} using the Modules for Experiments in Stellar Astrophysics (MESA) stellar evolution code \citep{Paxton_2011, Paxton_2019}. The evolutionary tracks were computed at solar metallicity with an initial rotation rate of $\Omega/\Omega_{\mathrm{crit}}=0.5$ using the INT/Mix1 scheme. The INT/Mix1 models assume dipolar magnetic field configuration, which includes internal magnetic field braking and the default rotational mixing scheme commonly used in MESA models. The spectroscopic luminosity of 63~Oph is calculated from $\mathcal{L}=T_{\mathrm{eff}}^{4}/g$, which avoids distance uncertainties present in the calculation of the bolometric luminosity $L$. For comparison, we show the sample of confirmed magnetic O-type stars discussed by \cite{grunhut_2017}, excluding Plaskett's star, a peculiar binary system whose parameters are not yet firmly established \citep{Grunhut_2022}. We highlight the location of $\zeta$~Ori~Aa in the sHRD as it is the most evolved confirmed magnetic O star \citep{bouret_2008, Blazere_2015}. We discuss additional evolved magnetic O star candidates in Section~\ref{sec:analogues}. Table~\ref{tab:magOstars} provides $\mathcal{L}$ values determined from literature $T_{\mathrm{eff}}$ and $\log g$ measurements. We calculated $\mathcal{L}_{\odot}$ using the recommended solar values from \cite{Prsa_2016}. Note that 63~Oph overlaps with HD~108 and HD~191612 in the sHRD ($T_{\mathrm{eff}}=35$\,kK, $\log g=3.5$). In Section~\ref{sec:discuss_transitional_object} we discuss how these two stars' spectroscopic, magnetic and rotational properties differ from 63~Oph. 

The \cite{Keszthelyi_2022} model tracks imply a stellar mass of $\sim48-50\,M_{\odot}$ and radius of $\sim20\,R_{\odot}$ assuming a dipolar magnetic field strength of $1-10$\,kG on the zero-age main sequence (ZAMS). \cite{Petit_2019} previously inferred $M_{\mathrm{evol}}=30\,M_{\odot}$ and $R_{\mathrm{evol}}=13.41\,R_{\odot}$ for 63~Oph using \cite{Brott_2011} stellar models and the BONNSAI Bayesian code \citep{Schneider_2014}. The mass and radius reported by \cite{Petit_2019} are discrepant with those we infer from the \cite{Keszthelyi_2022} models, likely due to differences between the two stellar evolution codes.

We note that the evolutionary tracks and isochrones in Figure~\ref{fig:sHRD} are primarily illustrative and are shown to provide context. Magnetic stellar evolutionary tracks have not yet been well calibrated against the magnetic O star sample. Additionally, the initial distributions of magnetic field strength and rotational velocity are poorly constrained for magnetic O-type stars. Massive star evolutionary tracks have varying degrees of sensitivity to modelling assumptions such as the magnetic field geometry, chemical mixing scheme, mass-loss prescription and implementation of convective overshoot \citep{Brott_2011, Ekstrom_2012, Keszthelyi_2019, Agrawal_2022, Keszthelyi_2022}. Possible binary interactions and stellar mergers add additional modelling complications. We leave detailed evolutionary modelling of 63~Oph for future work once the magnetic field has been fully characterized.

\section{Spectropolarimetric Analysis}\label{sec:specpol_analysis}
\subsection{Least-Squares Deconvolution}\label{sec:LSD}
We applied the least-squares deconvolution (LSD) procedure \citep{donati_1997, kochukhov_2010} to the coadded ESPaDOnS polarimetric spectra. This multiline technique increases the effective S/N of the Stokes~$V$ spectra, allowing for the detection of weak magnetic signatures \citep{donati_2009}. The LSD procedure cross-correlates a line mask with the observed spectrum, providing mean LSD Stokes~$I$, $V$ and diagnostic null $N$ profiles. 

We obtained a suitable line list from the VALD3 atomic line database \citep{piskunov_1995, ryabchikova_2015} via the `extract stellar' request, with parameters of $T_{\mathrm{eff}}=35,000$\,K, $\log g =4.0$ and a line depth threshold of 0.01. The LSD line mask was computed from the VALD line list using the \texttt{makeMask} function in the SpecpolFlow\footnote{\url{https://github.com/folsomcp/specpolFlow}} Python software package \citep{Folsom_2025}. For lines without experimentally measured values, the effective Land\'e factor $g_{\mathrm{eff}}$ was estimated from the electron configuration (when provided) for LS, JJ and JK coupling \citep{landi_2004}. For the remaining lines where the Land\'e factor could not be determined from the VALD data, we set $g_{\mathrm{eff}}=1.1$, which is the median value of all lines. The line mask depths for each observation were adjusted using the `tweak' fitting routine in the interactive IDL code described by \cite{grunhut_2017}. Balmer lines, emission lines, and strongly contaminated telluric regions were excluded from the mask, leaving $\sim 400$ lines with nonzero adjusted depths between 370 and 750\,nm.

It can be challenging to detect magnetic fields in O-type stars as their spectra contain relatively few strong spectral lines at optical wavelengths. As the longitudinal field strength of 63~Oph is relatively weak for a magnetic O star, choices in the construction of the line mask could impact the S/N of the LSD~Stokes~$V$ profile and, therefore, its detectability and associated $\langle B_{Z} \rangle$ measurement. Similar to \cite{grunhut_2017}, we tested several line mask constructions in addition to the optimal `cleaned and tweaked' line mask described above. From the optimal mask, we generated a line mask with a depth cutoff threshold of $d>0.05$ and masks consisting of only metal and only He lines. In addition, we tested the tailored 12-line mask originally utilized by \cite{donati_2006_HD191612} to obtain a detection in HD~191612. This line mask consists of He and CNO lines and has been successfully used to obtain detections in several magnetic O-type stars (e.g. \citealt{wade_2012}). Following \cite{grunhut_2017}, we refer to this mask as the ``Of?p mask."

\begin{figure*}
\centering
\includegraphics[width=\textwidth]{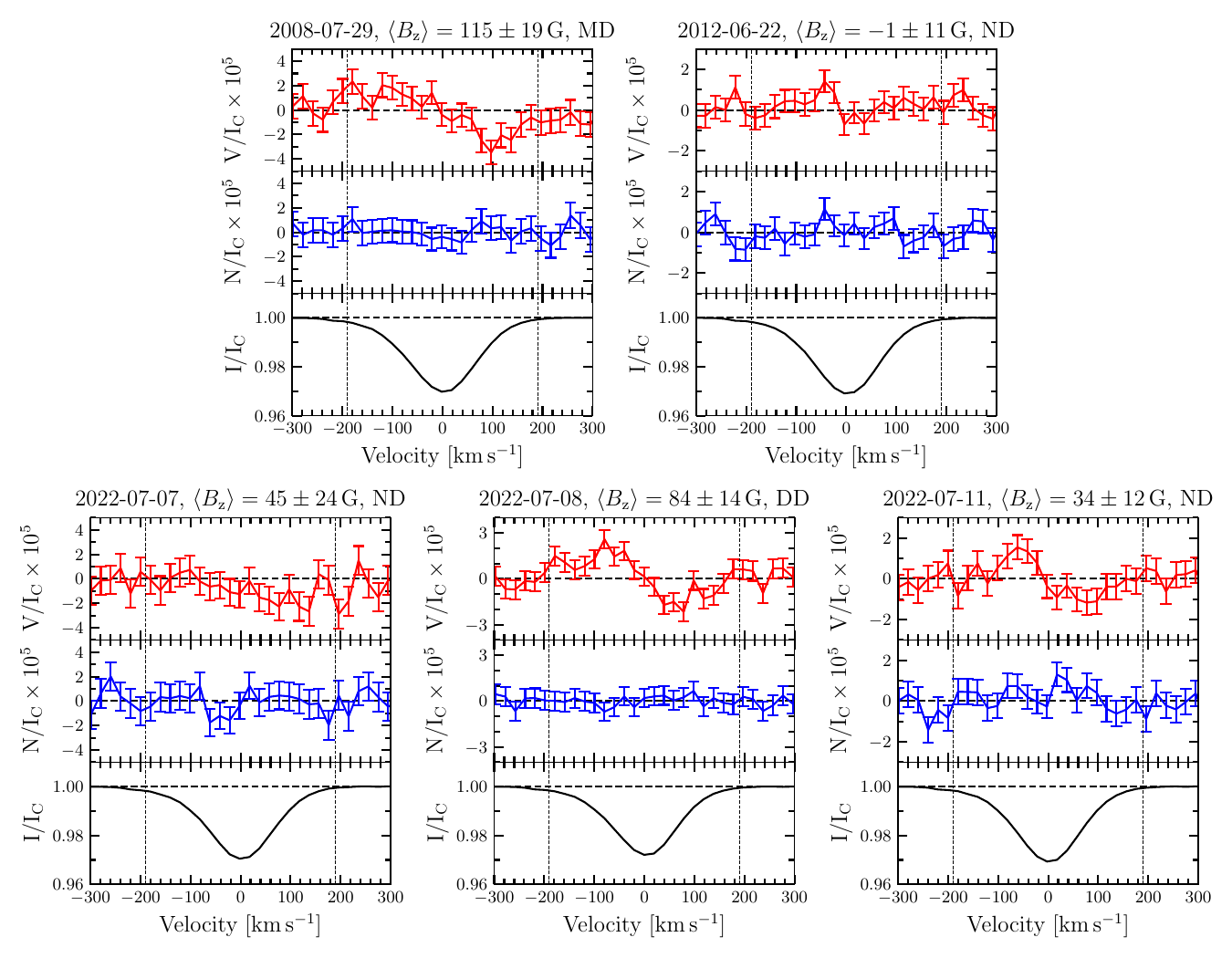}
    \caption{Continuum-normalized LSD profiles computed from the five ESPaDOnS observations using the optimal line mask. From top to bottom, LSD Stokes~$V$ (red), $N$ (blue) and $I$ (black). The dashed vertical lines indicate the velocity bounds used in the magnetic analysis to calculate $\langle B_{z}\rangle$ and the FAP.}
    \label{fig:lsd_plots}
\end{figure*}

The LSD profiles were computed with the {\sc iLSD} code \citep{kochukhov_2010}. We adopt LSD normalization parameters of $\lambda_{0}=500$\,nm, $g_{0}=1.2$ and $d_{0}=0.1$ for the wavelength, effective Land\'e factor and line depth respectively. Following \cite{grunhut_2017} we adopted an LSD velocity binning of 19.8\,km\,s$^{-1}$ and a regularization parameter of 0.2. The regularization parameter increases the S/N of the LSD Stokes~$V$ profile, allowing for the detection of weak magnetic signatures. The LSD profiles were shifted to the rest frame and normalized by a linear fit to the continuum of the LSD Stokes~$I$ profiles. The metal and He LSD profiles were computed simultaneously using the multiprofile fitting capability of {\sc iLSD}. LSD profiles computed with the optimal line mask are shown in Figure~\ref{fig:lsd_plots}. Figure~\ref{fig:LSD_mask_tests} shows the LSD profiles computed with the additional test line masks. We discuss the results of the line mask tests in Section~\ref{sec:line_mask_tests}. We refer to magnetic results computed from the optimal mask LSD profiles throughout the rest of the paper unless otherwise stated. 

\subsection{Diffuse Interstellar Band at 5797\,$\mathrm{\AA}$}\label{sec:DIB}
In Figures~\ref{fig:lsd_plots} and  \ref{fig:LSD_mask_tests} there is a weak absorption feature in the far blue wing ($\sim-220$\,km$\,\textrm{s}^{-1}$) of the LSD Stokes~$I$ profiles. It is seen most prominently in Stokes~$I$ profiles computed from the metal and Of?p masks. We determined the origin of this feature to be a diffuse interstellar band (DIB) located at 5797\,$\mathrm{\AA}$ near the C~\textsc{iv}~$\lambda$5801 line \citep{Herbig_1975, Fan_2019}. This DIB is known to impact O star LSD Stokes~$I$ profiles (e.g. \citealt{wade_2012, wade_ngc1624}). It is important to note that the DIB does not impact the circularly polarized Stokes~$V$ profile. We experimented with removing the C~\textsc{iv}~$\lambda$5801 from the line mask but found this gave higher false alarm probabilities (FAP). We determined that it is necessary to include the C~\textsc{iv}~$\lambda$5801 line to obtain magnetic detections, likely because it is relatively deep for a metal line and has a $g_{\mathrm{eff}}$ above the median mask value. Following previous studies we have kept C~\textsc{iv}~$\lambda$5801 included in the masks.

\subsection{Longitudinal Magnetic Field}\label{sec:Bz}
We computed surface-averaged longitudinal magnetic field measurements $\langle B_{\mathrm{z}}\rangle$ from the RV-corrected LSD Stokes~$V$ profiles using the first-moment method (Eq.~1 of \citealt{wade_2000}) as implemented in the \texttt{calc\_bz} routine in SpecpolFlow. The $\langle B_{\mathrm{z}}\rangle$ measurements were computed integrating between -190 and 190 km\,s$^{-1}$ and are recorded in Table~\ref{tab:mag_log}. The associated 1$\sigma$ uncertainties were calculated using formal propagation of the LSD Stokes~$V$ uncertainties. Our $\langle B_{\mathrm{z}}\rangle$ measurements for the 2008 and 2012 LSD~Stokes~$V$ profiles computed with the optimal mask are consistent with those of \cite{grunhut_2017}. The 2008-07-29 ($\langle B_{\mathrm{z}}\rangle = 115\pm19$\,G) and 2022-07-08 ($\langle B_{\mathrm{z}}\rangle=84\pm14$\,G) longitudinal magnetic fields are both detected at the $6\sigma$ level. Of the new observations, only the 2022-07-08 spectra has a $\langle B_{\mathrm{z}}\rangle$ measurement $>3\sigma$.

\subsection{False Alarm Probability}\label{sec:AFP}
Magnetic signatures in LSD~Stokes~$V$ profiles are typically diagnosed from the FAP metric described by \cite{Donati_1992}. An LSD Stokes~$V$ signature is determined to be definitely detected (DD) for an FAP$<10^{-5}$, marginally detected (MD) for an FAP between $10^{-5}$ and $10^{-3}$ and not detected (ND) for an FAP$>10^{-3}$ \citep{donati_1997}. While these thresholds are somewhat arbitrary, they tend to be regarded as reasonable, conservative markers. We computed the FAP for each LSD~Stokes~$V$ profile with the velocity bounds used in the $\langle B_{\mathrm{z}}\rangle$ calculation. We obtain a MD ($\textrm{FAP}=3.34\times 10^{-5}$) and ND in the 2008 and 2012 Stokes~$V$ profiles, respectively, consistent with \cite{grunhut_2017}. The 2022-07-08 LSD~Stokes~$V$ profile is DD. The 2022-07-07 and 2022-07-11 Stokes~$V$ profiles are both formally NDs. However, a coherent variation of Stokes $V$ suggestive of a weak magnetic signature consistent with those observed on 2008-07-29 and 2022-07-08 is apparent in the 2022-07-11 profile (Figure~\ref{fig:lsd_plots}). 

We performed a similar FAP analysis of the diagnostic null $N$ profiles. No signatures were detected in the $N$ profiles, which indicates that instrumental systematics are not the likely origin of the Stokes~$V$ signatures. The FAPs for the LSD Stokes~$V$ and $N$ profiles are provided in Table~\ref{tab:mag_log}. 

\subsection{Line Mask Tests}\label{sec:line_mask_tests}
In addition to the optimal line mask, we tested four other line masks: an adjusted depth threshold of $d>0.05$, metal line only, He line only, and the \cite{donati_2006_HD191612} Of?p mask. We measured longitudinal fields and FAPs (Table~\ref{tab:mag_log}) from the test mask LSD profiles (Figure~\ref{fig:LSD_mask_tests}) using the same procedures described in Sections~\ref{sec:Bz} and \ref{sec:AFP}. Typically, the optimal line mask provides the lowest FAPs and highest significance of the longitudinal field measurement ($|\langle B_{z} \rangle|/\sigma_{B}$). The optimal, $d>0.05$, and He masks all yield MDs for the 2008-07-29 observation. For the 2022-07-28 observation, the optimal, $d>0.05$ and metal masks are DDs and the He and Of?p masks are MDs. While test masks do not give any formal MDs or DDs for the 2022-07-11 observation, the Of?p mask gives $\mathrm{FAP}=2.29\times10^{-3}$ which is near the upper threshold of a MD. 

As seen in Figure~\ref{fig:LSD_mask_tests}, the He LSD profiles are broader compared to the metal-only LSD profile. The differences in profile shapes are reflected in the $\langle B_{z} \rangle$ measurements. For example, the 2022-07-08 metal mask gives $\langle B_{z} \rangle=187\pm42$\,G compared to $\langle B_{z} \rangle=76\pm14$\,G for the He mask. Overall, for MD or DD LSD Stokes~$V$ profiles we obtain $\langle B_{z} \rangle$ measurements of $\sim100-200$\,G. The improved FAPs of the optimal line mask compared to the $d>0.05$ mask suggest that there is detectable polarization signal in the shallow lines.   

\subsection{Dipolar Field Strength}\label{sec:Bd}
For an oblique dipole, the rotational variation of $\langle B_{z}\rangle$ can be expressed as a function of the limb-darkening coefficient, the dipolar field strength $B_{\mathrm{p}}$ and the inclination ($i$) and obliquity ($\beta$) angles \citep{preston_1967}. As we have only snapshot measurements of $\langle B_{z}\rangle$, we cannot accurately determine $i$ and $\beta$. We can establish a lower bound on $B_{\mathrm{p}}$ by taking the dipolar magnetic axis to be aligned with the line of sight at $\langle B_{z}^{\mathrm{max}}\rangle$. Adopting $u=0.3$ \citep{Claret_2000} a lower bound on $B_{\mathrm{p}}$ is given by $B_{\mathrm{p}}\geq3.53\,\langle B_{z}^{\mathrm{max}}\rangle$ (from Eq.~1 of \citealt{preston_1967}). 

The maximum $\langle B_{z} \rangle$ measurement ($115\pm19$\,G) comes from the 2008-07-29 observation and implies $B_{\mathrm{p}}\geq406\pm67$\,G. However, as the 2008-07-29 LSD Stokes~$V$ signature is a MD, we adopt $\langle B_{z}^{\mathrm{max}}\rangle=84\pm14$\,G measured from the 2022-07-08 definite detection. This sets a conservative lower bound of $B_{\mathrm{p}}\geq300\pm50$\,G. 

As we have determined a lower limit for $B_{\mathrm{p}}$, we can only set lower limits on the wind confinement parameter $\eta_{\star}$ and the Alfvén radius $R_{\mathrm{A}}$ \citep{ud-Doula_2002, udDoula_2008, petit_2013}. Additionally, the high uncertainties on the stellar mass and theoretical mass-loss rate lead to a percent uncertainty of $\sim50\%$ on the lower bound for $\eta_{\star}$. We leave full determination of 63~Oph's magnetospheric parameters for future work once $B_{\mathrm{p}}$ and $M_{\star}$ are better constrained. 

\section{Variability Analysis}\label{sec:var_analysis}
\subsection{Line Profile Variability}\label{sec:line_prof_var}
The variability of magnetic O-type stars is typically understood within a framework in which the dipole’s axis is tilted relative to the star’s rotation axis. This misalignment causes the magnetospheric density to vary along the line of sight during rotation, leading to spectroscopic variability, including variable Balmer and He line emission \citep{petit_2013, ud-Doula_2013}. In Figure~\ref{fig:espadons_grid} we plot selected H, He, and metal line profiles for the five ESPaDOnS observations. The Balmer lines show the strongest variability, and weak variability is seen in the He and metal lines. One H$\alpha$ profile is partially in emission and corresponds to the definite magnetic detection (2022-07-08). All other H and He line profiles are principally in absorption. We do not observe an obvious nebular contribution to the H$\alpha$ line profiles. 

Figure~\ref{fig:Hermes_grid} shows short-term variability observed in HERMES spectra over 8 days. The core and inner wings of the H$\alpha$ and H$\beta$ lines again show the strongest variability. The metal line profiles are relatively constant over the observing span.

The spectral behaviour of 63~Oph is noticeably different from the (ostensibly physically similar) Of?p stars HD~108 and HD~191612 which show strong emission in H$\alpha$, He \textsc{ii} $\lambda$4686 and He~\textsc{i}~$\lambda$5876 at certain phases during their rotational cycles \citep{Walborn_1972, GREGOR_2023}. Additionally, 63~Oph does not show the characteristic spectral signature of the Of?p phenomenon — C~\textsc{iii}~$\lambda\lambda$4647-4650-4651 emission with comparable strength to nearby N~\textsc{iii}~$\lambda\lambda$4634-4640-4642 \citep{Walborn_1972}.

\subsection{Period Analysis Methodology}\label{sec:period_analysis_method}
In Sections~ \ref{sec:Balmer_line_analysis}, \ref{sec:RV_main} and \ref{sec:photometry_main}, we present the period analysis performed on the Balmer line equivalent widths (EWs), RVs, and photometric time series. We are primarily interested in identifying coherent, low-frequency ($\lesssim1\,\mathrm{d}^{-1}$) signals that can arise from rotational modulation or a binary companion. High-precision light curves of evolved O-type stars show ubiquitous `red-noise' or stochastic-low-frequency variability \citep{Bowman_2020, Burssens_2020} which can make it challenging to confidently detect coherent, low-amplitude frequencies $\lesssim1\,\mathrm{d}^{-1}$.

We analyzed all of the time series using the \texttt{astropy} \citep{astropy_2022} implementation of the generalized Lomb-Scargle (GLS) periodogram \citep{Zechmeister_2009, vanderplas_2018}. The GLS method modifies the traditional Lomb-Scargle method \citep{Lomb_1976, Scargle_1982} by adding an offset term to the sinusoidal model fits and incorporating measurement uncertainties in the $\chi^{2}$ calculation. Each periodogram was computed with an oversampling factor of 10 over a frequency grid of [2/$\Delta T,\,1]\,\mathrm{d}^{-1}$ where $\Delta T$ is the length of the time series. The lower-frequency bound ensures that all periods are sampled at least twice in each time series.

We first searched for significant frequencies using the traditional FAP. We adopted a conservative threshold of $\mathrm{FAP}=0.01$, which was estimated for each periodogram using the \cite{Baluev_2008} approximation implemented in \texttt{astropy}. The FAP provides the probability that a set of observations with no periodic signal would give a peak of a given magnitude, assuming only Gaussian noise. There are limitations to the FAP criterion as it cannot distinguish spurious peaks that arise from aliasing in the window function or from low-frequency red noise. 

We then searched for periods using a standard iterative prewhitening technique with a S/N significance criterion. For each data set we computed the unnormalized periodogram. We converted the spectral power $P_{\mathrm{LS}}(\nu)$ to amplitude $A(\nu)$ using the relation ${A(\nu)=\sqrt{4P_{\mathrm{LS}}(\nu)/N}}$ where $N$ is the number of measurements \citep{aert_2021}. The dominant period in the periodogram was identified, and then, the best-fitting sinusoid with that period was subtracted from the time series. The S/N is given by the ratio between the amplitude of the sinusoidal fit and the periodogram's mean amplitude after prewhitening. This process was repeated until the commonly adopted threshold of ${\mathrm{S/N}>4}$ was no longer satisfied \citep{Breger_1993}. Typically, the S/N criterion identifies fewer periods than the FAP, as it accounts for the SLF variability present in the observations. However, in a short time series, prewhitening may not identify a coherent signal if the amplitude is similar to the level of the SLF variability. 

\begin{figure}
    \centering
    \includegraphics[width=\linewidth]{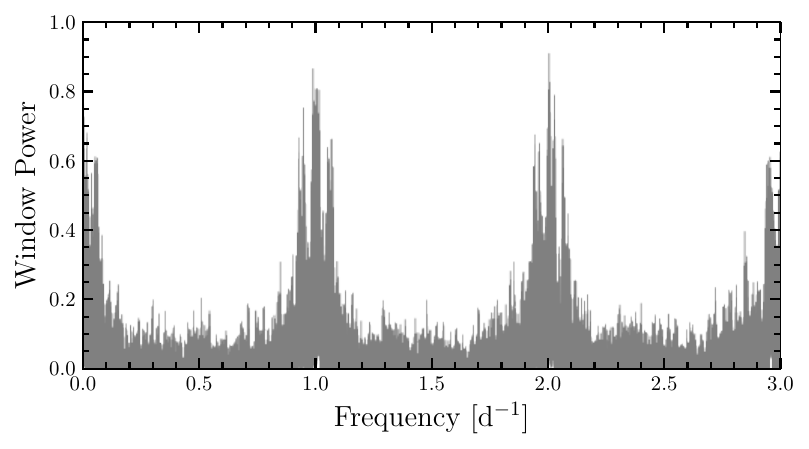}
    \caption{Power spectrum of the observing window for the EW and RV spectroscopic measurements. Strong peaks are seen around periods of 368.7, 1, and 0.5\,d due to yearly and daily aliasing.}
    \label{fig:obs_window}
\end{figure}

Identified frequencies were filtered according to the Loumos-Deeming criterion \citep{Loumos_1978} which gives that two frequencies are unresolved if they are separated by less than $1.5/\Delta T$. Obvious aliases were also filtered from inspection of the observing window periodograms which were computed using the method described in \cite{vanderplas_2018}. Figure~\ref{fig:obs_window} shows the power spectrum of the observing window for the EW and $\mathrm{RV}$ measurements.

\subsection{Balmer Line Variability}\label{sec:Balmer_line_analysis}
\subsubsection{H$\alpha$ Telluric Correction}\label{sec:tel_corr}
The degree of telluric contamination over the H$\alpha$ line region varies between spectra due to the elevation of the observatory and observing conditions. To reduce systematics related to telluric line variability in our analysis of H$\alpha$ we applied telluric corrections using \texttt{TelFit} \citep{Gullikson_2014}. \texttt{TelFit} acts as a Python wrapper for the \texttt{LBLRTM} code (Line By Line Radiative Transfer Model; \citealt{Clough_1992, Clough_2005}) and can perform least-squares fitting for temperature, pressure, and atmospheric abundances. For each spectrum, we retrieved the atmospheric profile closest to the time of observation from the Global Data Assimilation System Archive\footnote{\url{https://www.ready.noaa.gov/READYamet.php}} hosted by the National Oceanic and Atmospheric Administration's Air Resources Laboratory \citep{ROLPH2017210}. The observatory temperature and pressure were retrieved from spectrum FITS files when available. For the FIES observations, we set these initial parameters based on archival meteorological data\footnote{\url{https://www.not.iac.es/weather/}}. 

\begin{figure}
    \centering
    \includegraphics[width=\linewidth]{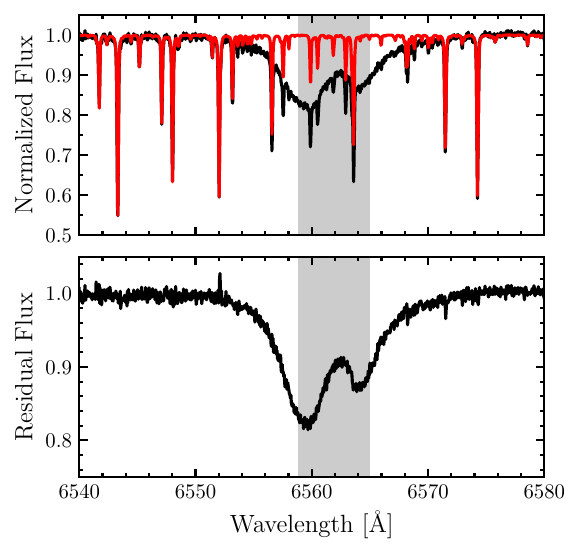}
    \caption{Example telluric correction to the strongly contaminated H$\alpha$ region of the 2022-08-30 HERMES spectra. The top panel shows the uncorrected flux with the model fit (red line). The bottom panel shows the residual flux after dividing by the telluric model. The shaded grey area indicates the region used to calculate the line-core H$\alpha$ EW measurements.}
    \label{fig:example_tel_fit}
\end{figure}

Similar to \cite{Gullikson_2014}, we fit for the humidity in four spectral regions dominated by water bands ($588-595$\,nm, $647-655$\,nm, $695-703$\,nm and $712-730$\,nm) and adopted the weighted average. The final telluric model was computed over the H$\alpha$ region and then divided out of the observed flux to produce a telluric corrected H$\alpha$ profile. Figure~\ref{fig:example_tel_fit} shows an example telluric correction. The residual telluric amplitudes are typically less than $3\%$ of the continuum.

\subsubsection{EW Measurements}\label{sec:EW_measurements}
To search for evidence of rotational modulation, we measured H$\alpha$ and H$\beta$ line-core EWs for all spectra. Each H$\alpha$ and H$\beta$ line profile was locally renormalized using a linear fit to the nearby continuum. The Balmer lines consist of an underlying photospheric absorption profile and a circumstellar emission component. To help increase the $\mathrm{S/N}$ of our EW measurements, we selected computation bounds centered on the variable emission near the line cores. The line regions outside of the bounds primarily consist of the stable absorption component. Restricting the measured regions to the line core also helps reduce the impact of systematics from normalization and residual telluric contamination. The H$\alpha$ EWs were computed between 6558.8 and 6565.8\,$\mathrm{\AA}$ and the H$\beta$ EWs were computed between 4859.0 and 4863.0\,$\mathrm{\AA}$.  The EWs and associated 1$\sigma$ uncertainties were calculated using the \texttt{specutils} Python package \citep{nicholas_earl_2024_14042033} and are recorded in Table~\ref{tab:obs_log}. The 2011-09-10 FIES spectrum was excluded from the H$\beta$ EW measurements due to an apparent instrumental artifact that prevented accurate normalization of the line profile.

As we expect any nebular contribution to be nonvariable, we assume it is not a significant source of systematic uncertainty in our EW analysis.

\subsubsection{EW Period Analysis}\label{sec:EW_period_analysis}
From the H$\alpha$ EW power spectrum, we identified two peaks at 19.7556\,d and 19.8675\,d with $\mathrm{FAP}<0.01$. The H$\beta$ EW power spectrum has one significant peak at 19.8675\,d. The next highest peak is at 19.7556\,d and is close to the FAP threshold. The prewhitening procedure identifies a period at $\sim19.76$\,d in both the H$\alpha$ and H$\beta$ EWs with a S/N of 7.0 and 6.5 respectively. The detection of the $\sim19.8$\,d period in H$\alpha$ and H$\beta$ with both significance criteria provides strong evidence that the period is real.

\begin{figure*}
    \centering
    \includegraphics[width=\linewidth]{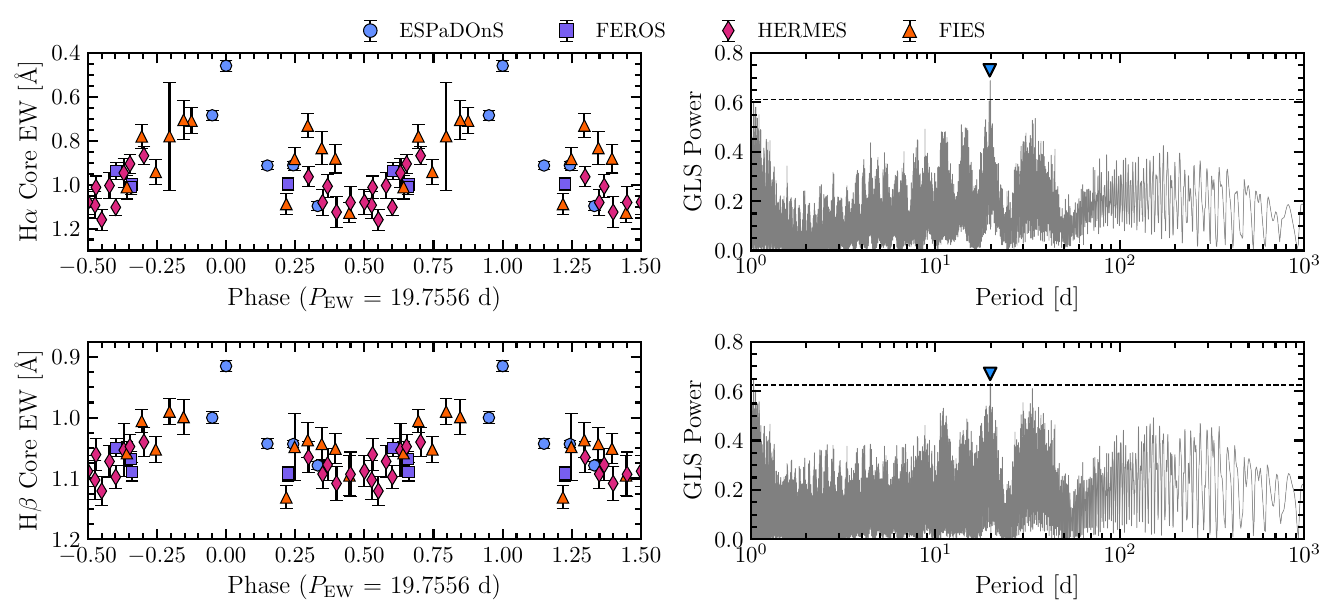}
    \caption{\textbf{Left}: Two cycles of H$\alpha$ and H$\beta$ core EW measurements phased to the most significant period in the H$\alpha$ EW periodogram ($P=19.7556$\,d). Error bars correspond to 1$\sigma$ uncertainties. \textbf{Right}: Corresponding GLS periodograms for the EW measurements. The dashed horizontal black line denotes the $\mathrm{FAP}=0.01$ level. Blue triangles denote the significant peaks.}
    \label{fig:Ha_Hbeta_EW}
\end{figure*}

We do not have sufficient observations to differentiate between the 19.7556\,d and 19.8675\,d periods. We adopt $P_{\mathrm{EW}}=19.7556$\,d as the fiducial period as it has the highest peak in the H$\alpha$ power spectrum. Figure~\ref{fig:Ha_Hbeta_EW} shows the H$\alpha$ and H$\beta$ EW measurements phased to $P_{\mathrm{EW}}$ as well as the corresponding GLS periodograms. The EWs were phased using an ephemeris of $\mathrm{HJD}_{0}=2459768.94994$ (2022-07-08) which corresponds to the date of maximum observed emission (minimum EW).

\subsubsection{Dynamic Spectra}\label{sec:Dynamic_sepctra}
\begin{figure*}
  \centering
  \includegraphics[width=0.49\textwidth]{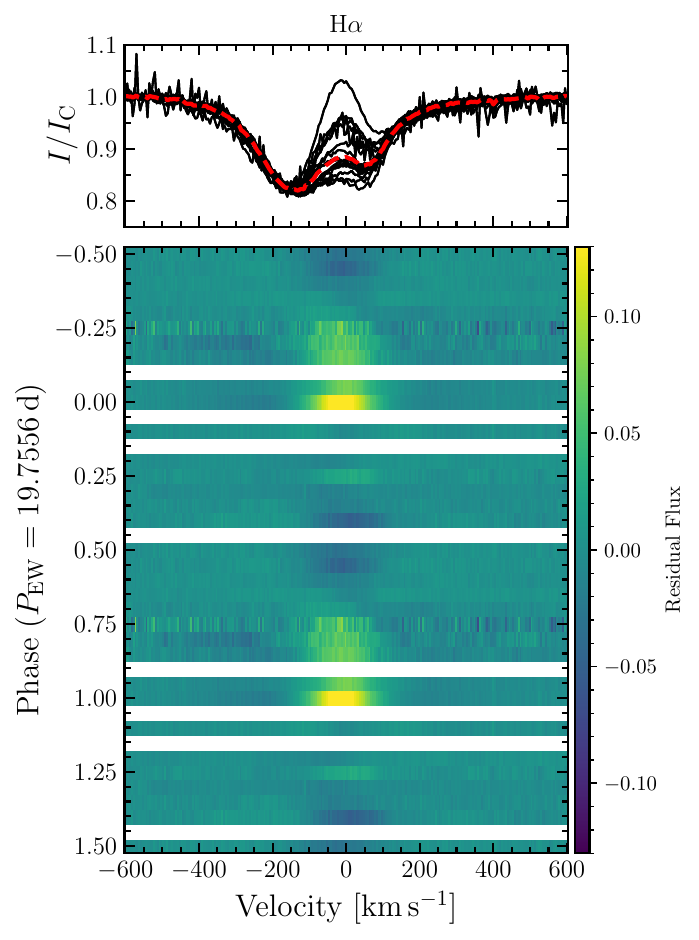}
  \hfill
  \includegraphics[width=0.49\textwidth]{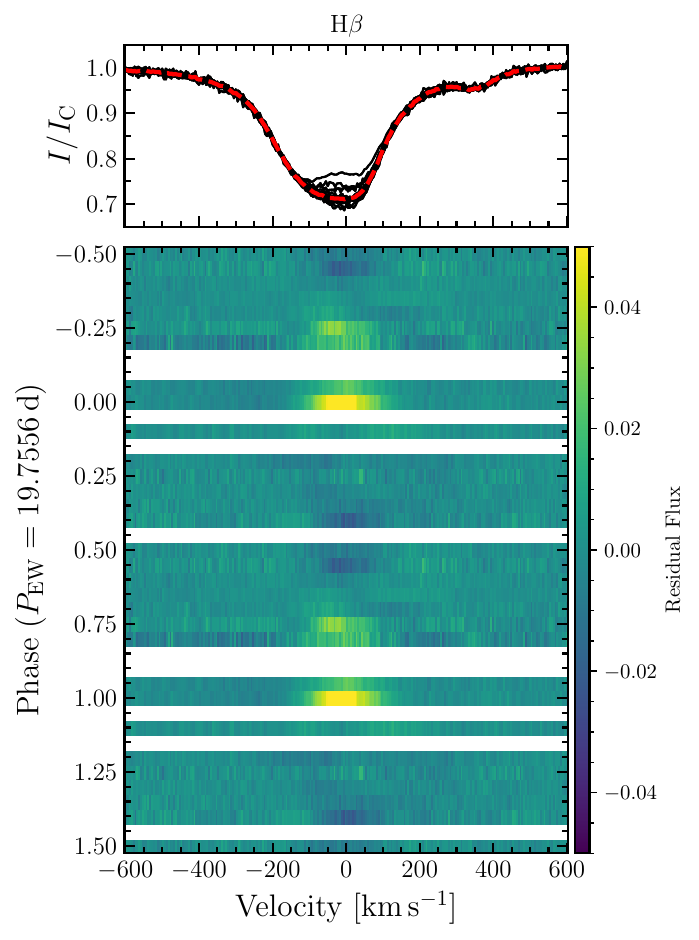}
  \caption{Dynamic plots of the H$\alpha$ and H$\beta$ line profiles phased to $P_{\mathrm{EW}}=19.7556\,$d. The red dashed lines denote the mean flux. The dynamic spectra of the residual mean subtracted flux is plotted over two cycles in phase.}
  \label{fig:dynamic_spectra}
\end{figure*}

In Figure~\ref{fig:dynamic_spectra} we plot the phased dynamic spectra of the H$\alpha$ and H$\beta$ line profiles to help visualize the periodicity. The normalized fluxes were interpolated onto a uniform velocity grid with a step size of $\Delta v=5\,\mathrm{km}\,\mathrm{s}^{-1}$, and then binned in phase with a bin size of $\Delta\phi=0.05$. A periodic emission feature is observed in the core of both the H$\alpha$ and H$\beta$ line profiles. The phasing of the dynamic spectra is relatively coherent despite the spectra being obtained over multiple period cycles. The phase gaps in the dynamic spectra highlight the need for additional observations around maximum emission. Ultimately, we conclude that the dynamic spectra support an interpretation of $P_{\mathrm{EW}}$ as a real, coherent period.

\subsubsection{Interpreting the Balmer Line Variability}\label{sec:interp_ew_var}
As in previous magnetic O-type star studies, we attribute the $\sim19.8$\,d Balmer line variability to a rotating magnetosphere. The single-waved EW variation given by $P_{\mathrm{EW}}$ implies a dipolar magnetic field geometry where only one magnetic pole is visible over the rotation cycle. The strictly positive $\langle B_{z}\rangle$ detections are consistent with this interpretation. While our period analysis identifies a single-waved variation, a double-waved variation at 2P$_{\mathrm{EW}}$ is not conclusively excluded due to insufficient phase coverage.

Although we interpret $P_{\mathrm{EW}}$ as the likely stellar rotation period, we note that $P_{\mathrm{EW}}$ is in $>3\sigma$ tension with $P_{\mathrm{rot}}^{\mathrm{max}}=13\pm$\,2\,d. As mentioned in Section~\ref{sec:rotation_macro} the most probable explanation is that the commonly employed GOF method provides an overestimation of $v_{\mathrm{eq}}\sin i$. This effect is most clearly seen in slowly rotating magnetic O stars where the rotation period and inclination angle can be inferred from magnetic measurements \citep{Sundqvist_2013}. Alternatively, it is possible that the Balmer lines do not primarily trace the surface rotation period. \cite{Buysschaert_2017} identified a dominant period of $\sim10$\,d in the P-Cygni H$\alpha$ profiles of $\zeta$~Ori~Aa, which is longer than the $\sim6.8$\,d surface rotation period inferred from $\langle B_{\mathrm{z}}\rangle$ measurements \citep{Blazere_2015}. The longer circumstellar period traced by H$\alpha$ is likely due to $\zeta$~Ori~Aa's weak dipole field strength ($B_{\mathrm{p}}=142$\,G) and possibly absent magnetosphere. Additional spectropolarimetric observations of 63~Oph will allow the $\langle B_{z}\rangle$ and H$\alpha$ periods to be inferred independently and may resolve this apparent discrepancy.

\subsection{Radial Velocities}\label{sec:RV_main}
\subsubsection{RV Measurement}\label{sec:RV_measurement}
As LSD is a cross-correlation technique, RVs can be derived using LSD profiles computed from Stokes~$I$ spectra. We generated a tailored line mask for the RV analysis as the IACOB spectra have a lower S/N than the ESPaDOnS spectra, and there are differences in the usable wavelength ranges between instruments. We excluded the C~\textsc{iv}~$\lambda$5801 line due to contamination from the nearby DIB (see Section~\ref{sec:DIB}). The full RV mask contained 63 lines between 3810 and 6690\,$\mathrm{\AA}$. All lines included in the mask had depths greater than 5\% of the continuum. As some He lines are observed to be variable, we also constructed metal-only (34 lines) and He only (29 lines) masks for a comparison.

\begin{figure*}
     \centering
     \includegraphics[width=\linewidth]{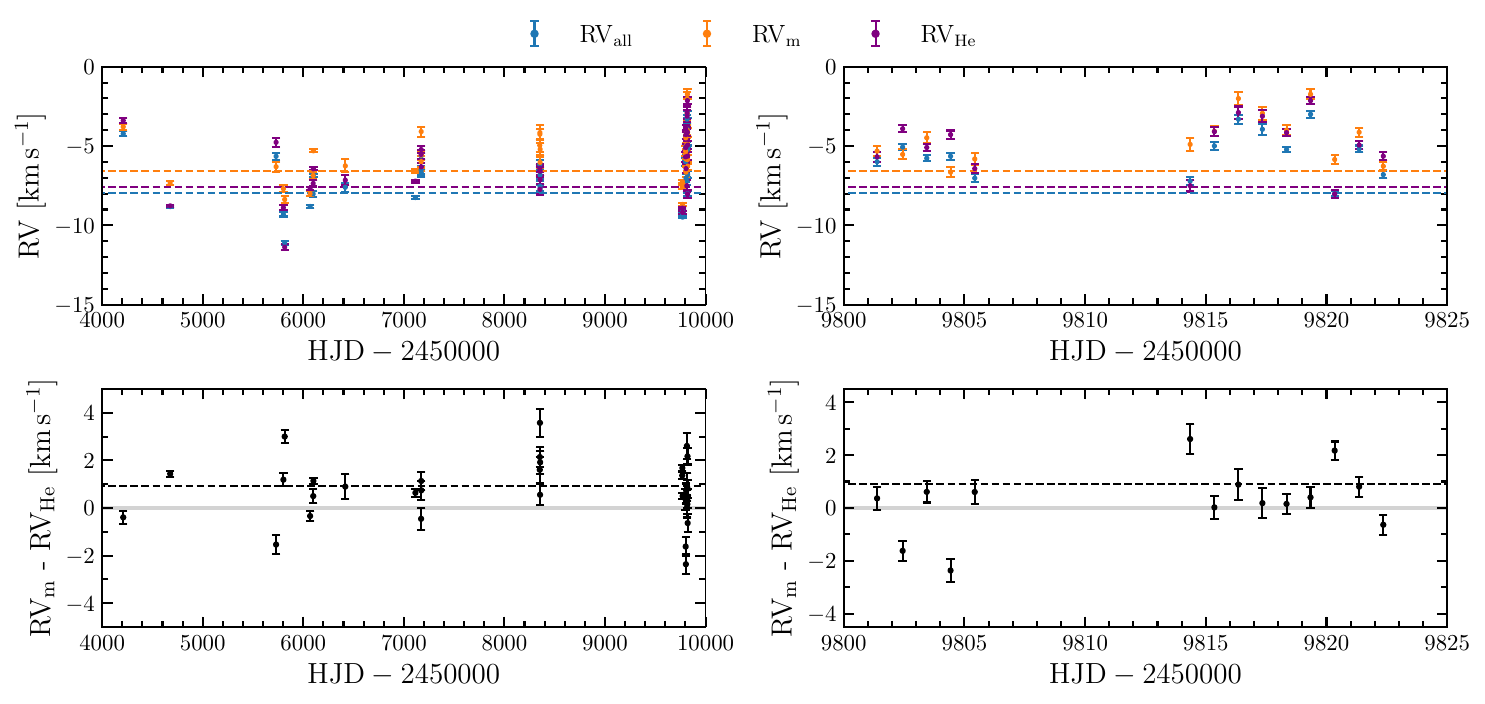}
     \caption{\textbf{Top}: RV measurements of 63~Oph computed from Gaussian fits to LSD~Stokes~$I$ profiles generated with the full RV line mask ($\mathrm{RV}_{\mathrm{all}}$, blue), metal mask ($\mathrm{RV}_{\mathrm{m}}$, orange) and He mask ($\mathrm{RV}_{\mathrm{He}}$, purple). The associated coloured dashed lines denote the weighted mean of each dataset. The left panel shows the full RV dataset obtained between 2007 and 2022. The right panel shows the most recent RV measurements from FIES and HERMES spectra. \textbf{Bottom}: Difference between metal line and He line RVs (RV$_{\mathrm{m}}$-RV$_{\mathrm{He}}$). The dashed black line denotes the weighted mean and the solid grey line denotes $\mathrm{RV}_{\mathrm{m}}-\mathrm{RV}_{\mathrm{He}}=0$.}
     \label{fig:RV_plot}
 \end{figure*}

We computed LSD Stokes~$I$ profiles for all 35 optical spectra using the same LSD depth and wavelength normalization parameters given in Section~\ref{sec:LSD}. The metal and He LSD profiles were computed simultaneously using {\sc iLSD}. We adopted LSD bin sizes equal to the median pixel velocity spacing of each spectra. Each LSD Stokes~$I$ profile was normalized by a linear fit to the continuum to correct for small offsets. The RV of each spectrum was computed from a nonlinear least-squares Gaussian fit to the associated Stokes~$I$ profile. Due to variable asymmetries in the Stokes~$I$ wings, each fit was performed over a velocity span of 100 km$\,\mathrm{s}^{-1}$ centered on the profile core. We performed the fits using the SciPy \texttt{curve\_fit} routine and estimated 1$\sigma$ formal uncertainties from the covariance matrix. We denote RVs computed with the all line, metal line, and He masks as $\mathrm{RV}_{\mathrm{all}}$, $\mathrm{RV}_{\mathrm{m}}$ and $\mathrm{RV}_{\mathrm{He}}$ respectively. The RV measurements and associated uncertainties are plotted in Figure~\ref{fig:RV_plot} and recorded in Table~\ref{tab:obs_log}.

 We note that \textit{Gaia} DR3 reports a median RV of $90.34\pm5.48$\,km\,s$^{-1}$ and RV amplitude of 45.53\,km\,s$^{-1}$ from 10 epoch measurements of 63~Oph \citep{Gaia_2023, Katz_2023}. These values are highly discrepant with respect to our RV measurements, likely due to the application of an incorrect temperature template ($\texttt{rv\_template\_teff}=6000$\,K). As it is challenging to determine accurate RVs for hot stars with \textit{Gaia}, only stars with $T_{\mathrm{eff}}\leq14,500$\,K should have RVs reported in DR3 \citep{blomme_2023}. We suspect that \textit{Gaia} RVs for 63~Oph were calculated due to an incorrect $T_{\mathrm{eff}}$ determination during data processing. Therefore, we ignore these results in our analysis.
 
\subsubsection{RV Variability}\label{sec:RV_Variability}
The weighted means of the three RV sets are similar: $\overline{\mathrm{RV}}_{\mathrm{all}}=-7.97\pm0.02$\,km\,s$^{-1}$, $\overline{\mathrm{RV}}_{\mathrm{m}}=-6.58\pm0.03$\,km\,s$^{-1}$ and $\overline{\mathrm{RV}}_{\mathrm{He}}=-7.58\pm0.03$\,km\,s$^{-1}$. The bottom plots of Figure~\ref{fig:RV_plot} show the difference $\mathrm{RV}_{\mathrm{m}}-\mathrm{RV}_{\mathrm{He}}$ for each observation. The observed offsets of up to a few kilometers per second between $\mathrm{RV}_{\mathrm{m}}$ and $\mathrm{RV}_{\mathrm{He}}$ are expected as it is common to observe RV differences between individual O star spectral lines (e.g. \citealt{Sana_2013}). These differences can be partially attributed to differences between line formation regions that can extend from the photosphere through the accelerating wind \citep{martins_2015_radial}. Varying density distributions and velocity flows in a rotating magnetosphere will add additional complexity to the spectral line formation. 

 The peak-to-peak amplitudes of the three RV sets are comparable: $\Delta\mathrm{RV}_{\mathrm{pp, all}}=8.1\pm0.3$, $\Delta\mathrm{RV}_{\mathrm{pp, m}}=6.9\pm0.3$ and $\Delta\mathrm{RV}_{\mathrm{pp, He}}=9.2\pm0.3$. Our range of RVs is generally consistent with historical snapshot RV measurements \citep{wilson_1953, conti_1977, garmany_1980}. Notably, we do not observe any long-term, large amplitude RV variability. 

We consider two criteria proposed by \cite{Sana_2013} to assess the significance of RV variability and help diagnose the presence of a spectroscopic binary companion. For a set of RV measurements, the null hypothesis of constant RV is rejected if for any two measurements 
\begin{equation}\label{eq:RV_varb}
    \frac{|v_{i}-v_{j}|}{\sqrt{\sigma_{i}^{2}+\sigma_{j}^{2}}} > 4
\end{equation}
where $v_{i}$, $v_{j}$ are the RV measurements, and $\sigma_{i}$, $\sigma_{j}$ are their associated $1\sigma$ errors. This significance criterion is satisfied by the peak-to-peak amplitudes in all three RV data sets, and also between RVs on some consecutive nights.

The second criterion is that a star can be classified as a spectroscopic binary if for a pair $v_{i}$, $v_{j}$ Equation~\ref{eq:RV_varb} is satisfied and
\begin{equation}\label{eq:spec_binary}
    |v_{i}-v_{j}|>\Delta\mathrm{RV}_{\mathrm{min}}
\end{equation}
where $\Delta\mathrm{RV}_{\mathrm{min}}$ is a minimum RV amplitude threshold. As the amplitude of intrinsic RV (and photometric) variability increases with luminosity for MS O-type stars (e.g. \citealt{Bowman_2020, Simon_2024}) the choice of $\Delta\mathrm{RV}_{\mathrm{min}}$ should depend on the star's spectral class. This threshold is hard to determine precisely as there has not yet been a sufficient population level characterization of O star short period ($\sim1-10$\,d) intrinsic RV variability. Massive OB star spectroscopic multiplicity studies have previously adopted conservative thresholds of $\Delta\mathrm{RV}_{\mathrm{min}}\approx15-20$\,km\,s$^{-1}$ (e.g. \citealt{Sana_2013, Dunstall_2015, Bodensteiner_2021, Bodensteiner_2025}). As the typically adopted thresholds are higher than our measured peak-to-peak amplitudes ($\Delta\mathrm{RV}_{\mathrm{pp}}<10$\,km\,$\mathrm{s}^{-1}$), we observe no evidence of a spectroscopic companion under the second criterion.

\subsubsection{RV Period Analysis}\label{sec:RV_period}
We performed a period analysis as described in Section~\ref{sec:period_analysis_method} on the RV datasets to search for evidence of coherent variability. While the amplitude threshold criterion (Eq.~\ref{eq:spec_binary}) does not provide evidence of a companion, it does not preclude the existence of one in a low-amplitude orbit. Other processes, such as pulsation or rotational modulation, may also induce coherent variability in the RVs. 

\begin{figure}
    \centering
    \includegraphics[width=\linewidth]{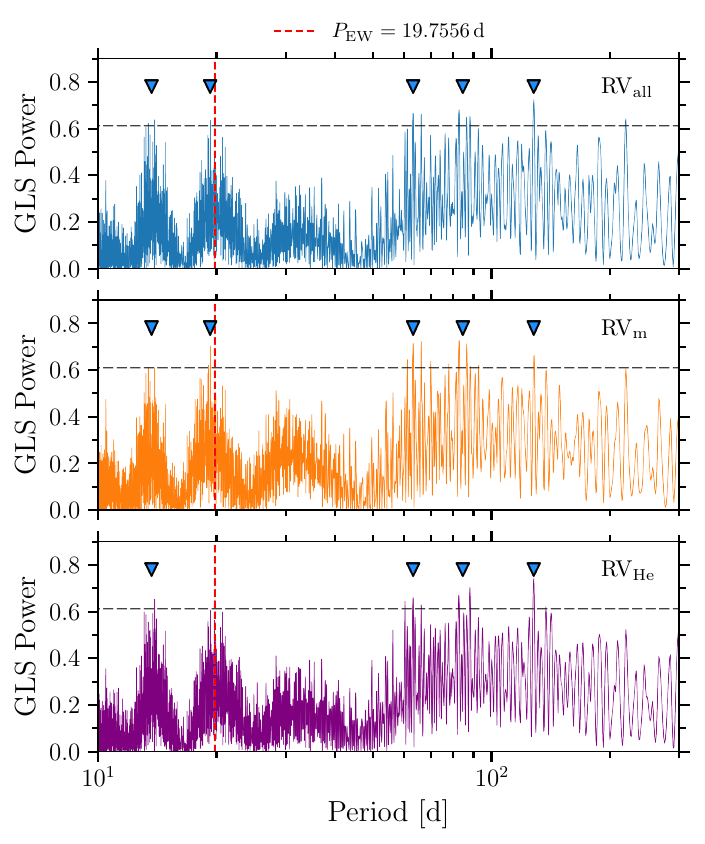}
    \caption{GLS periodograms of the $\mathrm{RV}_{\mathrm{all}}$, $\mathrm{RV}_{\mathrm{m}}$ and $\mathrm{RV}_{\mathrm{He}}$ measurements for 63~Oph. The dashed, horizontal black lines show the $\mathrm{FAP}=0.01$ level. Blue triangles denote significant periods around 13.7\,d, 19.3\,d, 63.1\,d, 84.3\,d and 127.6\,d. The dashed, vertical red lines denote the EW period $P_{\mathrm{EW}}$.}
    \label{fig:RV_LS_Period}
\end{figure}

Figure~\ref{fig:RV_LS_Period} shows the $\mathrm{RV}_{\mathrm{all}}$, $\mathrm{RV}_{\mathrm{m}}$ and $\mathrm{RV}_{\mathrm{He}}$ GLS periodograms over a range of $10-300$\,d. The three power spectra exhibit similar shapes and are characterized by clusters of peaks instead of a single dominant peak. The power spectra contain $8-12$ significant periods under the FAP criterion. The dominant periods in the $\mathrm{RV}_{\mathrm{all}}$, $\mathrm{RV}_{\mathrm{m}}$ and $\mathrm{RV}_{\mathrm{He}}$ power spectra are at $127.6$\,d, $82.4$\,d and $127.6$\,d respectively. The significant periods in common between the three power spectra can be roughly grouped into intervals centred around approximately 13.7\,d, 63.1\,d, 84.3\,d and 127.6\,d. There are significant peaks near 19.3\,d in the $\mathrm{RV}_{\mathrm{all}}$ and  $\mathrm{RV}_{\mathrm{m}}$ power spectra. No significant frequencies are detected in the power spectrum of $\mathrm{RV}_{\mathrm{m}}-\mathrm{RV}_{\mathrm{He}}$.

No significant periods are detected in the $\mathrm{RV}_{\mathrm{all}}$ or $\mathrm{RV}_{\mathrm{He}}$ measurements under the S/N criterion. A single period at 82.4\,d is detected in $\mathrm{RV}_{\mathrm{m}}$ with $\mathrm{S/N=5.0}$.

\subsubsection{Interpreting the RV Variability}\label{sec:interp_RV_var}
As the RV power spectra do not unambiguously identify a dominant period, we cannot confidently identify coherent periods and ascribe physical origins from the periodograms alone. The complexities seen in the RV power spectra are likely due to several compounding factors, including intrinsic SLF variability and window aliasing. There may also be additional systematics due to differing resolutions and small RV zero-point offsets between spectrographs. 

A cluster of peaks around $P_{\mathrm{EW}}$ is present in all three power spectra. The RVs of some magnetic massive stars are known to be modulated according to the rotation period (e.g. \citealt{Grunhut_2012}), which can be attributed to rotating surface features or asymmetries in the magnetosphere. More observations are needed to confirm if 63~Oph's RVs are modulated by rotation.

\subsection{Time Series Photometry}\label{sec:photometry_main}
\subsubsection{Photometry Period Analysis}
\begin{figure*}
    \centering
    \includegraphics[width=\linewidth]{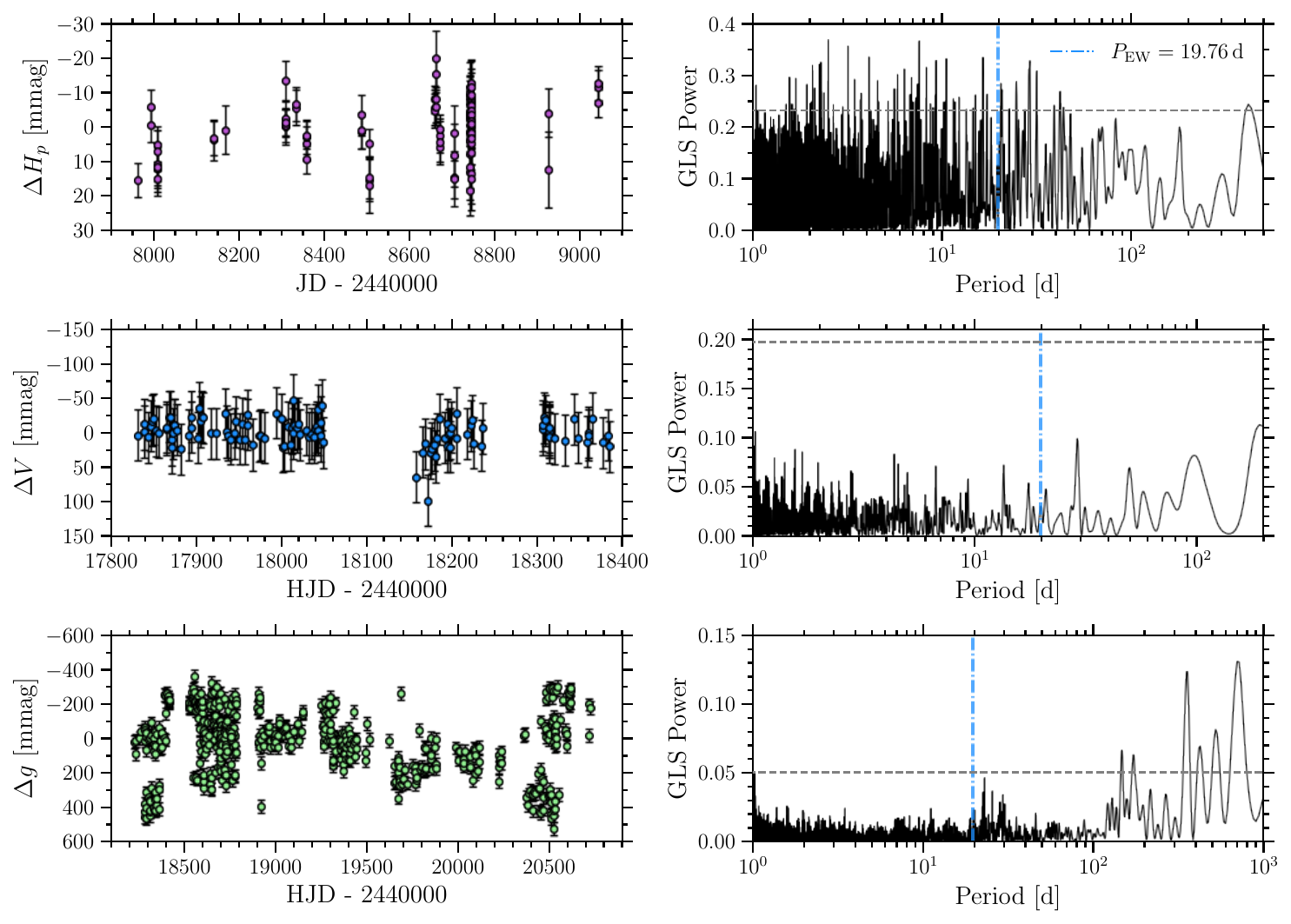}
    \caption{Light curves and corresponding GLS periodograms of 63~Oph. From top to bottom: Hipparcos, ASAS-SN $V$-band and ASAS-SN $g$-band. The dashed, grey horizontal lines denote the $\mathrm{FAP}=0.01$ level. The vertical, dashed-dotted blue lines denote $P_{\mathrm{EW}}$.}
    \label{fig:light_curves}
\end{figure*}

We performed a period analysis on the Hipparcos, ASAS-SN and K2 photometry (Section~\ref{sec:obs_photometry}) using the methodology described in Section~\ref{sec:period_analysis_method}. Figure~\ref{fig:light_curves} shows the Hipparcos, ASAS-SN $g$-band and ASAS-SN $V$-band light curves with corresponding GLS periodograms. The K2 light curve and periodogram are presented in Figure~\ref{fig:k2_slf}.

The majority of peaks in the Hipparcos and K2 power spectra are significant under the FAP criterion. None of the FAP identified periods are consistent with $P_{\mathrm{EW}}$. In contrast, there are no significant periods in the Hipparcos and K2 photometry under the S/N criterion. The dominant Hipparcos period is at 2.3\,d ($\mathrm{S/N}=3.7$) and the dominant K2 period is at 6.3\,d ($\mathrm{S/N}=3.3$). The extreme difference in the number of significant frequencies between the two criteria indicates that the space-based photometry is strongly impacted by SLF variability. We further discuss the SLF variability of the K2 light curve in Section~\ref{sec:K2_SLF}.

No significant periods are detected in the ASAS-SN $V$-band light curve with either criterion. The dominant period is at 191\,d ($\mathrm{S/N}=3.2$). In the $g$-band light curve periods are detected at approximately 172\,d, 357\,d, 424\,d, 522\,d and 698\,d with the FAP criterion. Two periods are detected at approximately 357\,d ($\mathrm{S/N}=7.8$) and 698\,d ($\mathrm{S/N}=6.9$) from iterative pre-whitening. The ASAS-SN $g$-band light curve has substantially higher-amplitude variations than the Hipparocs and $V$-band light curves. The long-period trends may be systematics from the machine learning extraction pipeline used on the saturated photometry (Section~\ref{sec:obs_photometry}). No significant periods are detected after prewhitening all the long-period variations.

\subsubsection{K2 SLF Variability}\label{sec:K2_SLF}
The K2 light curve appears qualitatively similar to the SLF dominated light curves of evolved O-type stars observed by \textit{TESS} (e.g. \citealt{Burssens_2020}). The origin of massive star SLF variability is still under investigation. Various origins have been proposed including sub-surface convection (e.g. \citealt{Cantiello_2009, Cantiello_2021, Schultz_2022}), core-excited internal gravity waves (e.g. \citealt{Rogers_2013, Thompson_2024, Bowman_2024}) and stellar winds \citep{Kritcka_2018, Kritcka_2021}. It is also possible that small-scale flows in the magnetosphere impact the SLF variability of magnetic O stars. However, short-timescale magnetohydrodynamic modelling has not yet been performed to investigate this effect.

\begin{figure*}
    \centering
    \includegraphics[width=\linewidth]{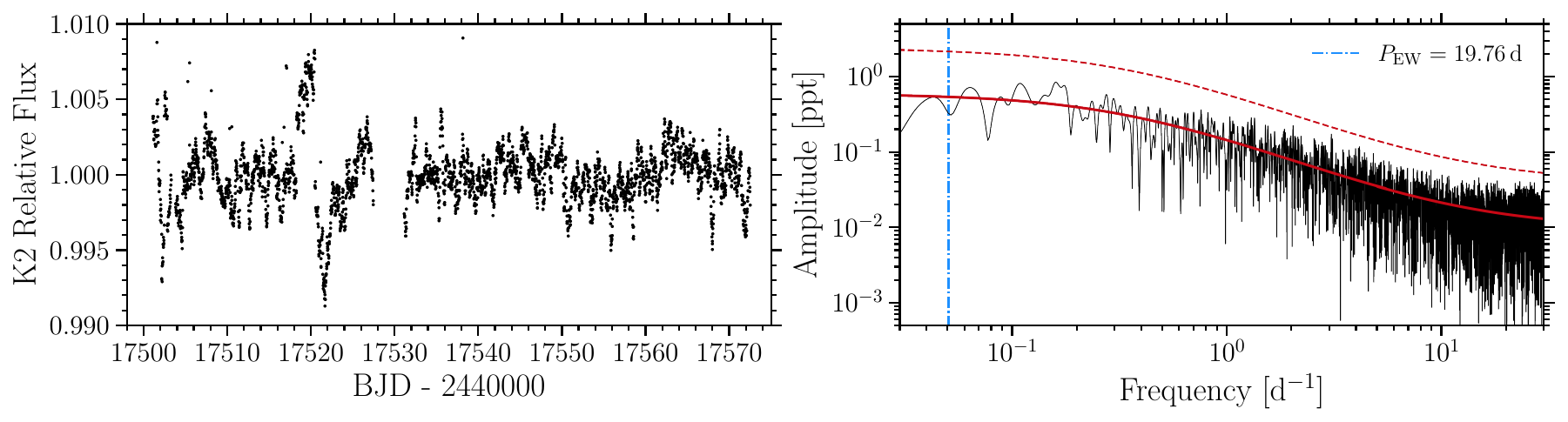}
    \caption{K2 light curve and corresponding amplitude spectrum for 63~Oph. The solid red line denotes the SLF fit to the amplitude spectrum, and the red dashed line denotes the $\mathrm{S/N}=4$ threshold for a significant peak detection. The vertical, dash-dotted blue line denotes the frequency corresponding to $P_{\mathrm{EW}}$.}
    \label{fig:k2_slf}
\end{figure*}

We characterize the K2 SLF variability using the commonly adopted (e.g. \citealt{Bowman_2020}) semi-Lorentzian function of the form 
\begin{equation}\label{eq:SLF}
    \alpha(\nu)=\frac{\alpha_{0}}{1+\bigg(\frac{\nu}{\nu_{\mathrm{char}}}\bigg)^{\gamma}}+C_{\mathrm{W}}
\end{equation}
where $\alpha_{0}$ is the characteristic amplitude, $\nu_{\mathrm{char}}$ is the characteristic frequency, $\gamma$ is the logarithmic amplitude gradient, and $C_{\mathrm{W}}$ is the instrumental white-noise term. A nonlinear least-squares fit of Equation~\ref{eq:SLF} to the K2 amplitude spectrum gives $\alpha_{0}=0.591\pm0.006$, $\nu_{\mathrm{char}}=0.352\pm0.008\,\mathrm{d}^{-1}$, $\gamma = 1.16\pm0.01$ and $C_{\mathrm{W}}=0.001\pm0.004$. Figure~\ref{fig:k2_slf} shows the K2 light curve and SLF fit to the amplitude spectrum. It is difficult to directly compare our results with the SLF analysis of \textit{TESS} O star light curves due to differences between the passbands. Additionally, $\alpha_{0}$ is a relative amplitude due to the halo extraction. However, the low $\nu_{\mathrm{char}}$ we obtain is generally consistent with the low values observed in the \textit{TESS} light curves of evolved O-type stars \citep{Bowman_2020}.

As the K2 power spectrum is frequency dependent, the SLF fit provides a better measure of the S/N than an interval average. We searched for significant periods in the power spectrum by looking for peaks with an amplitude greater than four times the SLF fit (dashed red line Fig.~\ref{fig:k2_slf}). No significant periods are identified, and there is no strong peak around the frequency corresponding to $P_{\mathrm{EW}}$.

\subsubsection{Interpreting the Photometric Variability}
Except for the $g$-band long-period trends, no periods were detected in the four light curves under the S/N criterion. We observe no evidence of binary eclipses or ellipsoidal variability, which would indicate a close binary companion. None of the RV periods discussed in Section~\ref{sec:RV_period} are detected in the photometry. 

Interestingly, there is also no clear evidence for the $P_{\mathrm{EW}}\approx19.8\,$d periodicity in the photometry. The shape and amplitude of a light curve generated by a dipolar magnetosphere depend on several parameters, including the stellar mass-loss rate, magnetic geometry (inclination and obliquity angles) and magnetic field strength \citep{owocki_2016, Munoz_2020}. Rotational modulation is observed in ground-based photometry of some strongly magnetic O-type stars, such as HD~108 \citep{Barannikov_2007} and HD~191612 \citep{Munoz_2022}. However, there are magnetic O stars in which rotational modulation has only been detected in high-precision space-photometry (e.g. HD~148937; \citealt{Frost_2024}).

The nondetection of $P_{\mathrm{EW}}$ in the ASAS-SN and Hipparcos photometry is likely because the magnetospheric occultations have low photometric amplitude. This interpretation is supported by the relatively weak Balmer line EW variations. Additional factors may contribute to the nondetection in the K2 light curve, including competing SLF variability, short observing duration, and systematics from the halo photometry extraction.

\section{Discussion}\label{sec:discussion}
\subsection{Magnetic Fields of Evolved O-type Stars}\label{sec:analogues}
To help inform our interpretation of 63~Oph's properties we have sought to identify similar objects in the literature. Throughout our analysis, we have made repeated comparisons to the magnetic O supergiant $\zeta$~Ori~Aa as it is the most evolved O-type star with a confirmed magnetic field \citep{bouret_2008, Blazere_2015} and its spectroscopic and photometric variability has been well studied \citep{Buysschaert_2017}. While efforts have been made to detect magnetic fields in other evolved O-type stars, none present as clear evidence for a large-scale surface field as $\zeta$~Ori~Aa. 

The best magnetic field constraints on evolved O-type stars have been set using high-resolution spectropolarimetric observations. \cite{david_uraz_2014} analyzed ESPaDOnS observations of eight O-type giants, bright giants, and supergiants to evaluate a proposed relationship between large-scale, dipolar fields and UV discrete absorption components (DACs). No magnetic fields were detected in the individual or nightly averaged LSD~Stokes~$V$ profiles, and a Bayesian analysis of the Stokes~$V$ profiles determined maximum dipolar fields strengths of $\sim60-360$\,G at 95\% confidence. An additional $\sim50$ stars with luminosity classes (LC) III-I were studied during the MiMeS survey and subsequent Bayesian analysis with no confirmed detections \citep{grunhut_2017, Petit_2019}. 

In addition to 63~Oph, the other evolved O star magnetic candidate identified by MiMeS is $\delta$~Ori~A (HD 36486; O9.5~II+B2~V+B0~IV), a member of the six component multiple stellar system $\delta$~Ori (e.g. \citealt{harvin_2002, Mayer_2010, Shenar_2015}). The triple system $\delta$~Ori~A consists of a $\sim6$\,d eclipsing binary {Aa1~(O9.5 II)~+~Aa2~(B2~V)} and long-period interferometric binary companion Ab~(B0~IV). In a recent analysis, \cite{Oplivstilova_2023} found that the O9.5~II primary is evolved from the ZAMS ($\log g=3.5\pm0.1$) but has not yet reached overflow. \cite{grunhut_2017} obtained a DD of $\delta$~Ori~A in a single Narval observation, which was supported by a Bayesian analysis of the LSD~Stokes~$V$ profile \citep{Petit_2019}. However, the multiple components were not disentangled in the LSD analysis. As the tertiary Ab companion has a close angular separation of $\sim0.3^{\prime\prime}$ from the Aa binary \citep{Tokovinin_2020}, it is blended in the Narval spectra, making up $\sim25\%$ of the flux in the $V$-band \citep{Shenar_2015, Oplivstilova_2023}. Additional study is needed to confidently determine the magnetic status of $\delta$~Ori~A's components.

An interesting comparison to 63~Oph is the O giant $\xi$~Per (HD~24912; $V=4.1$\,mag; O7.5~III(n)((f)) \citealt{sota_2011}). The star has similar physical parameters to 63~Oph ($T_{\mathrm{eff}}=35.9\pm0.5\,$kK, $\log g=3.67\pm0.04$, $\log \mathcal{L}/\mathcal{L}_{\odot}=3.94\pm0.05$; \citealt{holgado_2018}). $\xi$~Per has been proposed as a magnetic O star candidate due to observed DACs \citep{Henrichs_1994, deJong_2001}, nonthermal radio emission \citep{Schnerr_2007} and corotating surface features \citep{Ramiaramanantsoa_2014}. However, no magnetic detection has been obtained from multiple polarimetric studies \citep{deJong_2001, Henrichs_2009, david_uraz_2014, grunhut_2017}. The maximum $B_{\mathrm{p}}$ for $\xi$~Per has been constrained to be less than 60\,G at 95\% confidence \citep{david_uraz_2014, Petit_2019} which is a factor of five less than the minimum $B_{\mathrm{p}}$ we infer for 63~Oph.

Several authors have proposed that the well-studied supergiant $\zeta$~Puppis (HD 66811; $V=2.3$\,mag; O4~I(n)fp \citealt{sota_2014}) could possess surface magnetic fields due to observed H$\alpha$ variability \citep{Mihalas_1980, Moffat_1981}, DACs \citep{Howarth_1995} and photometric variability interpreted as surface spots \citep{Ramiaramanantsoa_2018}. As in the case of $\xi$~Per, multiple polarimetric studies of $\zeta$~Pup have failed to obtain a magnetic detection \citep{Chesneau_2002, Barker_1981, david_uraz_2014, Hubrig_2016}. The maximum $B_{\mathrm{p}}$ for $\zeta$~Pup has been constrained to be less than $\sim 100$\,G at 95\% confidence \citep{david_uraz_2014, Petit_2019}. 

There have been several studies that have claimed magnetic detections in O-type stars at the $\sim3\sigma$ level in low-resolution Focal Reducer and low dispersion Spectrograph (FORS) observations (see \cite{Bagnulo_2012, grunhut_2017} and references therein). \cite{Bagnulo_2012, Bagnulo_2013} showed that, when analyzing FORS data, there are additional uncertainties above the photon noise due to the data reduction process and instrumental instabilities. The authors argue that repeated $\langle B_{z}\rangle$ measurements at the $5-6\sigma$ level are required to confidently claim magnetic detections using FORS data. A handful of magnetic O-type stars have been detected with FORS under this criterion (e.g. \citealt{bagnulo_2015, Castro_2015}), and most have additional confirmation through high-resolution spectropolarimetry. These confirmed detections are included in the sample discussed by \cite{grunhut_2017} and are shown in Figure~\ref{fig:sHRD}.

One interesting FORS magnetic candidate is HD~226868, the O9.7~Iab supergiant primary component of the famous high-mass X-ray binary Cyg X-1 (e.g. \citealt{Webster_1972, Miller_Jones_2021, Ramachandran_2025}). \cite{Karitskaya_2009, Karitskaya_2010} reported two photospheric $\langle B_{z}\rangle$ detections greater than 5$\sigma$ from 13 FORS1 observations. \cite{Bagnulo_2012} obtained two significant detections at the 5$\sigma$ level from reanalysis of the FORS1 observations, although the authors caution that instrumental flexures and possible linear polarization crosstalk may impact the data. \cite{Hubrig_2019} reported no $\langle B_{z}\rangle$ detections from four additional FORS2 observations. Additional observations are needed to confirm the magnetic status of the supergiant.

\subsection{Is 63~Oph a Rare Transitional Object?}\label{sec:discuss_transitional_object}
Our findings convincingly demonstrate that 63~Oph is a magnetic O-type star, and its properties make it stand out among the larger sample of such stars. While our $T_{\mathrm{eff}}$ and $\log g$ measurements place 63~Oph near HD~108 and HD~191612 in the sHRD, there are key differences in 63~Oph's spectral and magnetic properties. The $B_{\mathrm{p}}$ strengths of HD~108 and HD~191612 are on the order of $2-4$\,kG, and both stars have undergone significant magnetic braking with respective rotation periods of $\sim55$\,yrs and $537.6$\,d \citep{Wade_2011, Shultz_2017, GREGOR_2023}. While we are only able to set a lower limit on $B_{\mathrm{p}}$ for 63~Oph, the spectral characteristics, nondetection of photometric rotational modulation, moderate rotation period, and challenging field detections all point toward a weaker field strength than HD~108 and HD~191612. 

More generally, we have shown that 63~Oph does not exhibit the strong spectral variations and photometric rotational modulation that characterize most magnetic O-type stars, particularly the Of?p spectral class. As discussed in Section~\ref{sec:analogues} there are no known direct analogues to 63~Oph with convincing magnetic detections. For these reasons, we propose that 63~Oph is an example of a previously elusive transitional type object between the strongly magnetic stars near the MS and $\zeta$~Ori~Aa type magnetic supergiants. This makes 63~Oph an essential object for the study of fossil-field decay mechanisms among the most massive stars.

\cite{holgado_2018} identified a subset of stars (including 63~Oph) in the IACOB O star sample for which \texttt{FASTWIND} models were unable to simultaneously reproduce both the H$\alpha$ and He\textsc{ii}~$\lambda4686$ diagnostic wind lines (designated quality flag Q3). This discrepancy indicates a clumped or nonspherically symmetric wind and is found predominately in the spectra of O star giants and supergiants. In the case of 63~Oph, we attribute the H$\alpha$ variability and, therefore, its Q3 classification to a rotating magnetosphere. This raises the following question: Is there a subpopulation of evolved magnetic O stars hidden in the IACOB sample that show spectroscopic evidence of magnetically structured circumstellar material?

To investigate, we crossmatched the current list of Q3 IACOB stars with LCs III-I (58 targets) against the MiMeS O star sample (97 targets). We identified 12 stars in common, only four of which have multiple MiMeS observations (including 63~Oph and $\zeta$~Ori~Aa). The upper limits on the 95.4\% credible region for the dipole field strength ($B_{\mathrm{p}}^{95}$) range from $\sim30$ to 500\,G \citep{Petit_2019}. Assuming that 63~Oph and $\zeta$~Ori~Aa possess $B_{\mathrm{p}}$ strengths typical for their LCs, we can consider $B_{\mathrm{p}}$ to be well constrained for an O (bright) giant if $B_{\mathrm{p}}^{95}<300$\,G and in an O supergiant if $B_{\mathrm{p}}^{95}<150$\,G. Under this criterion, only eight Q3 stars (with LC I-III) have well-constrained magnetic fields, two of which are magnetic detections. We conclude that the Q3 stars have been poorly studied with high-resolution spectropolarimetry and suggest that their further study provides promising opportunities for discovering additional evolved magnetic O stars.

\subsection{Follow-up Observations}
Additional spectroscopic and polarimetric observations are needed to fully characterize the magnetic and variability properties of 63~Oph. Verifying the rotation period and determining the inclination angle from additional magnetic measurements will provide an opportunity to test methodologies for measuring line broadening in evolved massive stars. Once the magnetic field of 63~Oph has been fully characterized, the evolutionary modelling should be revisited using models that incorporate the effects of surface magnetic fields \citep{Keszthelyi_2022}. 

Further insights into 63~Oph's evolutionary history may be gained from the study of its associated, predominantly bipolar nebula (Figure~\ref{fig:63Oph_nebula}). Recently, modelling by \cite{Frost_2024} demonstrated that the magnetic O star component in the massive binary HD~148937 could be produced through a stellar merger. A merger scenario is supported by the young, complex, chemically enriched bipolar nebula that surrounds HD~148937 \citep{Mahy_2017, Beomdu_2024}. Considering these results, a detailed kinematic and abundance study of 63~Oph's nebular region is warranted to help constrain its origins and test the merger hypothesis. 

\section{Conclusions}\label{sec:Conclusions}
\begin{enumerate}
\item We confirm 63~Oph to be a magnetic O-type star based on one DD and one MD of the LSD~Stokes~$V$ profiles computed from ESPaDOnS spectropolarimetric observations. The associated longitudinal fields are both detected at $\sim6\sigma$.
\item The clearest magnetic field detection (lowest FAP) coincides with the strongest H$\alpha$ emission. This suggests that the emission component is related to the influence of the magnetic field rather than changes in the global mass-loss rate.
\item Assuming a dipolar magnetic field geometry, we set a conservative lower bound of $B_{\textrm{p}}\geq300\pm50$\,G on the dipole field strength.
\item An analysis of archival and newly obtained IACOB spectra revealed a dominant $\sim19.8$\,d period in H$\alpha$ and H$\beta$ EW measurements ($\mathrm{S/N}>6$). We attribute the EW periodicity to rotational modulation under the ORM.
\item There is no clear evidence of a close binary companion in the RV measurements or time series photometry.
\item We report spectroscopic stellar parameters for 63~Oph and plot its location in the sHRD. We propose that 63~Oph is an elusive transitional object between strongly magnetic stars near the MS and $\zeta$~Ori~Aa type magnetic massive supergiants.
\item The sample of IACOB stars with peculiar wind properties (classified Q3) should be further investigated for hidden, evolved magnetic stars similar to 63~Oph.
\end{enumerate}

\section{Acknowledgments}
We thank Colin Folsom for helpful discussions and Jason Grunhut for providing his LSD line masks for a comparison with our own. We thank the anonymous referees for their comments.

J.A.B. acknowledges support through a Postgraduate Doctoral Scholarship (PGS D) from the Natural Sciences and Engineering Research Council (NSERC) of Canada. G.A.W. acknowledges support in the form of a Discovery Grant from the Natural Sciences and Engineering Research Council (NSERC) of Canada. G.H. and S.S-D. acknowledge the support from the State Research Agency (AEI) of the Spanish Ministry of Science and Innovation and Universities (MCIU) and the European Regional Development Fund (FEDER) under grant PID2021-122397NB-C21/PID2022-136640NB- C22/10.13039/501100011033.

Based on observations obtained at the Canada-France-Hawaii Telescope (CFHT) which is operated by the National Research Council (NRC) of Canada, the Institut National des Sciences de l'Univers of the Centre National de la Recherche Scientifique (CNRS) of France, and the University of Hawaii. The observations at the CFHT were performed with care and respect from the summit of Maunakea which is a significant cultural and historic site. The IACOB spectroscopic database is based on observations made with the Nordic Optical Telescope operated by the Nordic Optical Telescope Scientific Association, and the Mercator Telescope, operated on the island of La Palma by the Flemish Community, at the Spanish Observatorio del Roque de los Muchachos of the Instituto de Astrofísica de Canarias. Based on observations obtained with the HERMES spectrograph, which is supported by the Research Foundation - Flanders (FWO), Belgium, the Research Council of KU Leuven, Belgium, the Fonds National de la Recherche Scientifique (F.R.S.-FNRS), Belgium, the Royal Observatory of Belgium, the Observatoire de Genève, Switzerland and the Thüringer Landessternwarte Tautenburg, Germany. Based on observations collected at the European Organisation for Astronomical Research in the Southern Hemisphere under ESO programmes 079.D-0564(A) and 089.D-0975(A). This work has made use of data from the European Space Agency (ESA) mission {\it Gaia} (\url{https://www.cosmos.esa.int/gaia}), processed by the {\it Gaia} Data Processing and Analysis Consortium (DPAC,
\url{https://www.cosmos.esa.int/web/gaia/dpac/consortium}). Funding for the DPAC has been provided by national institutions, in particular the institutions participating in the {\it Gaia} Multilateral Agreement. This paper includes data collected by the Kepler mission and obtained from the MAST data archive at the Space Telescope Science Institute (STScI). Funding for the Kepler mission is provided by the NASA Science Mission Directorate. STScI is operated by the Association of Universities for Research in Astronomy, Inc., under NASA contract NAS 5–26555. 

This research used the facilities of the Canadian Astronomy Data Centre operated by the National Research Council of Canada with the support of the Canadian Space Agency. This work has made use of the VALD database, operated at Uppsala University, the Institute of Astronomy RAS in Moscow, and the University of Vienna. This research has made use of the SIMBAD database, operated at CDS, Strasbourg, France, and NASA’s Astrophysics Data System (ADS). This research has made use of the SVO Filter Profile Service ``Carlos Rodrigo", funded by MCIN/AEI/10.13039/501100011033/ through grant PID2023-146210NB-I00.

\vspace{5mm}
\facilities{ADS, ASAS-SN, CFHT (ESPaDOnS), Gaia, Kepler, MAST, Mercator1.2m (HERMES), Max Planck:2.2m (FEROS), NOT (FIES)}

\software{Astropy \citep{astropy_2022}, halophot \citep{White_2017}, IACOB-BROAD \citep{Simon_2014}, IACOB-GBAT \citep{simon_diaz_2011}, {\sc iLSD} \citep{kochukhov_2010}, Lightkurve \citep{lightkurve_2018}, Matplotlib \citep{Hunter_2007}, NumPy \citep{harris_2020}, SAOImage DS9 \citep{smithsonian_2000}, SciPy \citep{2020SciPy-NMeth}, SpecpolFlow (\url{https://github.com/folsomcp/specpolFlow}), Specutils \citep{nicholas_earl_2024_14042033}, TelFit \citep{Gullikson_2014}.}

\clearpage
\appendix

\setcounter{table}{0}
\setcounter{figure}{0}

\renewcommand{\thetable}{\Alph{section}\arabic{table}}
\renewcommand{\thefigure}{\Alph{section}\arabic{figure}}

\renewcommand{\theHtable}{\Alph{section}\arabic{table}}
\renewcommand{\theHfigure}{\Alph{section}\arabic{figure}}
\FloatBarrier
\section{Figures and Tables}\label{append:A}
\begin{figure*}[htb!]
    \centering
    \includegraphics[width=0.9\linewidth]{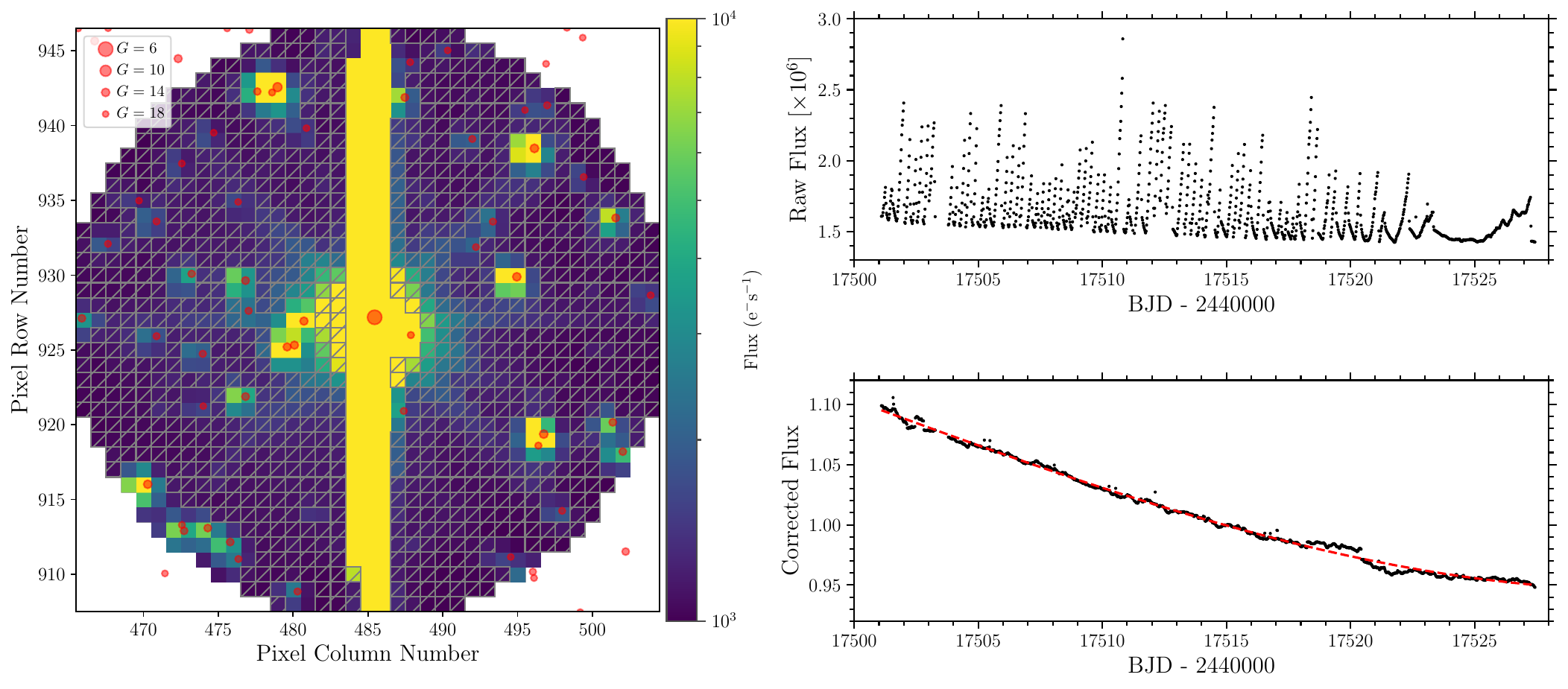}
    \caption{Illustration of the K2 photometry extraction for C9a. \textbf{Left}: The target pixel file and halo mask (grey hatching). Red circles denote contaminating stars from the $Gaia$ catalogue ($G<18$\,mag). The flux colour bar is logarithmically scaled between $10^{3}$ and $10^{4}$ for visibility. \textbf{Top Right}: Raw flux extracted with the halo mask. The dominant cyclical pattern is due to the roll motion of the spacecraft. \textbf{Bottom Right}: Normalized corrected flux calculated with \texttt{halophot}. The red dashed line shows the cubic polynomial fit used to remove the long-term trend.}
    \label{fig:K2_reduction}
\end{figure*}

\begin{figure}
    \centering
    \includegraphics[width=\linewidth]{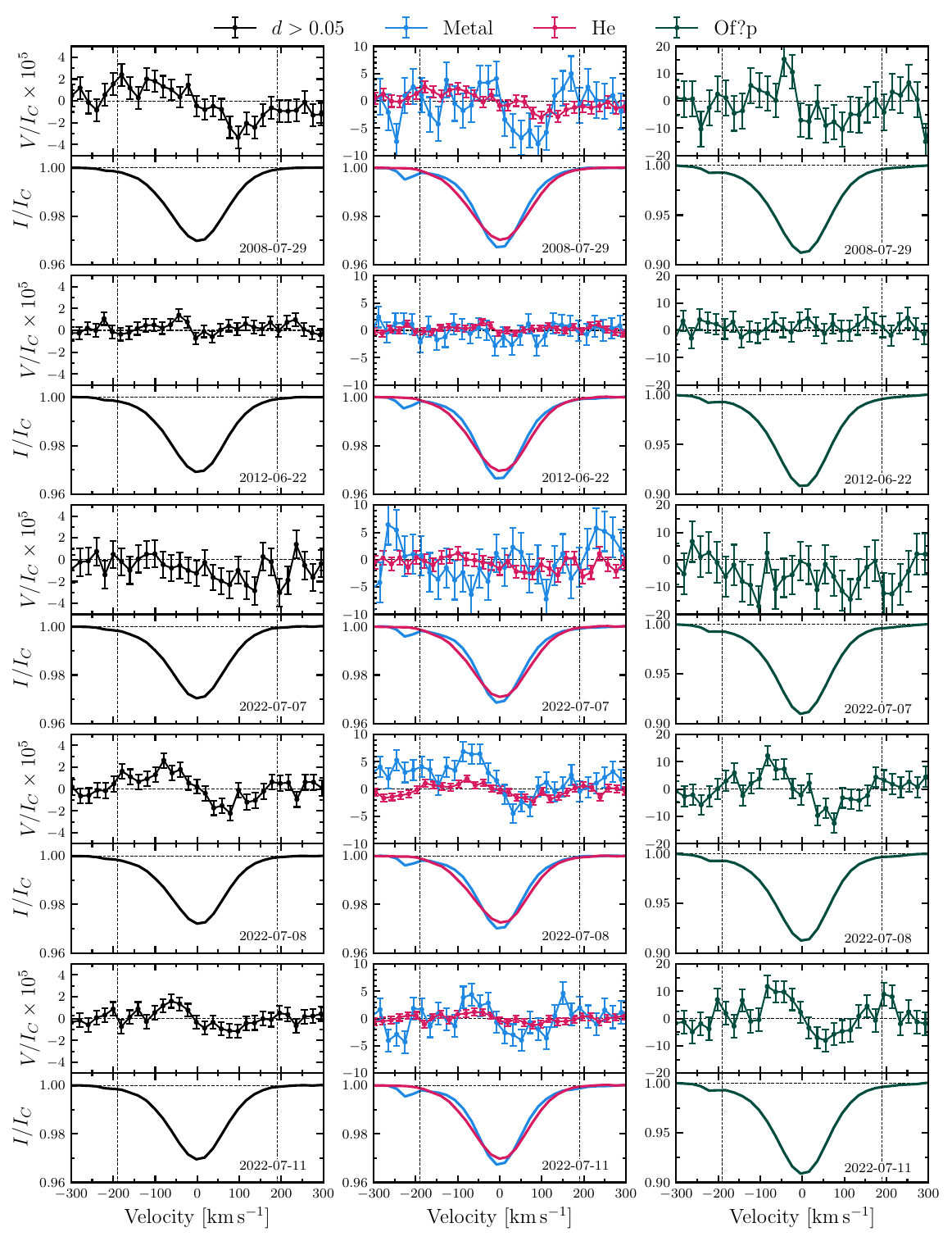}
    \caption{Continuum-normalized LSD~Stokes~$I$ and $V$ profiles computed using the test line masks described in Section~\ref{sec:LSD}. The LSD profiles have been shifted to the rest frame. Vertical dashed lines denote the integration bounds used to measure $\langle B_{z}\rangle$ and the FAP (Table~\ref{tab:mag_log}). The absorption feature in the blue wing of the Stokes~$I$ profiles is due to a DIB at 5797~$\mathrm{\AA}$ (Sec.~\ref{sec:DIB}).}
    \label{fig:LSD_mask_tests}
\end{figure}

\begin{figure*}
\includegraphics[width=\linewidth]{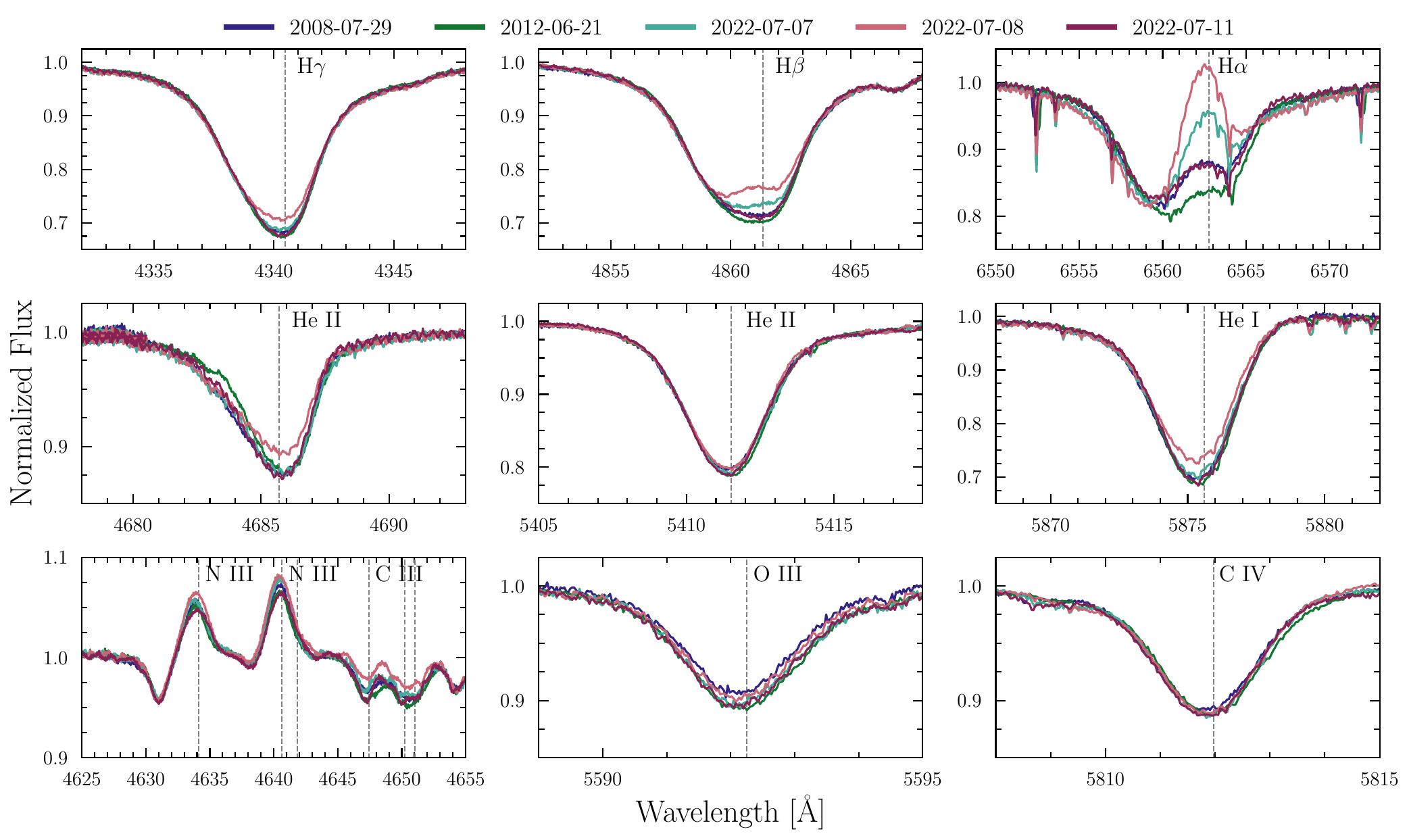}
    \caption{Line variability in the ESPaDOnS spectra of 63~Oph. The spectra are in the heliocentric rest frame and not corrected for atmospheric telluric absorption. The dashed vertical lines denote the air rest wavelengths.}
    \label{fig:espadons_grid}
\end{figure*}

\begin{figure*}
\includegraphics[width=\linewidth]{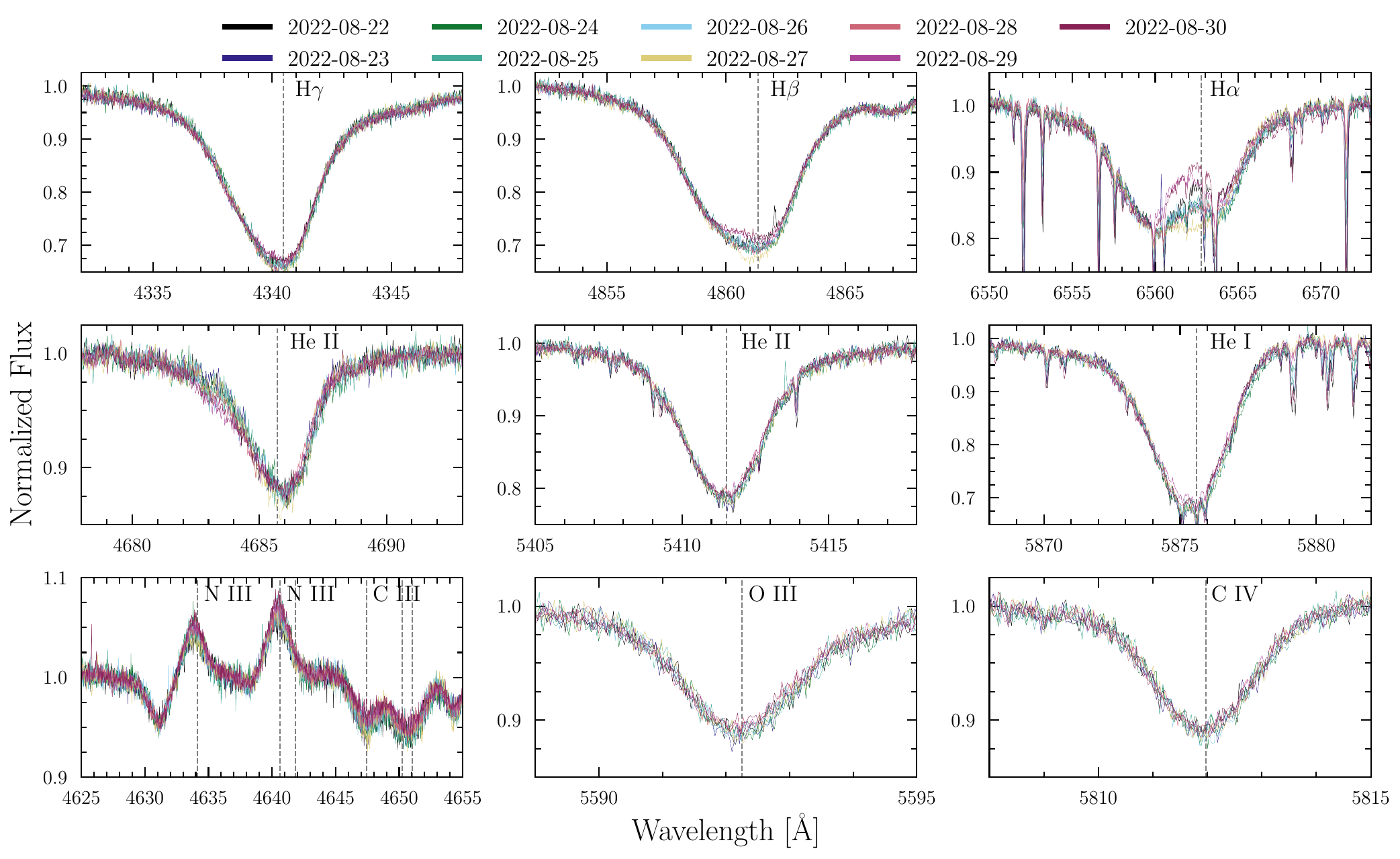}
    \caption{Line variability over an eight day time span in the HERMES spectra of 63~Oph. The spectra are in the heliocentric rest frame and not corrected for atmospheric telluric absorption. The dashed vertical lines denote the air rest wavelengths.}
    \label{fig:Hermes_grid}
\end{figure*}

Figure~\ref{fig:K2_reduction} illustrates the reduction of the K2 photometry for C9a. Figure~\ref{fig:LSD_mask_tests} shows the LSD profiles computed using the test line masks described in Section~\ref{sec:LSD}. Figures \ref{fig:espadons_grid} and \ref{fig:Hermes_grid} show example line profile variability in the ESPaDOnS and HERMES spectra respectively. Table~\ref{tab:obs_log} provides a log of the Stokes~$I$ spectra and corresponding RV and line-core EW measurements. Table~\ref{tab:magOstars} summarizes the physical parameters of the confirmed magnetic O-type stars plotted in Figure~\ref{fig:sHRD}.

\FloatBarrier

\begin{table*}
\scriptsize
\centering
\caption{Log of optical spectra and measurements for 63~Oph.}
\begin{tabular}{cccccccccccccccc}
\hline
Date&HJD\,$-$&Instrument&$t_\textrm{exp}$&S/N&$\mathrm{RV}_{\mathrm{all}}$&$\mathrm{RV}_{\mathrm{He}}$&$\mathrm{RV}_{\mathrm{m}}$&H$\alpha$ Core EW&H$\beta$ Core EW\\
&2\,450\,000&&[s]&(500\,nm)&[km\,s$^{-1}$]&[km\,s$^{-1}$]&[km\,s$^{-1}$]&[$\mathrm{\AA}$]&[$\mathrm{\AA}$]\\
\hline
2007-04-19&4209.7834&FEROS&299&196&$-4.23\pm0.16$&$-3.8\pm0.2$&$-3.39\pm0.18$&$0.94\pm0.04$&$1.050\pm0.015$\\
2008-07-29&4676.7997&ESPaDOnS&2800&1120&$-8.8\pm0.1$&$-7.33\pm0.11$&$-8.8\pm0.1$&$0.912\pm0.016$&$1.04\pm0.01$\\
2011-06-17&5729.5163&HERMES&900&186&$-5.7\pm0.2$&$-6.3\pm0.3$&$-4.8\pm0.3$&$1.01\pm0.05$&$1.06\pm0.03$\\
2011-08-28&5802.3755&FIES&272&214&$-9.3\pm0.13$&$-7.7\pm0.2$&$-8.88\pm0.17$&$1.09\pm0.05$&$1.131\pm0.019$\\
2011-09-10&5815.3597&FIES&238&268&$-11.10\pm0.14$&$-8.4\pm0.2$&$-11.37\pm0.19$&$0.71\pm0.06$&$-$\\
2012-05-20&6067.9310&FEROS&400&385&$-8.81\pm0.12$&$-8.01\pm0.16$&$-7.67\pm0.11$&$0.998\pm0.019$&$1.089\pm0.015$\\
2012-06-20&6098.8473&FEROS&400&349&$-8.02\pm0.17$&$-6.9\pm0.2$&$-7.36\pm0.19$&$1.00\pm0.03$&$1.092\pm0.012$\\
2012-06-22&6100.9523&ESPaDOnS&11520&1973&$-6.9\pm0.1$&$-5.3\pm0.1$&$-6.4\pm0.1$&$1.10\pm0.02$&$1.08\pm0.01$\\
2013-05-05&6417.7448&HERMES&360&152&$-7.6\pm0.3$&$-6.3\pm0.4$&$-7.2\pm0.3$&$1.01\pm0.05$&$1.08\pm0.03$\\
2015-04-02&7114.9224&FEROS&400&420&$-8.2\pm0.1$&$-6.58\pm0.14$&$-7.2\pm0.1$&$1.01\pm0.03$&$1.068\pm0.012$\\
2015-05-29&7171.6082&HERMES&900&145&$-6.5\pm0.2$&$-6.0\pm0.4$&$-5.5\pm0.3$&$1.09\pm0.06$&$1.10\pm0.03$\\
2015-05-30&7172.6433&HERMES&476&151&$-5.6\pm0.2$&$-4.1\pm0.3$&$-5.2\pm0.2$&$1.00\pm0.06$&$1.07\pm0.03$\\
2015-05-31&7173.6641&HERMES&705&120&$-6.7\pm0.2$&$-5.6\pm0.3$&$-6.3\pm0.3$&$0.95\pm0.07$&$1.05\pm0.04$\\
2018-08-20&8351.4458&FIES&900&123&$-6.2\pm0.3$&$-5.0\pm0.4$&$-6.6\pm0.4$&$0.88\pm0.05$&$1.05\pm0.06$\\
2018-08-21&8352.3777&FIES&500&143&$-6.48\pm0.19$&$-4.3\pm0.3$&$-6.4\pm0.3$&$0.73\pm0.05$&$1.04\pm0.03$\\
2018-08-22&8353.3775&FIES&361&158&$-7.5\pm0.3$&$-4.2\pm0.5$&$-7.7\pm0.3$&$0.83\pm0.07$&$1.04\pm0.03$\\
2018-08-23&8354.3523&FIES&391&184&$-6.8\pm0.2$&$-6.0\pm0.3$&$-6.6\pm0.3$&$0.88\pm0.07$&$1.05\pm0.03$\\
2018-08-24&8355.3553&FIES&315&151&$-7.7\pm0.2$&$-5.2\pm0.4$&$-7.1\pm0.3$&$1.13\pm0.05$&$1.09\pm0.03$\\
2022-07-07&9767.9648&ESPaDOnS&6360&1051&$-9.2\pm0.1$&$-7.57\pm0.12$&$-8.9\pm0.1$&$0.68\pm0.02$&$1.01\pm0.01$\\
2022-07-08&9768.9499&ESPaDOnS&6360&1937&$-9.0\pm0.1$&$-7.2\pm0.1$&$-8.9\pm0.1$&$0.46\pm0.02$&$0.92\pm0.01$\\
2022-07-11&9771.9016&ESPaDOnS&6360&1671&$-9.5\pm0.1$&$-8.7\pm0.1$&$-9.2\pm0.1$&$0.913\pm0.018$&$1.04\pm0.01$\\
2022-08-09&9801.3812&FIES&941&167&$-6.0\pm0.3$&$-5.3\pm0.3$&$-5.7\pm0.3$&$1.01\pm0.05$&$1.06\pm0.02$\\
2022-08-10&9802.4408&FIES&358&237&$-5.07\pm0.19$&$-5.5\pm0.3$&$-3.9\pm0.2$&$0.78\pm0.06$&$1.005\pm0.019$\\
2022-08-11&9803.4395&FIES&385&222&$-5.77\pm0.18$&$-4.5\pm0.3$&$-5.1\pm0.2$&$0.94\pm0.06$&$1.05\pm0.02$\\
2022-08-12&9804.4254&FIES&449&120&$-5.6\pm0.2$&$-6.6\pm0.3$&$-4.3\pm0.3$&$0.8\pm0.3$&$0.99\pm0.02$\\
2022-08-13&9805.4277&FIES&274&178&$-7.0\pm0.2$&$-5.8\pm0.4$&$-6.4\pm0.3$&$0.70\pm0.09$&$1.00\pm0.03$\\
2022-08-22&9814.3510&HERMES&450&210&$-7.2\pm0.3$&$-4.9\pm0.4$&$-7.5\pm0.4$&$0.96\pm0.04$&$1.07\pm0.03$\\
2022-08-23&9815.3564&HERMES&400&167&$-5.0\pm0.2$&$-4.1\pm0.3$&$-4.1\pm0.3$&$1.08\pm0.06$&$1.09\pm0.02$\\
2022-08-24&9816.3511&HERMES&300&175&$-3.3\pm0.3$&$-2.0\pm0.4$&$-2.9\pm0.4$&$1.12\pm0.07$&$1.11\pm0.03$\\
2022-08-25&9817.3460&HERMES&300&157&$-3.9\pm0.4$&$-2.9\pm0.4$&$-3.1\pm0.4$&$1.08\pm0.07$&$1.09\pm0.04$\\
2022-08-26&9818.3454&HERMES&300&163&$-5.22\pm0.17$&$-4.0\pm0.3$&$-4.2\pm0.2$&$1.08\pm0.06$&$1.09\pm0.03$\\
2022-08-27&9819.3392&HERMES&400&118&$-3.0\pm0.2$&$-1.7\pm0.3$&$-2.1\pm0.2$&$1.16\pm0.05$&$1.12\pm0.02$\\
2022-08-28&9820.3401&HERMES&500&210&$-8.00\pm0.18$&$-5.9\pm0.3$&$-8.0\pm0.2$&$1.10\pm0.04$&$1.097\pm0.019$\\
2022-08-29&9821.3542&HERMES&500&191&$-5.2\pm0.2$&$-4.1\pm0.3$&$-4.9\pm0.3$&$0.90\pm0.04$&$1.05\pm0.02$\\
2022-08-30&9822.3537&HERMES&500&204&$-6.8\pm0.2$&$-6.3\pm0.3$&$-5.6\pm0.3$&$0.87\pm0.04$&$1.04\pm0.02$\\
\hline
\end{tabular}
\label{tab:obs_log}
\end{table*}

\begin{table*}
\centering
\caption{Spectral type and stellar parameters for confirmed Galactic magnetic O-type stars.}
\begin{tabular}{lccccccc}
\hline
\hline
ID&Spectral&Reference&$\log\mathcal{L}$&$T_{\mathrm{eff}}$&$\log g$&Reference\\
&Type&&[$\mathcal{L}_{\odot}$]&[kK]&[cgs]&\\
\hline
CPD-28$^{\circ}$~2561&O6.5 f?p&\cite{sota_2014}&$3.57\pm0.14$&$35\pm2$&$4.0\pm0.1$&\cite{Wade_2015_cpd}\\

HD 108&O6.5-8.5 f?p&\cite{jesus_2019}&$4.07\pm0.22$&$35\pm2$&$3.5\pm0.2$&\cite{martins_2010}\\

HD 37022&O7 f?p var&\cite{jesus_2019}&$3.66\pm0.11$&$39\pm1$&$4.1\pm0.1$&\cite{simon_2006}\\

HD 54879&O9.7~V&\cite{sota_2011}&$3.33\pm0.10$&$30.5\pm0.5$&$4.0\pm0.1$&\cite{shenar_2017}\\ 

HD 57682&O9.2 IV&\cite{sota_2014}&$3.54\pm0.21$&$34.5\pm1.0$&$4.0\pm0.2$&\cite{Grunhut_2009}\\ 

HD 148937&O6 f?p&\cite{sota_2014}&$3.67\pm0.10$&$37.2^{+0.9}_{-0.4}$&$4.00\pm0.09$&\cite{Frost_2024}\\ 

HD 191612&O6-8 f?p var&\cite{sota_2011}&$4.07\pm0.11$&$35\pm1$&$3.5\pm0.1$&\cite{howarth_2007}\\ 

NGC 1624-2&O6.5-8f?cp&\cite{wade_ngc1624}&$3.57\pm0.22$&$35\pm2$&$4.0\pm0.2$&\cite{wade_ngc1624}\\ 

Tr 16-22&O8.5 Vp&\cite{sota_2014}&$3.52\pm0.22$&$34\pm2$&$4.0\pm0.2$$^{\dagger}$&\cite{Gagne_2011}\\

$\zeta$~Ori Aa&O9.2~Ib~varNwk&\cite{sota_2014}&$4.02\pm0.12$&$29.5\pm1.0$&$3.25\pm0.10$&\cite{bouret_2008}\\ 
\hline
\end{tabular}
\label{tab:magOstars}
\tablecomments{$^{\dagger}$ Following \cite{petit_2013} we assume $\log g=4.0\pm0.2$.}
\tablecomments{The spectroscopic luminosity $\mathcal{L}:=T_{\mathrm{eff}}^{4}/g$ is calculated from the literature $T_{\mathrm{eff}}$ and $\log g$ values.}
\tablecomments{For binary systems parameters are only given for the magnetic component.}
\end{table*}

\clearpage
\bibliography{new.ms}{}

\begin{thebibliography}{}
\expandafter\ifx\csname natexlab\endcsname\relax\def\natexlab#1{#1}\fi
\providecommand{\url}[1]{\href{#1}{#1}}
\providecommand{\dodoi}[1]{doi:~\href{http://doi.org/#1}{\nolinkurl{#1}}}
\providecommand{\doeprint}[1]{\href{http://ascl.net/#1}{\nolinkurl{http://ascl.net/#1}}}
\providecommand{\doarXiv}[1]{\href{https://arxiv.org/abs/#1}{\nolinkurl{https://arxiv.org/abs/#1}}}

\bibitem[{{Aerts}(2021)}]{aert_2021}
{Aerts}, C. 2021, Reviews of Modern Physics, 93, 015001,
  \dodoi{10.1103/RevModPhys.93.015001}

\bibitem[{{Agrawal} {et~al.}(2022){Agrawal}, {Sz{\'e}csi}, {Stevenson},
  {Eldridge}, \& {Hurley}}]{Agrawal_2022}
{Agrawal}, P., {Sz{\'e}csi}, D., {Stevenson}, S., {Eldridge}, J.~J., \&
  {Hurley}, J. 2022, \mnras, 512, 5717, \dodoi{10.1093/mnras/stac930}

\bibitem[{{Aigrain} {et~al.}(2016){Aigrain}, {Parviainen}, \&
  {Pope}}]{Aigrain_2016}
{Aigrain}, S., {Parviainen}, H., \& {Pope}, B.~J.~S. 2016, \mnras, 459, 2408,
  \dodoi{10.1093/mnras/stw706}

\bibitem[{{Astropy Collaboration} {et~al.}(2022){Astropy Collaboration},
  {Price-Whelan}, {Lim}, {Earl}, {Starkman}, {Bradley}, {Shupe}, {Patil},
  {Corrales}, {Brasseur}, {N{\"o}the}, {Donath}, {Tollerud}, {Morris},
  {Ginsburg}, {Vaher}, {Weaver}, {Tocknell}, {Jamieson}, {van Kerkwijk},
  {Robitaille}, {Merry}, {Bachetti}, {G{\"u}nther}, {Aldcroft},
  {Alvarado-Montes}, {Archibald}, {B{\'o}di}, {Bapat}, {Barentsen},
  {Baz{\'a}n}, {Biswas}, {Boquien}, {Burke}, {Cara}, {Cara}, {Conroy},
  {Conseil}, {Craig}, {Cross}, {Cruz}, {D'Eugenio}, {Dencheva}, {Devillepoix},
  {Dietrich}, {Eigenbrot}, {Erben}, {Ferreira}, {Foreman-Mackey}, {Fox},
  {Freij}, {Garg}, {Geda}, {Glattly}, {Gondhalekar}, {Gordon}, {Grant},
  {Greenfield}, {Groener}, {Guest}, {Gurovich}, {Handberg}, {Hart},
  {Hatfield-Dodds}, {Homeier}, {Hosseinzadeh}, {Jenness}, {Jones}, {Joseph},
  {Kalmbach}, {Karamehmetoglu}, {Ka{\l}uszy{\'n}ski}, {Kelley}, {Kern},
  {Kerzendorf}, {Koch}, {Kulumani}, {Lee}, {Ly}, {Ma}, {MacBride}, {Maljaars},
  {Muna}, {Murphy}, {Norman}, {O'Steen}, {Oman}, {Pacifici}, {Pascual},
  {Pascual-Granado}, {Patil}, {Perren}, {Pickering}, {Rastogi}, {Roulston},
  {Ryan}, {Rykoff}, {Sabater}, {Sakurikar}, {Salgado}, {Sanghi}, {Saunders},
  {Savchenko}, {Schwardt}, {Seifert-Eckert}, {Shih}, {Jain}, {Shukla}, {Sick},
  {Simpson}, {Singanamalla}, {Singer}, {Singhal}, {Sinha}, {Sip{\H{o}}cz},
  {Spitler}, {Stansby}, {Streicher}, {{\v{S}}umak}, {Swinbank}, {Taranu},
  {Tewary}, {Tremblay}, {de Val-Borro}, {Van Kooten}, {Vasovi{\'c}}, {Verma},
  {de Miranda Cardoso}, {Williams}, {Wilson}, {Winkel}, {Wood-Vasey}, {Xue},
  {Yoachim}, {Zhang}, {Zonca}, \& {Astropy Project
  Contributors}}]{astropy_2022}
{Astropy Collaboration}, {Price-Whelan}, A.~M., {Lim}, P.~L., {et~al.} 2022,
  \apj, 935, 167, \dodoi{10.3847/1538-4357/ac7c74}

\bibitem[{{Bagnulo} {et~al.}(2013){Bagnulo}, {Fossati}, {Kochukhov}, \&
  {Landstreet}}]{Bagnulo_2013}
{Bagnulo}, S., {Fossati}, L., {Kochukhov}, O., \& {Landstreet}, J.~D. 2013,
  \aap, 559, A103, \dodoi{10.1051/0004-6361/201322319}

\bibitem[{{Bagnulo} {et~al.}(2015){Bagnulo}, {Fossati}, {Landstreet}, \&
  {Izzo}}]{bagnulo_2015}
{Bagnulo}, S., {Fossati}, L., {Landstreet}, J.~D., \& {Izzo}, C. 2015, \aap,
  583, A115, \dodoi{10.1051/0004-6361/201526497}

\bibitem[{{Bagnulo} {et~al.}(2012){Bagnulo}, {Landstreet}, {Fossati}, \&
  {Kochukhov}}]{Bagnulo_2012}
{Bagnulo}, S., {Landstreet}, J.~D., {Fossati}, L., \& {Kochukhov}, O. 2012,
  \aap, 538, A129, \dodoi{10.1051/0004-6361/201118098}

\bibitem[{{Bailer-Jones} {et~al.}(2021){Bailer-Jones}, {Rybizki}, {Fouesneau},
  {Demleitner}, \& {Andrae}}]{bailer_jones_2021}
{Bailer-Jones}, C.~A.~L., {Rybizki}, J., {Fouesneau}, M., {Demleitner}, M., \&
  {Andrae}, R. 2021, \aj, 161, 147, \dodoi{10.3847/1538-3881/abd806}

\bibitem[{{Bailer-Jones} {et~al.}(2018){Bailer-Jones}, {Rybizki}, {Fouesneau},
  {Mantelet}, \& {Andrae}}]{Bailer_Jones_2018}
{Bailer-Jones}, C.~A.~L., {Rybizki}, J., {Fouesneau}, M., {Mantelet}, G., \&
  {Andrae}, R. 2018, \aj, 156, 58, \dodoi{10.3847/1538-3881/aacb21}

\bibitem[{{Baluev}(2008)}]{Baluev_2008}
{Baluev}, R.~V. 2008, \mnras, 385, 1279,
  \dodoi{10.1111/j.1365-2966.2008.12689.x}

\bibitem[{{Barannikov}(2007)}]{Barannikov_2007}
{Barannikov}, A.~A. 2007, Information Bulletin on Variable Stars, 5756, 1

\bibitem[{{Barentsen} {et~al.}(2020){Barentsen}, {Colon}, {Barclay}, {Debie},
  {O'Leary}, {Coughlin}, {Cody}, {Caldwell}, {Lapp}, {Leno}, {Gully},
  {Mullally}, {MinchinWeb}, {Hedges}, {Winlu}, {Hartley}, {Vin{\'\i}cius},
  {Foivos}, {Lee}, {Doig}, {Arnould}, {Fijam}, {Jessie-Dotson}, {Tsutsumi},
  {Saunders}, {Abrahamsen}, {De Val-Borro}, {Bowling}, {Neal}, \&
  {Kl{\"a}rner}}]{Barensten_2020}
{Barentsen}, G., {Colon}, K., {Barclay}, T., {et~al.} 2020,
  {KeplerGO/KeplerScienceWebsite}, v20200203,  Zenodo,
  \dodoi{10.5281/zenodo.593417}

\bibitem[{{Barker} {et~al.}(1981){Barker}, {Landstreet}, {Marlborough},
  {Thompson}, \& {Maza}}]{Barker_1981}
{Barker}, P.~K., {Landstreet}, J.~D., {Marlborough}, J.~M., {Thompson}, I., \&
  {Maza}, J. 1981, \apj, 250, 300, \dodoi{10.1086/159375}

\bibitem[{{Bessell}(2000)}]{Bessell_2000}
{Bessell}, M.~S. 2000, \pasp, 112, 961, \dodoi{10.1086/316598}

\bibitem[{{Blaha} \& {Humphreys}(1989)}]{Blaha_1989}
{Blaha}, C., \& {Humphreys}, R.~M. 1989, \aj, 98, 1598, \dodoi{10.1086/115244}

\bibitem[{{Blaz{\`e}re} {et~al.}(2015){Blaz{\`e}re}, {Neiner}, {Tkachenko},
  {Bouret}, \& {Rivinius}}]{Blazere_2015}
{Blaz{\`e}re}, A., {Neiner}, C., {Tkachenko}, A., {Bouret}, J.~C., \&
  {Rivinius}, T. 2015, \aap, 582, A110, \dodoi{10.1051/0004-6361/201526855}

\bibitem[{{Blomme} {et~al.}(2023){Blomme}, {Fr{\'e}mat}, {Sartoretti},
  {Guerrier}, {Panuzzo}, {Katz}, {Seabroke}, {Th{\'e}venin}, {Cropper},
  {Benson}, {Damerdji}, {Haigron}, {Marchal}, {Smith}, {Baker}, {Chemin},
  {David}, {Dolding}, {Gosset}, {Jan{\ss}en}, {Jasniewicz}, {Lobel}, {Plum},
  {Samaras}, {Snaith}, {Soubiran}, {Vanel}, {Zwitter}, {Brouillet}, {Caffau},
  {Crifo}, {Fabre}, {Fragkoudi}, {Huckle}, {Jean-Antoine Piccolo}, {Lasne},
  {Leclerc}, {Mastrobuono-Battisti}, {Royer}, {Viala}, \&
  {Zorec}}]{blomme_2023}
{Blomme}, R., {Fr{\'e}mat}, Y., {Sartoretti}, P., {et~al.} 2023, \aap, 674, A7,
  \dodoi{10.1051/0004-6361/202243685}

\bibitem[{{Bodensteiner} {et~al.}(2021){Bodensteiner}, {Sana}, {Wang},
  {Langer}, {Mahy}, {Banyard}, {de Koter}, {de Mink}, {Evans}, {G{\"o}tberg},
  {Patrick}, {Schneider}, \& {Tramper}}]{Bodensteiner_2021}
{Bodensteiner}, J., {Sana}, H., {Wang}, C., {et~al.} 2021, \aap, 652, A70,
  \dodoi{10.1051/0004-6361/202140507}

\bibitem[{{Bodensteiner} {et~al.}(2025){Bodensteiner}, {Shenar}, {Sana},
  {Britavskiy}, {Crowther}, {Langer}, {Lennon}, {Mahy}, {Patrick},
  {Villase{\~n}or}, {Abdul-Masih}, {Bowman}, {de Koter}, {de Mink}, {Deshmukh},
  {Fabry}, {Gilkis}, {G{\"o}tberg}, {Holgado}, {Izzard}, {Janssens}, {Kalari},
  {Keszthelyi}, {Kub{\'a}t}, {Mandel}, {Maravelias}, {Oskinova}, {Pauli},
  {Ramachandran}, {Rocha}, {Renzo}, {Sander}, {Schneider}, {Schootemeijer},
  {Sen}, {Stoop}, {Toonen}, {van Loon}, {Valli}, {Vigna-G{\'o}mez}, {Vink},
  {Wang}, \& {Xu}}]{Bodensteiner_2025}
{Bodensteiner}, J., {Shenar}, T., {Sana}, H., {et~al.} 2025, \aap, 698, A38,
  \dodoi{10.1051/0004-6361/202452623}

\bibitem[{{Borucki} {et~al.}(2010){Borucki}, {Koch}, {Basri}, {Batalha},
  {Brown}, {Caldwell}, {Caldwell}, {Christensen-Dalsgaard}, {Cochran},
  {DeVore}, {Dunham}, {Dupree}, {Gautier}, {Geary}, {Gilliland}, {Gould},
  {Howell}, {Jenkins}, {Kondo}, {Latham}, {Marcy}, {Meibom}, {Kjeldsen},
  {Lissauer}, {Monet}, {Morrison}, {Sasselov}, {Tarter}, {Boss}, {Brownlee},
  {Owen}, {Buzasi}, {Charbonneau}, {Doyle}, {Fortney}, {Ford}, {Holman},
  {Seager}, {Steffen}, {Welsh}, {Rowe}, {Anderson}, {Buchhave}, {Ciardi},
  {Walkowicz}, {Sherry}, {Horch}, {Isaacson}, {Everett}, {Fischer}, {Torres},
  {Johnson}, {Endl}, {MacQueen}, {Bryson}, {Dotson}, {Haas}, {Kolodziejczak},
  {Van Cleve}, {Chandrasekaran}, {Twicken}, {Quintana}, {Clarke}, {Allen},
  {Li}, {Wu}, {Tenenbaum}, {Verner}, {Bruhweiler}, {Barnes}, \&
  {Prsa}}]{Borucki_2010}
{Borucki}, W.~J., {Koch}, D., {Basri}, G., {et~al.} 2010, Science, 327, 977,
  \dodoi{10.1126/science.1185402}

\bibitem[{{Bouret} {et~al.}(2008){Bouret}, {Donati}, {Martins}, {Escolano},
  {Marcolino}, {Lanz}, \& {Howarth}}]{bouret_2008}
{Bouret}, J.~C., {Donati}, J.~F., {Martins}, F., {et~al.} 2008, \mnras, 389,
  75, \dodoi{10.1111/j.1365-2966.2008.13575.x}

\bibitem[{{Bowman} {et~al.}(2020){Bowman}, {Burssens}, {Sim{\'o}n-D{\'\i}az},
  {Edelmann}, {Rogers}, {Horst}, {R{\"o}pke}, \& {Aerts}}]{Bowman_2020}
{Bowman}, D.~M., {Burssens}, S., {Sim{\'o}n-D{\'\i}az}, S., {et~al.} 2020,
  \aap, 640, A36, \dodoi{10.1051/0004-6361/202038224}

\bibitem[{{Bowman} {et~al.}(2024){Bowman}, {Van Daele}, {Michielsen}, \& {Van
  Reeth}}]{Bowman_2024}
{Bowman}, D.~M., {Van Daele}, P., {Michielsen}, M., \& {Van Reeth}, T. 2024,
  \aap, 692, A49, \dodoi{10.1051/0004-6361/202451419}

\bibitem[{{Breger} {et~al.}(1993){Breger}, {Stich}, {Garrido}, {Martin},
  {Jiang}, {Li}, {Hube}, {Ostermann}, {Paparo}, \& {Scheck}}]{Breger_1993}
{Breger}, M., {Stich}, J., {Garrido}, R., {et~al.} 1993, \aap, 271, 482

\bibitem[{{Brott} {et~al.}(2011){Brott}, {de Mink}, {Cantiello}, {Langer}, {de
  Koter}, {Evans}, {Hunter}, {Trundle}, \& {Vink}}]{Brott_2011}
{Brott}, I., {de Mink}, S.~E., {Cantiello}, M., {et~al.} 2011, \aap, 530, A115,
  \dodoi{10.1051/0004-6361/201016113}

\bibitem[{{Burssens} {et~al.}(2020){Burssens}, {Sim{\'o}n-D{\'\i}az}, {Bowman},
  {Holgado}, {Michielsen}, {de Burgos}, {Castro}, {Barb{\'a}}, \&
  {Aerts}}]{Burssens_2020}
{Burssens}, S., {Sim{\'o}n-D{\'\i}az}, S., {Bowman}, D.~M., {et~al.} 2020,
  \aap, 639, A81, \dodoi{10.1051/0004-6361/202037700}

\bibitem[{{Buysschaert} {et~al.}(2017){Buysschaert}, {Neiner}, {Richardson},
  {Ramiaramanantsoa}, {David-Uraz}, {Pablo}, {Oksala}, {Moffat}, {Mennickent},
  {Legeza}, {Aerts}, {Kuschnig}, {Whittaker}, {Popowicz}, {Handler}, {Wade}, \&
  {Weiss}}]{Buysschaert_2017}
{Buysschaert}, B., {Neiner}, C., {Richardson}, N.~D., {et~al.} 2017, \aap, 602,
  A91, \dodoi{10.1051/0004-6361/201630318}

\bibitem[{{Cantiello} {et~al.}(2021){Cantiello}, {Lecoanet}, {Jermyn}, \&
  {Grassitelli}}]{Cantiello_2021}
{Cantiello}, M., {Lecoanet}, D., {Jermyn}, A.~S., \& {Grassitelli}, L. 2021,
  \apj, 915, 112, \dodoi{10.3847/1538-4357/ac03b0}

\bibitem[{{Cantiello} {et~al.}(2009){Cantiello}, {Langer}, {Brott}, {de Koter},
  {Shore}, {Vink}, {Voegler}, {Lennon}, \& {Yoon}}]{Cantiello_2009}
{Cantiello}, M., {Langer}, N., {Brott}, I., {et~al.} 2009, \aap, 499, 279,
  \dodoi{10.1051/0004-6361/200911643}

\bibitem[{{Castro} {et~al.}(2015){Castro}, {Fossati}, {Hubrig},
  {Sim{\'o}n-D{\'\i}az}, {Sch{\"o}ller}, {Ilyin}, {Carrol}, {Langer}, {Morel},
  {Schneider}, {Przybilla}, {Herrero}, {de Koter}, {Oskinova}, {Reisenegger},
  {Sana}, \& {BOB Collaboration}}]{Castro_2015}
{Castro}, N., {Fossati}, L., {Hubrig}, S., {et~al.} 2015, \aap, 581, A81,
  \dodoi{10.1051/0004-6361/201425354}

\bibitem[{{Chesneau} \& {Moffat}(2002)}]{Chesneau_2002}
{Chesneau}, O., \& {Moffat}, A.~F.~J. 2002, \pasp, 114, 612,
  \dodoi{10.1086/341683}

\bibitem[{{Chini} {et~al.}(2012){Chini}, {Hoffmeister}, {Nasseri}, {Stahl}, \&
  {Zinnecker}}]{chini_2012}
{Chini}, R., {Hoffmeister}, V.~H., {Nasseri}, A., {Stahl}, O., \& {Zinnecker},
  H. 2012, \mnras, 424, 1925, \dodoi{10.1111/j.1365-2966.2012.21317.x}

\bibitem[{{Claret}(2000)}]{Claret_2000}
{Claret}, A. 2000, \aap, 363, 1081

\bibitem[{{Clough} {et~al.}(1992){Clough}, {Iacono}, \& {Moncet}}]{Clough_1992}
{Clough}, S.~A., {Iacono}, M.~J., \& {Moncet}, J.-L. 1992, \jgr, 97, 15,761,
  \dodoi{10.1029/92JD01419}

\bibitem[{{Clough} {et~al.}(2005){Clough}, {Shephard}, {Mlawer}, {Delamere},
  {Iacono}, {Cady-Pereira}, {Boukabara}, \& {Brown}}]{Clough_2005}
{Clough}, S.~A., {Shephard}, M.~W., {Mlawer}, E.~J., {et~al.} 2005, \jqsrt, 91,
  233, \dodoi{10.1016/j.jqsrt.2004.05.058}

\bibitem[{{Conti} {et~al.}(1977){Conti}, {Leep}, \& {Lorre}}]{conti_1977}
{Conti}, P.~S., {Leep}, E.~M., \& {Lorre}, J.~J. 1977, \apj, 214, 759,
  \dodoi{10.1086/155305}

\bibitem[{{David-Uraz} {et~al.}(2014){David-Uraz}, {Wade}, {Petit}, {ud-Doula},
  {Sundqvist}, {Grunhut}, {Shultz}, {Neiner}, {Alecian}, {Henrichs}, {Bouret},
  \& {MiMeS Collaboration}}]{david_uraz_2014}
{David-Uraz}, A., {Wade}, G.~A., {Petit}, V., {et~al.} 2014, \mnras, 444, 429,
  \dodoi{10.1093/mnras/stu1458}

\bibitem[{{de Jong} {et~al.}(2001){de Jong}, {Henrichs}, {Kaper}, {Nichols},
  {Bjorkman}, {Bohlender}, {Cao}, {Gordon}, {Hill}, {Jiang}, {Kolka},
  {Morrison}, {Neff}, {O'Neal}, {Scheers}, \& {Telting}}]{deJong_2001}
{de Jong}, J.~A., {Henrichs}, H.~F., {Kaper}, L., {et~al.} 2001, \aap, 368,
  601, \dodoi{10.1051/0004-6361:20000570}

\bibitem[{{Donati} {et~al.}(2006{\natexlab{a}}){Donati}, {Catala},
  {Landstreet}, \& {Petit}}]{donati_2006}
{Donati}, J.~F., {Catala}, C., {Landstreet}, J.~D., \& {Petit}, P.
  2006{\natexlab{a}}, in Astronomical Society of the Pacific Conference Series,
  Vol. 358, Solar Polarization 4, ed. R.~{Casini} \& B.~W. {Lites}, 362

\bibitem[{{Donati} {et~al.}(2006{\natexlab{b}}){Donati}, {Howarth}, {Bouret},
  {Petit}, {Catala}, \& {Landstreet}}]{donati_2006_HD191612}
{Donati}, J.~F., {Howarth}, I.~D., {Bouret}, J.~C., {et~al.}
  2006{\natexlab{b}}, \mnras, 365, L6, \dodoi{10.1111/j.1745-3933.2005.00115.x}

\bibitem[{{Donati} \& {Landstreet}(2009)}]{donati_2009}
{Donati}, J.~F., \& {Landstreet}, J.~D. 2009, \araa, 47, 333,
  \dodoi{10.1146/annurev-astro-082708-101833}

\bibitem[{{Donati} {et~al.}(1997){Donati}, {Semel}, {Carter}, {Rees}, \&
  {Collier Cameron}}]{donati_1997}
{Donati}, J.~F., {Semel}, M., {Carter}, B.~D., {Rees}, D.~E., \& {Collier
  Cameron}, A. 1997, \mnras, 291, 658, \dodoi{10.1093/mnras/291.4.658}

\bibitem[{{Donati} {et~al.}(1992){Donati}, {Semel}, \& {Rees}}]{Donati_1992}
{Donati}, J.~F., {Semel}, M., \& {Rees}, D.~E. 1992, \aap, 265, 669

\bibitem[{{Ducati}(2002)}]{Ducati_2002}
{Ducati}, J.~R. 2002, {VizieR Online Data Catalog: Catalogue of Stellar
  Photometry in Johnson's 11-color system.}, CDS/ADC Collection of Electronic
  Catalogues, 2237, 0 (2002)

\bibitem[{{Dunstall} {et~al.}(2015){Dunstall}, {Dufton}, {Sana}, {Evans},
  {Howarth}, {Sim{\'o}n-D{\'\i}az}, {de Mink}, {Langer}, {Ma{\'\i}z
  Apell{\'a}niz}, \& {Taylor}}]{Dunstall_2015}
{Dunstall}, P.~R., {Dufton}, P.~L., {Sana}, H., {et~al.} 2015, \aap, 580, A93,
  \dodoi{10.1051/0004-6361/201526192}

\bibitem[{Earl {et~al.}(2024)Earl, Tollerud, O'Steen, brechmos, Kerzendorf,
  Busko, shaileshahuja, D'Avella, Lim, Robitaille, Ginsburg, Homeier, Sipőcz,
  Averbukh, Tocknell, Cherinka, Ogaz, Geda, Conroy, Davies, Günther, Barbary,
  Foster, Droettboom, Nguyen, Bray, Casey, Cruz, Ferguson, \&
  Crawford}]{nicholas_earl_2024_14042033}
Earl, N., Tollerud, E., O'Steen, R., {et~al.} 2024, astropy/specutils: v1.19.0,
  v1.19.0,  Zenodo, \dodoi{10.5281/zenodo.14042033}

\bibitem[{{Ekstr{\"o}m} {et~al.}(2008){Ekstr{\"o}m}, {Meynet}, {Maeder}, \&
  {Barblan}}]{Ekstrom_2008}
{Ekstr{\"o}m}, S., {Meynet}, G., {Maeder}, A., \& {Barblan}, F. 2008, \aap,
  478, 467, \dodoi{10.1051/0004-6361:20078095}

\bibitem[{{Ekstr{\"o}m} {et~al.}(2012){Ekstr{\"o}m}, {Georgy}, {Eggenberger},
  {Meynet}, {Mowlavi}, {Wyttenbach}, {Granada}, {Decressin}, {Hirschi},
  {Frischknecht}, {Charbonnel}, \& {Maeder}}]{Ekstrom_2012}
{Ekstr{\"o}m}, S., {Georgy}, C., {Eggenberger}, P., {et~al.} 2012, \aap, 537,
  A146, \dodoi{10.1051/0004-6361/201117751}

\bibitem[{{ESA}(1997)}]{ESA_1997}
{ESA}, ed. 1997, ESA Special Publication, Vol. 1200, {The HIPPARCOS and TYCHO
  catalogues. Astrometric and photometric star catalogues derived from the ESA
  HIPPARCOS Space Astrometry Mission}

\bibitem[{{Fan} {et~al.}(2019){Fan}, {Hobbs}, {Dahlstrom}, {Welty}, {York},
  {Rachford}, {Snow}, {Sonnentrucker}, {Baskes}, \& {Zhao}}]{Fan_2019}
{Fan}, H., {Hobbs}, L.~M., {Dahlstrom}, J.~A., {et~al.} 2019, \apj, 878, 151,
  \dodoi{10.3847/1538-4357/ab1b74}

\bibitem[{{Ferrario} {et~al.}(2009){Ferrario}, {Pringle}, {Tout}, \&
  {Wickramasinghe}}]{Ferrario_2009}
{Ferrario}, L., {Pringle}, J.~E., {Tout}, C.~A., \& {Wickramasinghe}, D.~T.
  2009, \mnras, 400, L71, \dodoi{10.1111/j.1745-3933.2009.00765.x}

\bibitem[{{Ferrario} \& {Wickramasinghe}(2006)}]{Ferrario_2006}
{Ferrario}, L., \& {Wickramasinghe}, D. 2006, \mnras, 367, 1323,
  \dodoi{10.1111/j.1365-2966.2006.10058.x}

\bibitem[{{Folsom} {et~al.}(2025){Folsom}, {Erba}, {Petit}, {Seadrow},
  {Stanley}, {Natan}, {Zaire}, {Oksala}, {Villadiego Forero}, {Moore}, \&
  {Catalan Olais}}]{Folsom_2025}
{Folsom}, C.~P., {Erba}, C., {Petit}, V., {et~al.} 2025, Journal of Open Source
  Software, 10, 7891, \dodoi{10.21105/joss.07891}

\bibitem[{{Fossati} {et~al.}(2016){Fossati}, {Schneider}, {Castro}, {Langer},
  {Sim{\'o}n-D{\'\i}az}, {M{\"u}ller}, {de Koter}, {Morel}, {Petit}, {Sana}, \&
  {Wade}}]{Fossati_2016}
{Fossati}, L., {Schneider}, F.~R.~N., {Castro}, N., {et~al.} 2016, \aap, 592,
  A84, \dodoi{10.1051/0004-6361/201628259}

\bibitem[{{Frost} {et~al.}(2024){Frost}, {Sana}, {Mahy}, {Wade}, {Barron}, {Le
  Bouquin}, {M{\'e}rand}, {Schneider}, {Shenar}, {Barb{\'a}}, {Bowman},
  {Fabry}, {Farhang}, {Marchant}, {Morrell}, \& {Smoker}}]{Frost_2024}
{Frost}, A.~J., {Sana}, H., {Mahy}, L., {et~al.} 2024, Science, 384, 214,
  \dodoi{10.1126/science.adg7700}

\bibitem[{{Gagn{\'e}} {et~al.}(2011){Gagn{\'e}}, {Fehon}, {Savoy}, {Cohen},
  {Townsley}, {Broos}, {Povich}, {Corcoran}, {Walborn}, {Remage Evans},
  {Moffat}, {Naz{\'e}}, \& {Oskinova}}]{Gagne_2011}
{Gagn{\'e}}, M., {Fehon}, G., {Savoy}, M.~R., {et~al.} 2011, \apjs, 194, 5,
  \dodoi{10.1088/0067-0049/194/1/5}

\bibitem[{{Gaia Collaboration} {et~al.}(2018){Gaia Collaboration}, {Brown},
  {Vallenari}, {Prusti}, {de Bruijne}, {Babusiaux}, {Bailer-Jones}, {Biermann},
  {Evans}, {Eyer}, {Jansen}, {Jordi}, {Klioner}, {Lammers}, {Lindegren},
  {Luri}, {Mignard}, {Panem}, {Pourbaix}, {Randich}, {Sartoretti}, {Siddiqui},
  {Soubiran}, {van Leeuwen}, {Walton}, {Arenou}, {Bastian}, {Cropper},
  {Drimmel}, {Katz}, {Lattanzi}, {Bakker}, {Cacciari}, {Casta{\~n}eda},
  {Chaoul}, {Cheek}, {De Angeli}, {Fabricius}, {Guerra}, {Holl}, {Masana},
  {Messineo}, {Mowlavi}, {Nienartowicz}, {Panuzzo}, {Portell}, {Riello},
  {Seabroke}, {Tanga}, {Th{\'e}venin}, {Gracia-Abril}, {Comoretto},
  {Garcia-Reinaldos}, {Teyssier}, {Altmann}, {Andrae}, {Audard},
  {Bellas-Velidis}, {Benson}, {Berthier}, {Blomme}, {Burgess}, {Busso},
  {Carry}, {Cellino}, {Clementini}, {Clotet}, {Creevey}, {Davidson}, {De
  Ridder}, {Delchambre}, {Dell'Oro}, {Ducourant},
  {Fern{\'a}ndez-Hern{\'a}ndez}, {Fouesneau}, {Fr{\'e}mat}, {Galluccio},
  {Garc{\'\i}a-Torres}, {Gonz{\'a}lez-N{\'u}{\~n}ez}, {Gonz{\'a}lez-Vidal},
  {Gosset}, {Guy}, {Halbwachs}, {Hambly}, {Harrison}, {Hern{\'a}ndez},
  {Hestroffer}, {Hodgkin}, {Hutton}, {Jasniewicz}, {Jean-Antoine-Piccolo},
  {Jordan}, {Korn}, {Krone-Martins}, {Lanzafame}, {Lebzelter}, {L{\"o}ffler},
  {Manteiga}, {Marrese}, {Mart{\'\i}n-Fleitas}, {Moitinho}, {Mora}, {Muinonen},
  {Osinde}, {Pancino}, {Pauwels}, {Petit}, {Recio-Blanco}, {Richards},
  {Rimoldini}, {Robin}, {Sarro}, {Siopis}, {Smith}, {Sozzetti}, {S{\"u}veges},
  {Torra}, {van Reeven}, {Abbas}, {Abreu Aramburu}, {Accart}, {Aerts},
  {Altavilla}, {{\'A}lvarez}, {Alvarez}, {Alves}, {Anderson}, {Andrei},
  {Anglada Varela}, {Antiche}, {Antoja}, {Arcay}, {Astraatmadja}, {Bach},
  {Baker}, {Balaguer-N{\'u}{\~n}ez}, {Balm}, {Barache}, {Barata}, {Barbato},
  {Barblan}, {Barklem}, {Barrado}, {Barros}, {Barstow}, {Bartholom{\'e}
  Mu{\~n}oz}, {Bassilana}, {Becciani}, {Bellazzini}, {Berihuete}, {Bertone},
  {Bianchi}, {Bienaym{\'e}}, {Blanco-Cuaresma}, {Boch}, {Boeche}, {Bombrun},
  {Borrachero}, {Bossini}, {Bouquillon}, {Bourda}, {Bragaglia}, {Bramante},
  {Breddels}, {Bressan}, {Brouillet}, {Br{\"u}semeister}, {Brugaletta},
  {Bucciarelli}, {Burlacu}, {Busonero}, {Butkevich}, {Buzzi}, {Caffau},
  {Cancelliere}, {Cannizzaro}, {Cantat-Gaudin}, {Carballo}, {Carlucci},
  {Carrasco}, {Casamiquela}, {Castellani}, {Castro-Ginard}, {Charlot},
  {Chemin}, {Chiavassa}, {Cocozza}, {Costigan}, {Cowell}, {Crifo}, {Crosta},
  {Crowley}, {Cuypers}, {Dafonte}, {Damerdji}, {Dapergolas}, {David}, {David},
  {de Laverny}, {De Luise}, {De March}, {de Martino}, {de Souza}, {de Torres},
  {Debosscher}, {del Pozo}, {Delbo}, {Delgado}, {Delgado}, {Di Matteo},
  {Diakite}, {Diener}, {Distefano}, {Dolding}, {Drazinos}, {Dur{\'a}n},
  {Edvardsson}, {Enke}, {Eriksson}, {Esquej}, {Eynard Bontemps}, {Fabre},
  {Fabrizio}, {Faigler}, {Falc{\~a}o}, {Farr{\`a}s Casas}, {Federici},
  {Fedorets}, {Fernique}, {Figueras}, {Filippi}, {Findeisen}, {Fonti},
  {Fraile}, {Fraser}, {Fr{\'e}zouls}, {Gai}, {Galleti}, {Garabato},
  {Garc{\'\i}a-Sedano}, {Garofalo}, {Garralda}, {Gavel}, {Gavras}, {Gerssen},
  {Geyer}, {Giacobbe}, {Gilmore}, {Girona}, {Giuffrida}, {Glass}, {Gomes},
  {Granvik}, {Gueguen}, {Guerrier}, {Guiraud}, {Guti{\'e}rrez-S{\'a}nchez},
  {Haigron}, {Hatzidimitriou}, {Hauser}, {Haywood}, {Heiter}, {Helmi}, {Heu},
  {Hilger}, {Hobbs}, {Hofmann}, {Holland}, {Huckle}, {Hypki}, {Icardi},
  {Jan{\ss}en}, {Jevardat de Fombelle}, {Jonker}, {Juh{\'a}sz}, {Julbe},
  {Karampelas}, {Kewley}, {Klar}, {Kochoska}, {Kohley}, {Kolenberg},
  {Kontizas}, {Kontizas}, {Koposov}, {Kordopatis}, {Kostrzewa-Rutkowska},
  {Koubsky}, {Lambert}, {Lanza}, {Lasne}, {Lavigne}, {Le Fustec}, {Le
  Poncin-Lafitte}, {Lebreton}, {Leccia}, {Leclerc}, {Lecoeur-Taibi},
  {Lenhardt}, {Leroux}, {Liao}, {Licata}, {Lindstr{\o}m}, {Lister}, {Livanou},
  {Lobel}, {L{\'o}pez}, {Managau}, {Mann}, {Mantelet}, {Marchal}, {Marchant},
  {Marconi}, {Marinoni}, {Marschalk{\'o}}, {Marshall}, {Martino}, {Marton},
  {Mary}, {Massari}, {Matijevi{\v{c}}}, {Mazeh}, {McMillan}, {Messina},
  {Michalik}, {Millar}, {Molina}, {Molinaro}, {Moln{\'a}r}, {Montegriffo},
  {Mor}, {Morbidelli}, {Morel}, {Morris}, {Mulone}, {Muraveva}, {Musella},
  {Nelemans}, {Nicastro}, {Noval}, {O'Mullane}, {Ord{\'e}novic},
  {Ord{\'o}{\~n}ez-Blanco}, {Osborne}, {Pagani}, {Pagano}, {Pailler},
  {Palacin}, {Palaversa}, {Panahi}, {Pawlak}, {Piersimoni}, {Pineau}, {Plachy},
  {Plum}, {Poggio}, {Poujoulet}, {Pr{\v{s}}a}, {Pulone}, {Racero}, {Ragaini},
  {Rambaux}, {Ramos-Lerate}, {Regibo}, {Reyl{\'e}}, {Riclet}, {Ripepi}, {Riva},
  {Rivard}, {Rixon}, {Roegiers}, {Roelens}, {Romero-G{\'o}mez}, {Rowell},
  {Royer}, {Ruiz-Dern}, {Sadowski}, {Sagrist{\`a} Sell{\'e}s}, {Sahlmann},
  {Salgado}, {Salguero}, {Sanna}, {Santana-Ros}, {Sarasso}, {Savietto},
  {Schultheis}, {Sciacca}, {Segol}, {Segovia}, {S{\'e}gransan}, {Shih},
  {Siltala}, {Silva}, {Smart}, {Smith}, {Solano}, {Solitro}, {Sordo}, {Soria
  Nieto}, {Souchay}, {Spagna}, {Spoto}, {Stampa}, {Steele},
  {Steidelm{\"u}ller}, {Stephenson}, {Stoev}, {Suess}, {Surdej}, {Szabados},
  {Szegedi-Elek}, {Tapiador}, {Taris}, {Tauran}, {Taylor}, {Teixeira},
  {Terrett}, {Teyssandier}, {Thuillot}, {Titarenko}, {Torra Clotet}, {Turon},
  {Ulla}, {Utrilla}, {Uzzi}, {Vaillant}, {Valentini}, {Valette}, {van Elteren},
  {Van Hemelryck}, {van Leeuwen}, {Vaschetto}, {Vecchiato}, {Veljanoski},
  {Viala}, {Vicente}, {Vogt}, {von Essen}, {Voss}, {Votruba}, {Voutsinas},
  {Walmsley}, {Weiler}, {Wertz}, {Wevers}, {Wyrzykowski}, {Yoldas},
  {{\v{Z}}erjal}, {Ziaeepour}, {Zorec}, {Zschocke}, {Zucker}, {Zurbach}, \&
  {Zwitter}}]{Gaia_2018}
{Gaia Collaboration}, {Brown}, A.~G.~A., {Vallenari}, A., {et~al.} 2018, \aap,
  616, A1, \dodoi{10.1051/0004-6361/201833051}

\bibitem[{{Gaia Collaboration} {et~al.}(2023){Gaia Collaboration}, {Vallenari},
  {Brown}, {Prusti}, {de Bruijne}, {Arenou}, {Babusiaux}, {Biermann},
  {Creevey}, {Ducourant}, {Evans}, {Eyer}, {Guerra}, {Hutton}, {Jordi},
  {Klioner}, {Lammers}, {Lindegren}, {Luri}, {Mignard}, {Panem}, {Pourbaix},
  {Randich}, {Sartoretti}, {Soubiran}, {Tanga}, {Walton}, {Bailer-Jones},
  {Bastian}, {Drimmel}, {Jansen}, {Katz}, {Lattanzi}, {van Leeuwen}, {Bakker},
  {Cacciari}, {Casta{\~n}eda}, {De Angeli}, {Fabricius}, {Fouesneau},
  {Fr{\'e}mat}, {Galluccio}, {Guerrier}, {Heiter}, {Masana}, {Messineo},
  {Mowlavi}, {Nicolas}, {Nienartowicz}, {Pailler}, {Panuzzo}, {Riclet}, {Roux},
  {Seabroke}, {Sordo}, {Th{\'e}venin}, {Gracia-Abril}, {Portell}, {Teyssier},
  {Altmann}, {Andrae}, {Audard}, {Bellas-Velidis}, {Benson}, {Berthier},
  {Blomme}, {Burgess}, {Busonero}, {Busso}, {C{\'a}novas}, {Carry}, {Cellino},
  {Cheek}, {Clementini}, {Damerdji}, {Davidson}, {de Teodoro}, {Nu{\~n}ez
  Campos}, {Delchambre}, {Dell'Oro}, {Esquej}, {Fern{\'a}ndez-Hern{\'a}ndez},
  {Fraile}, {Garabato}, {Garc{\'\i}a-Lario}, {Gosset}, {Haigron}, {Halbwachs},
  {Hambly}, {Harrison}, {Hern{\'a}ndez}, {Hestroffer}, {Hodgkin}, {Holl},
  {Jan{\ss}en}, {Jevardat de Fombelle}, {Jordan}, {Krone-Martins}, {Lanzafame},
  {L{\"o}ffler}, {Marchal}, {Marrese}, {Moitinho}, {Muinonen}, {Osborne},
  {Pancino}, {Pauwels}, {Recio-Blanco}, {Reyl{\'e}}, {Riello}, {Rimoldini},
  {Roegiers}, {Rybizki}, {Sarro}, {Siopis}, {Smith}, {Sozzetti}, {Utrilla},
  {van Leeuwen}, {Abbas}, {{\'A}brah{\'a}m}, {Abreu Aramburu}, {Aerts},
  {Aguado}, {Ajaj}, {Aldea-Montero}, {Altavilla}, {{\'A}lvarez}, {Alves},
  {Anders}, {Anderson}, {Anglada Varela}, {Antoja}, {Baines}, {Baker},
  {Balaguer-N{\'u}{\~n}ez}, {Balbinot}, {Balog}, {Barache}, {Barbato},
  {Barros}, {Barstow}, {Bartolom{\'e}}, {Bassilana}, {Bauchet}, {Becciani},
  {Bellazzini}, {Berihuete}, {Bernet}, {Bertone}, {Bianchi}, {Binnenfeld},
  {Blanco-Cuaresma}, {Blazere}, {Boch}, {Bombrun}, {Bossini}, {Bouquillon},
  {Bragaglia}, {Bramante}, {Breedt}, {Bressan}, {Brouillet}, {Brugaletta},
  {Bucciarelli}, {Burlacu}, {Butkevich}, {Buzzi}, {Caffau}, {Cancelliere},
  {Cantat-Gaudin}, {Carballo}, {Carlucci}, {Carnerero}, {Carrasco},
  {Casamiquela}, {Castellani}, {Castro-Ginard}, {Chaoul}, {Charlot}, {Chemin},
  {Chiaramida}, {Chiavassa}, {Chornay}, {Comoretto}, {Contursi}, {Cooper},
  {Cornez}, {Cowell}, {Crifo}, {Cropper}, {Crosta}, {Crowley}, {Dafonte},
  {Dapergolas}, {David}, {David}, {de Laverny}, {De Luise}, {De March}, {De
  Ridder}, {de Souza}, {de Torres}, {del Peloso}, {del Pozo}, {Delbo},
  {Delgado}, {Delisle}, {Demouchy}, {Dharmawardena}, {Di Matteo}, {Diakite},
  {Diener}, {Distefano}, {Dolding}, {Edvardsson}, {Enke}, {Fabre}, {Fabrizio},
  {Faigler}, {Fedorets}, {Fernique}, {Fienga}, {Figueras}, {Fournier},
  {Fouron}, {Fragkoudi}, {Gai}, {Garcia-Gutierrez}, {Garcia-Reinaldos},
  {Garc{\'\i}a-Torres}, {Garofalo}, {Gavel}, {Gavras}, {Gerlach}, {Geyer},
  {Giacobbe}, {Gilmore}, {Girona}, {Giuffrida}, {Gomel}, {Gomez},
  {Gonz{\'a}lez-N{\'u}{\~n}ez}, {Gonz{\'a}lez-Santamar{\'\i}a},
  {Gonz{\'a}lez-Vidal}, {Granvik}, {Guillout}, {Guiraud},
  {Guti{\'e}rrez-S{\'a}nchez}, {Guy}, {Hatzidimitriou}, {Hauser}, {Haywood},
  {Helmer}, {Helmi}, {Sarmiento}, {Hidalgo}, {Hilger}, {H{\l}adczuk}, {Hobbs},
  {Holland}, {Huckle}, {Jardine}, {Jasniewicz}, {Jean-Antoine Piccolo},
  {Jim{\'e}nez-Arranz}, {Jorissen}, {Juaristi Campillo}, {Julbe}, {Karbevska},
  {Kervella}, {Khanna}, {Kontizas}, {Kordopatis}, {Korn}, {K{\'o}sp{\'a}l},
  {Kostrzewa-Rutkowska}, {Kruszy{\'n}ska}, {Kun}, {Laizeau}, {Lambert},
  {Lanza}, {Lasne}, {Le Campion}, {Lebreton}, {Lebzelter}, {Leccia}, {Leclerc},
  {Lecoeur-Taibi}, {Liao}, {Licata}, {Lindstr{\o}m}, {Lister}, {Livanou},
  {Lobel}, {Lorca}, {Loup}, {Madrero Pardo}, {Magdaleno Romeo}, {Managau},
  {Mann}, {Manteiga}, {Marchant}, {Marconi}, {Marcos}, {Marcos Santos},
  {Mar{\'\i}n Pina}, {Marinoni}, {Marocco}, {Marshall}, {Martin Polo},
  {Mart{\'\i}n-Fleitas}, {Marton}, {Mary}, {Masip}, {Massari},
  {Mastrobuono-Battisti}, {Mazeh}, {McMillan}, {Messina}, {Michalik}, {Millar},
  {Mints}, {Molina}, {Molinaro}, {Moln{\'a}r}, {Monari}, {Mongui{\'o}},
  {Montegriffo}, {Montero}, {Mor}, {Mora}, {Morbidelli}, {Morel}, {Morris},
  {Muraveva}, {Murphy}, {Musella}, {Nagy}, {Noval}, {Oca{\~n}a}, {Ogden},
  {Ordenovic}, {Osinde}, {Pagani}, {Pagano}, {Palaversa}, {Palicio},
  {Pallas-Quintela}, {Panahi}, {Payne-Wardenaar}, {Pe{\~n}alosa Esteller},
  {Penttil{\"a}}, {Pichon}, {Piersimoni}, {Pineau}, {Plachy}, {Plum}, {Poggio},
  {Pr{\v{s}}a}, {Pulone}, {Racero}, {Ragaini}, {Rainer}, {Raiteri}, {Rambaux},
  {Ramos}, {Ramos-Lerate}, {Re Fiorentin}, {Regibo}, {Richards}, {Rios Diaz},
  {Ripepi}, {Riva}, {Rix}, {Rixon}, {Robichon}, {Robin}, {Robin}, {Roelens},
  {Rogues}, {Rohrbasser}, {Romero-G{\'o}mez}, {Rowell}, {Royer}, {Ruz Mieres},
  {Rybicki}, {Sadowski}, {S{\'a}ez N{\'u}{\~n}ez}, {Sagrist{\`a} Sell{\'e}s},
  {Sahlmann}, {Salguero}, {Samaras}, {Sanchez Gimenez}, {Sanna},
  {Santove{\~n}a}, {Sarasso}, {Schultheis}, {Sciacca}, {Segol}, {Segovia},
  {S{\'e}gransan}, {Semeux}, {Shahaf}, {Siddiqui}, {Siebert}, {Siltala},
  {Silvelo}, {Slezak}, {Slezak}, {Smart}, {Snaith}, {Solano}, {Solitro},
  {Souami}, {Souchay}, {Spagna}, {Spina}, {Spoto}, {Steele},
  {Steidelm{\"u}ller}, {Stephenson}, {S{\"u}veges}, {Surdej}, {Szabados},
  {Szegedi-Elek}, {Taris}, {Taylor}, {Teixeira}, {Tolomei}, {Tonello}, {Torra},
  {Torra}, {Torralba Elipe}, {Trabucchi}, {Tsounis}, {Turon}, {Ulla}, {Unger},
  {Vaillant}, {van Dillen}, {van Reeven}, {Vanel}, {Vecchiato}, {Viala},
  {Vicente}, {Voutsinas}, {Weiler}, {Wevers}, {Wyrzykowski}, {Yoldas}, {Yvard},
  {Zhao}, {Zorec}, {Zucker}, \& {Zwitter}}]{Gaia_2023}
{Gaia Collaboration}, {Vallenari}, A., {Brown}, A.~G.~A., {et~al.} 2023, \aap,
  674, A1, \dodoi{10.1051/0004-6361/202243940}

\bibitem[{{Garmany} {et~al.}(1980){Garmany}, {Conti}, \&
  {Massey}}]{garmany_1980}
{Garmany}, C.~D., {Conti}, P.~S., \& {Massey}, P. 1980, \apj, 242, 1063,
  \dodoi{10.1086/158537}

\bibitem[{{Georgy} {et~al.}(2017){Georgy}, {Meynet}, {Ekstr{\"o}m}, {Wade},
  {Petit}, {Keszthelyi}, \& {Hirschi}}]{Georgy_2017}
{Georgy}, C., {Meynet}, G., {Ekstr{\"o}m}, S., {et~al.} 2017, \aap, 599, L5,
  \dodoi{10.1051/0004-6361/201730401}

\bibitem[{{Gilliland} {et~al.}(2010){Gilliland}, {Jenkins}, {Borucki},
  {Bryson}, {Caldwell}, {Clarke}, {Dotson}, {Haas}, {Hall}, {Klaus}, {Koch},
  {McCauliff}, {Quintana}, {Twicken}, \& {van Cleve}}]{Gilliland_2010}
{Gilliland}, R.~L., {Jenkins}, J.~M., {Borucki}, W.~J., {et~al.} 2010, \apjl,
  713, L160, \dodoi{10.1088/2041-8205/713/2/L160}

\bibitem[{{Grunhut}(2012)}]{Grunhut_2012_norm}
{Grunhut}, J.~H. 2012, PhD thesis, Queens University, Canada

\bibitem[{{Grunhut} {et~al.}(2009){Grunhut}, {Wade}, {Marcolino}, {Petit},
  {Henrichs}, {Cohen}, {Alecian}, {Bohlender}, {Bouret}, {Kochukhov}, {Neiner},
  {St-Louis}, \& {Townsend}}]{Grunhut_2009}
{Grunhut}, J.~H., {Wade}, G.~A., {Marcolino}, W.~L.~F., {et~al.} 2009, \mnras,
  400, L94, \dodoi{10.1111/j.1745-3933.2009.00771.x}

\bibitem[{{Grunhut} {et~al.}(2012){Grunhut}, {Wade}, {Sundqvist}, {ud-Doula},
  {Neiner}, {Ignace}, {Marcolino}, {Rivinius}, {Fullerton}, {Kaper},
  {Mauclaire}, {Buil}, {Garrel}, {Ribeiro}, \& {Ubaud}}]{Grunhut_2012}
{Grunhut}, J.~H., {Wade}, G.~A., {Sundqvist}, J.~O., {et~al.} 2012, \mnras,
  426, 2208, \dodoi{10.1111/j.1365-2966.2012.21799.x}

\bibitem[{{Grunhut} {et~al.}(2017){Grunhut}, {Wade}, {Neiner}, {Oksala},
  {Petit}, {Alecian}, {Bohlender}, {Bouret}, {Henrichs}, {Hussain},
  {Kochukhov}, \& {MiMeS Collaboration}}]{grunhut_2017}
{Grunhut}, J.~H., {Wade}, G.~A., {Neiner}, C., {et~al.} 2017, \mnras, 465,
  2432, \dodoi{10.1093/mnras/stw2743}

\bibitem[{{Grunhut} {et~al.}(2022){Grunhut}, {Wade}, {Folsom}, {Neiner},
  {Kochukhov}, {Alecian}, {Shultz}, {Petit}, {MiMeS Collaboration}, \&
  {BinaMIcS Collaboration}}]{Grunhut_2022}
{Grunhut}, J.~H., {Wade}, G.~A., {Folsom}, C.~P., {et~al.} 2022, \mnras, 512,
  1944, \dodoi{10.1093/mnras/stab3320}

\bibitem[{{Gullikson} {et~al.}(2014){Gullikson}, {Dodson-Robinson}, \&
  {Kraus}}]{Gullikson_2014}
{Gullikson}, K., {Dodson-Robinson}, S., \& {Kraus}, A. 2014, \aj, 148, 53,
  \dodoi{10.1088/0004-6256/148/3/53}

\bibitem[{Harris {et~al.}(2020)Harris, Millman, van~der Walt, Gommers,
  Virtanen, Cournapeau, Wieser, Taylor, Berg, Smith, Kern, Picus, Hoyer, van
  Kerkwijk, Brett, Haldane, del R{\'{i}}o, Wiebe, Peterson,
  G{\'{e}}rard-Marchant, Sheppard, Reddy, Weckesser, Abbasi, Gohlke, \&
  Oliphant}]{harris_2020}
Harris, C.~R., Millman, K.~J., van~der Walt, S.~J., {et~al.} 2020, Nature, 585,
  357, \dodoi{10.1038/s41586-020-2649-2}

\bibitem[{{Hart} {et~al.}(2023){Hart}, {Shappee}, {Hey}, {Kochanek}, {Stanek},
  {Lim}, {Dobbs}, {Tucker}, {Jayasinghe}, {Beacom}, {Boright}, {Holoien},
  {Ong}, {Prieto}, {Thompson}, \& {Will}}]{Hart_2023}
{Hart}, K., {Shappee}, B.~J., {Hey}, D., {et~al.} 2023, arXiv e-prints,
  arXiv:2304.03791, \dodoi{10.48550/arXiv.2304.03791}

\bibitem[{{Harvin} {et~al.}(2002){Harvin}, {Gies}, {Bagnuolo}, {Penny}, \&
  {Thaller}}]{harvin_2002}
{Harvin}, J.~A., {Gies}, D.~R., {Bagnuolo}, Jr., W.~G., {Penny}, L.~R., \&
  {Thaller}, M.~L. 2002, \apj, 565, 1216, \dodoi{10.1086/324705}

\bibitem[{{Henrichs} {et~al.}(1994){Henrichs}, {Kaper}, \&
  {Nichols}}]{Henrichs_1994}
{Henrichs}, H.~F., {Kaper}, L., \& {Nichols}, J.~S. 1994, \aap, 285, 565

\bibitem[{{Henrichs} {et~al.}(2009){Henrichs}, {Schnerr}, {de Jong}, {Kaper},
  {Donati}, \& {Catala}}]{Henrichs_2009}
{Henrichs}, H.~F., {Schnerr}, R.~S., {de Jong}, J.~A., {et~al.} 2009, in IAU
  Symposium, Vol. 259, Cosmic Magnetic Fields: From Planets, to Stars and
  Galaxies, ed. K.~G. {Strassmeier}, A.~G. {Kosovichev}, \& J.~E. {Beckman},
  383--384, \dodoi{10.1017/S1743921309030774}

\bibitem[{{Herbig}(1975)}]{Herbig_1975}
{Herbig}, G.~H. 1975, \apj, 196, 129, \dodoi{10.1086/153400}

\bibitem[{{Herrero} {et~al.}(1992){Herrero}, {Kudritzki}, {Vilchez}, {Kunze},
  {Butler}, \& {Haser}}]{Herrero_1992}
{Herrero}, A., {Kudritzki}, R.~P., {Vilchez}, J.~M., {et~al.} 1992, \aap, 261,
  209

\bibitem[{{Holgado}(2019)}]{holgado_2019}
{Holgado}, G. 2019, PhD thesis, Astrophysical Institute of the Canaries;
  University of La Laguna, Spain

\bibitem[{{Holgado} {et~al.}(2018){Holgado}, {Sim{\'o}n-D{\'\i}az},
  {Barb{\'a}}, {Puls}, {Herrero}, {Castro}, {Garcia}, {Ma{\'\i}z
  Apell{\'a}niz}, {Negueruela}, \& {Sab{\'\i}n-Sanjuli{\'a}n}}]{holgado_2018}
{Holgado}, G., {Sim{\'o}n-D{\'\i}az}, S., {Barb{\'a}}, R.~H., {et~al.} 2018,
  \aap, 613, A65, \dodoi{10.1051/0004-6361/201731543}

\bibitem[{{Howarth} {et~al.}(1995){Howarth}, {Prinja}, \&
  {Massa}}]{Howarth_1995}
{Howarth}, I.~D., {Prinja}, R.~K., \& {Massa}, D. 1995, \apjl, 452, L65,
  \dodoi{10.1086/309710}

\bibitem[{{Howarth} {et~al.}(2007){Howarth}, {Walborn}, {Lennon}, {Puls},
  {Naz{\'e}}, {Annuk}, {Antokhin}, {Bohlender}, {Bond}, {Donati}, {Georgiev},
  {Gies}, {Harmer}, {Herrero}, {Kolka}, {McDavid}, {Morel}, {Negueruela},
  {Rauw}, \& {Reig}}]{howarth_2007}
{Howarth}, I.~D., {Walborn}, N.~R., {Lennon}, D.~J., {et~al.} 2007, \mnras,
  381, 433, \dodoi{10.1111/j.1365-2966.2007.12178.x}

\bibitem[{{Howell} {et~al.}(2014){Howell}, {Sobeck}, {Haas}, {Still},
  {Barclay}, {Mullally}, {Troeltzsch}, {Aigrain}, {Bryson}, {Caldwell},
  {Chaplin}, {Cochran}, {Huber}, {Marcy}, {Miglio}, {Najita}, {Smith},
  {Twicken}, \& {Fortney}}]{Howell_2014}
{Howell}, S.~B., {Sobeck}, C., {Haas}, M., {et~al.} 2014, \pasp, 126, 398,
  \dodoi{10.1086/676406}

\bibitem[{{Hubrig} {et~al.}(2016){Hubrig}, {Kholtygin}, {Ilyin},
  {Sch{\"o}ller}, \& {Oskinova}}]{Hubrig_2016}
{Hubrig}, S., {Kholtygin}, A., {Ilyin}, I., {Sch{\"o}ller}, M., \& {Oskinova},
  L.~M. 2016, \apj, 822, 104, \dodoi{10.3847/0004-637X/822/2/104}

\bibitem[{{Hubrig} {et~al.}(2019){Hubrig}, {Kholtygin}, {Sidoli},
  {Sch{\"o}ller}, \& {J{\"a}rvinen}}]{Hubrig_2019}
{Hubrig}, S., {Kholtygin}, A.~F., {Sidoli}, L., {Sch{\"o}ller}, M., \&
  {J{\"a}rvinen}, S.~P. 2019, in IAU Symposium, Vol. 346, High-mass X-ray
  Binaries: Illuminating the Passage from Massive Binaries to Merging Compact
  Objects, ed. L.~M. {Oskinova}, E.~{Bozzo}, T.~{Bulik}, \& D.~R. {Gies},
  40--44, \dodoi{10.1017/S1743921318007585}

\bibitem[{{Hubrig} {et~al.}(2008){Hubrig}, {Sch{\"o}ller}, {Schnerr},
  {Gonz{\'a}lez}, {Ignace}, \& {Henrichs}}]{hubrig_2008}
{Hubrig}, S., {Sch{\"o}ller}, M., {Schnerr}, R.~S., {et~al.} 2008, \aap, 490,
  793, \dodoi{10.1051/0004-6361:200810171}

\bibitem[{Hunter(2007)}]{Hunter_2007}
Hunter, J.~D. 2007, Computing in Science \& Engineering, 9, 90,
  \dodoi{10.1109/MCSE.2007.55}

\bibitem[{{Karitskaya} {et~al.}(2009){Karitskaya}, {Bochkarev}, {Hubrig},
  {Gnedin}, {Pogodin}, {Yudin}, {Agafonov}, \& {Sharova}}]{Karitskaya_2009}
{Karitskaya}, E.~A., {Bochkarev}, N.~G., {Hubrig}, S., {et~al.} 2009, arXiv
  e-prints, arXiv:0908.2719, \dodoi{10.48550/arXiv.0908.2719}

\bibitem[{{Karitskaya} {et~al.}(2010){Karitskaya}, {Bochkarev}, {Hubrig},
  {Gnedin}, {Pogodin}, {Yudin}, {Agafonov}, \& {Sharova}}]{Karitskaya_2010}
---. 2010, Information Bulletin on Variable Stars, 5950, 1

\bibitem[{{Katz} {et~al.}(2023){Katz}, {Sartoretti}, {Guerrier}, {Panuzzo},
  {Seabroke}, {Th{\'e}venin}, {Cropper}, {Benson}, {Blomme}, {Haigron},
  {Marchal}, {Smith}, {Baker}, {Chemin}, {Damerdji}, {David}, {Dolding},
  {Fr{\'e}mat}, {Gosset}, {Jan{\ss}en}, {Jasniewicz}, {Lobel}, {Plum},
  {Samaras}, {Snaith}, {Soubiran}, {Vanel}, {Zwitter}, {Antoja}, {Arenou},
  {Babusiaux}, {Brouillet}, {Caffau}, {Di Matteo}, {Fabre}, {Fabricius},
  {Fragkoudi}, {Haywood}, {Huckle}, {Hottier}, {Lasne}, {Leclerc},
  {Mastrobuono-Battisti}, {Royer}, {Teyssier}, {Zorec}, {Crifo}, {Jean-Antoine
  Piccolo}, {Turon}, \& {Viala}}]{Katz_2023}
{Katz}, D., {Sartoretti}, P., {Guerrier}, A., {et~al.} 2023, \aap, 674, A5,
  \dodoi{10.1051/0004-6361/202244220}

\bibitem[{{Kaufer} {et~al.}(1999){Kaufer}, {Stahl}, {Tubbesing},
  {N{\o}rregaard}, {Avila}, {Francois}, {Pasquini}, \&
  {Pizzella}}]{kaufer_1999}
{Kaufer}, A., {Stahl}, O., {Tubbesing}, S., {et~al.} 1999, The Messenger, 95, 8

\bibitem[{{Kervella} {et~al.}(2019){Kervella}, {Arenou}, {Mignard}, \&
  {Th{\'e}venin}}]{Kervella_2019}
{Kervella}, P., {Arenou}, F., {Mignard}, F., \& {Th{\'e}venin}, F. 2019, \aap,
  623, A72, \dodoi{10.1051/0004-6361/201834371}

\bibitem[{{Keszthelyi}(2023)}]{Keszthelyi_2023}
{Keszthelyi}, Z. 2023, Galaxies, 11, 40, \dodoi{10.3390/galaxies11020040}

\bibitem[{{Keszthelyi} {et~al.}(2019){Keszthelyi}, {Meynet}, {Georgy}, {Wade},
  {Petit}, \& {David-Uraz}}]{Keszthelyi_2019}
{Keszthelyi}, Z., {Meynet}, G., {Georgy}, C., {et~al.} 2019, \mnras, 485, 5843,
  \dodoi{10.1093/mnras/stz772}

\bibitem[{{Keszthelyi} {et~al.}(2022){Keszthelyi}, {de Koter}, {G{\"o}tberg},
  {Meynet}, {Brands}, {Petit}, {Carrington}, {David-Uraz}, {Geen}, {Georgy},
  {Hirschi}, {Puls}, {Ramalatswa}, {Shultz}, \& {ud-Doula}}]{Keszthelyi_2022}
{Keszthelyi}, Z., {de Koter}, A., {G{\"o}tberg}, Y., {et~al.} 2022, \mnras,
  517, 2028, \dodoi{10.1093/mnras/stac2598}

\bibitem[{{Kochanek} {et~al.}(2017){Kochanek}, {Shappee}, {Stanek}, {Holoien},
  {Thompson}, {Prieto}, {Dong}, {Shields}, {Will}, {Britt}, {Perzanowski}, \&
  {Pojma{\'n}ski}}]{Konachek_2017}
{Kochanek}, C.~S., {Shappee}, B.~J., {Stanek}, K.~Z., {et~al.} 2017, \pasp,
  129, 104502, \dodoi{10.1088/1538-3873/aa80d9}

\bibitem[{{Kochukhov} {et~al.}(2010){Kochukhov}, {Makaganiuk}, \&
  {Piskunov}}]{kochukhov_2010}
{Kochukhov}, O., {Makaganiuk}, V., \& {Piskunov}, N. 2010, \aap, 524, A5,
  \dodoi{10.1051/0004-6361/201015429}

\bibitem[{{Koen} \& {Eyer}(2002)}]{Koen_2002}
{Koen}, C., \& {Eyer}, L. 2002, \mnras, 331, 45,
  \dodoi{10.1046/j.1365-8711.2002.05150.x}

\bibitem[{{Krti{\v{c}}ka} \& {Feldmeier}(2018)}]{Kritcka_2018}
{Krti{\v{c}}ka}, J., \& {Feldmeier}, A. 2018, \aap, 617, A121,
  \dodoi{10.1051/0004-6361/201731614}

\bibitem[{{Krti{\v{c}}ka} \& {Feldmeier}(2021)}]{Kritcka_2021}
---. 2021, \aap, 648, A79, \dodoi{10.1051/0004-6361/202040148}

\bibitem[{{Landi Degl'Innocenti} \& {Landolfi}(2004)}]{landi_2004}
{Landi Degl'Innocenti}, E., \& {Landolfi}, M. 2004, {Polarization in Spectral
  Lines}, Vol. 307 (Kluwer Academic Publishers),
  \dodoi{10.1007/978-1-4020-2415-3}

\bibitem[{{Langer} \& {Kudritzki}(2014)}]{Langer_2014}
{Langer}, N., \& {Kudritzki}, R.~P. 2014, \aap, 564, A52,
  \dodoi{10.1051/0004-6361/201423374}

\bibitem[{{Lightkurve Collaboration} {et~al.}(2018){Lightkurve Collaboration},
  {Cardoso}, {Hedges}, {Gully-Santiago}, {Saunders}, {Cody}, {Barclay}, {Hall},
  {Sagear}, {Turtelboom}, {Zhang}, {Tzanidakis}, {Mighell}, {Coughlin}, {Bell},
  {Berta-Thompson}, {Williams}, {Dotson}, \& {Barentsen}}]{lightkurve_2018}
{Lightkurve Collaboration}, {Cardoso}, J.~V.~d.~M., {Hedges}, C., {et~al.}
  2018, {Lightkurve: Kepler and TESS time series analysis in Python},
  Astrophysics Source Code Library.
\newblock \doeprint{1812.013}

\bibitem[{{Lim} {et~al.}(2024){Lim}, {Naz{\'e}}, {Chang}, \&
  {Hutsem{\'e}kers}}]{Beomdu_2024}
{Lim}, B., {Naz{\'e}}, Y., {Chang}, S.-J., \& {Hutsem{\'e}kers}, D. 2024, \apj,
  961, 72, \dodoi{10.3847/1538-4357/ad12c4}

\bibitem[{{Lindegren} {et~al.}(2018){Lindegren}, {Hern{\'a}ndez}, {Bombrun},
  {Klioner}, {Bastian}, {Ramos-Lerate}, {de Torres}, {Steidelm{\"u}ller},
  {Stephenson}, {Hobbs}, {Lammers}, {Biermann}, {Geyer}, {Hilger}, {Michalik},
  {Stampa}, {McMillan}, {Casta{\~n}eda}, {Clotet}, {Comoretto}, {Davidson},
  {Fabricius}, {Gracia}, {Hambly}, {Hutton}, {Mora}, {Portell}, {van Leeuwen},
  {Abbas}, {Abreu}, {Altmann}, {Andrei}, {Anglada}, {Balaguer-N{\'u}{\~n}ez},
  {Barache}, {Becciani}, {Bertone}, {Bianchi}, {Bouquillon}, {Bourda},
  {Br{\"u}semeister}, {Bucciarelli}, {Busonero}, {Buzzi}, {Cancelliere},
  {Carlucci}, {Charlot}, {Cheek}, {Crosta}, {Crowley}, {de Bruijne}, {de
  Felice}, {Drimmel}, {Esquej}, {Fienga}, {Fraile}, {Gai}, {Garralda},
  {Gonz{\'a}lez-Vidal}, {Guerra}, {Hauser}, {Hofmann}, {Holl}, {Jordan},
  {Lattanzi}, {Lenhardt}, {Liao}, {Licata}, {Lister}, {L{\"o}ffler},
  {Marchant}, {Martin-Fleitas}, {Messineo}, {Mignard}, {Morbidelli}, {Poggio},
  {Riva}, {Rowell}, {Salguero}, {Sarasso}, {Sciacca}, {Siddiqui}, {Smart},
  {Spagna}, {Steele}, {Taris}, {Torra}, {van Elteren}, {van Reeven}, \&
  {Vecchiato}}]{Lindegren_2018}
{Lindegren}, L., {Hern{\'a}ndez}, J., {Bombrun}, A., {et~al.} 2018, \aap, 616,
  A2, \dodoi{10.1051/0004-6361/201832727}

\bibitem[{{Lindegren} {et~al.}(2021){Lindegren}, {Klioner}, {Hern{\'a}ndez},
  {Bombrun}, {Ramos-Lerate}, {Steidelm{\"u}ller}, {Bastian}, {Biermann}, {de
  Torres}, {Gerlach}, {Geyer}, {Hilger}, {Hobbs}, {Lammers}, {McMillan},
  {Stephenson}, {Casta{\~n}eda}, {Davidson}, {Fabricius}, {Gracia-Abril},
  {Portell}, {Rowell}, {Teyssier}, {Torra}, {Bartolom{\'e}}, {Clotet},
  {Garralda}, {Gonz{\'a}lez-Vidal}, {Torra}, {Abbas}, {Altmann}, {Anglada
  Varela}, {Balaguer-N{\'u}{\~n}ez}, {Balog}, {Barache}, {Becciani}, {Bernet},
  {Bertone}, {Bianchi}, {Bouquillon}, {Brown}, {Bucciarelli}, {Busonero},
  {Butkevich}, {Buzzi}, {Cancelliere}, {Carlucci}, {Charlot}, {Cioni},
  {Crosta}, {Crowley}, {del Peloso}, {del Pozo}, {Drimmel}, {Esquej}, {Fienga},
  {Fraile}, {Gai}, {Garcia-Reinaldos}, {Guerra}, {Hambly}, {Hauser},
  {Jan{\ss}en}, {Jordan}, {Kostrzewa-Rutkowska}, {Lattanzi}, {Liao}, {Licata},
  {Lister}, {L{\"o}ffler}, {Marchant}, {Masip}, {Mignard}, {Mints}, {Molina},
  {Mora}, {Morbidelli}, {Murphy}, {Pagani}, {Panuzzo}, {Pe{\~n}alosa Esteller},
  {Poggio}, {Re Fiorentin}, {Riva}, {Sagrist{\`a} Sell{\'e}s}, {Sanchez
  Gimenez}, {Sarasso}, {Sciacca}, {Siddiqui}, {Smart}, {Souami}, {Spagna},
  {Steele}, {Taris}, {Utrilla}, {van Reeven}, \&
  {Vecchiato}}]{Lindegren_2021_astrometric}
{Lindegren}, L., {Klioner}, S.~A., {Hern{\'a}ndez}, J., {et~al.} 2021, \aap,
  649, A2, \dodoi{10.1051/0004-6361/202039709}

\bibitem[{{Lomb}(1976)}]{Lomb_1976}
{Lomb}, N.~R. 1976, \apss, 39, 447, \dodoi{10.1007/BF00648343}

\bibitem[{{Loumos} \& {Deeming}(1978)}]{Loumos_1978}
{Loumos}, G.~L., \& {Deeming}, T.~J. 1978, \apss, 56, 285,
  \dodoi{10.1007/BF01879560}

\bibitem[{{Lozinskaya} {et~al.}(1983){Lozinskaya}, {Larkina}, \&
  {Putilina}}]{lozinskaya_1983}
{Lozinskaya}, T.~A., {Larkina}, V.~V., \& {Putilina}, E.~V. 1983, Soviet
  Astronomy Letters, 9, 344

\bibitem[{{Mahy} {et~al.}(2017){Mahy}, {Hutsem{\'e}kers}, {Naz{\'e}}, {Royer},
  {Lebouteiller}, \& {Waelkens}}]{Mahy_2017}
{Mahy}, L., {Hutsem{\'e}kers}, D., {Naz{\'e}}, Y., {et~al.} 2017, \aap, 599,
  A61, \dodoi{10.1051/0004-6361/201629585}

\bibitem[{{Ma{\'\i}z Apell{\'a}niz}(2022)}]{Jesus_2022_parallax}
{Ma{\'\i}z Apell{\'a}niz}, J. 2022, \aap, 657, A130,
  \dodoi{10.1051/0004-6361/202142365}

\bibitem[{{Ma{\'\i}z Apell{\'a}niz} {et~al.}(2023){Ma{\'\i}z Apell{\'a}niz},
  {Holgado}, {Pantaleoni Gonz{\'a}lez}, \& {Caballero}}]{Jesus_2023}
{Ma{\'\i}z Apell{\'a}niz}, J., {Holgado}, G., {Pantaleoni Gonz{\'a}lez}, M., \&
  {Caballero}, J.~A. 2023, \aap, 677, A137, \dodoi{10.1051/0004-6361/202346759}

\bibitem[{{Ma{\'\i}z Apell{\'a}niz} {et~al.}(2021){Ma{\'\i}z Apell{\'a}niz},
  {Pantaleoni Gonz{\'a}lez}, \& {Barb{\'a}}}]{Jesus_2021_parallax}
{Ma{\'\i}z Apell{\'a}niz}, J., {Pantaleoni Gonz{\'a}lez}, M., \& {Barb{\'a}},
  R.~H. 2021, \aap, 649, A13, \dodoi{10.1051/0004-6361/202140418}

\bibitem[{{Ma{\'\i}z Apell{\'a}niz} {et~al.}(2019){Ma{\'\i}z Apell{\'a}niz},
  {Trigueros P{\'a}ez}, {Negueruela}, {Barb{\'a}}, {Sim{\'o}n-D{\'\i}az},
  {Lorenzo}, {Sota}, {Gamen}, {Fari{\~n}a}, {Salas}, {Caballero}, {Morrell},
  {Pellerin}, {Alfaro}, {Herrero}, {Arias}, \& {Marco}}]{jesus_2019}
{Ma{\'\i}z Apell{\'a}niz}, J., {Trigueros P{\'a}ez}, E., {Negueruela}, I.,
  {et~al.} 2019, \aap, 626, A20, \dodoi{10.1051/0004-6361/201935359}

\bibitem[{{Manzari} {et~al.}(2024){Manzari}, {Park}, {Safdi}, \&
  {Savoray}}]{Manzari_2024}
{Manzari}, C.~A., {Park}, Y., {Safdi}, B.~R., \& {Savoray}, I. 2024, \prl, 133,
  211002, \dodoi{10.1103/PhysRevLett.133.211002}

\bibitem[{{Markova} {et~al.}(2018){Markova}, {Puls}, \&
  {Langer}}]{Markova_2018}
{Markova}, N., {Puls}, J., \& {Langer}, N. 2018, \aap, 613, A12,
  \dodoi{10.1051/0004-6361/201731361}

\bibitem[{{Mart{\'\i}nez-Sebasti{\'a}n}
  {et~al.}(2025){Mart{\'\i}nez-Sebasti{\'a}n}, {Sim{\'o}n-D{\'\i}az}, {Jin},
  {Keszthelyi}, {Holgado}, {Langer}, \& {Puls}}]{Martinez_2025}
{Mart{\'\i}nez-Sebasti{\'a}n}, C., {Sim{\'o}n-D{\'\i}az}, S., {Jin}, H.,
  {et~al.} 2025, \aap, 693, L10, \dodoi{10.1051/0004-6361/202452622}

\bibitem[{{Martins}(2018)}]{Martins_2018}
{Martins}, F. 2018, \aap, 616, A135, \dodoi{10.1051/0004-6361/201833050}

\bibitem[{{Martins} {et~al.}(2010){Martins}, {Donati}, {Marcolino}, {Bouret},
  {Wade}, {Escolano}, {Howarth}, \& {Mimes Collaboration}}]{martins_2010}
{Martins}, F., {Donati}, J.~F., {Marcolino}, W.~L.~F., {et~al.} 2010, \mnras,
  407, 1423, \dodoi{10.1111/j.1365-2966.2010.17005.x}

\bibitem[{{Martins} {et~al.}(2015{\natexlab{a}}){Martins}, {Marcolino},
  {Hillier}, {Donati}, \& {Bouret}}]{martins_2015_radial}
{Martins}, F., {Marcolino}, W., {Hillier}, D.~J., {Donati}, J.~F., \& {Bouret},
  J.~C. 2015{\natexlab{a}}, \aap, 574, A142,
  \dodoi{10.1051/0004-6361/201423882}

\bibitem[{{Martins} {et~al.}(2017){Martins}, {Sim{\'o}n-D{\'\i}az},
  {Barb{\'a}}, {Gamen}, \& {Ekstr{\"o}m}}]{martins_2017}
{Martins}, F., {Sim{\'o}n-D{\'\i}az}, S., {Barb{\'a}}, R.~H., {Gamen}, R.~C.,
  \& {Ekstr{\"o}m}, S. 2017, \aap, 599, A30,
  \dodoi{10.1051/0004-6361/201629548}

\bibitem[{{Martins} {et~al.}(2015{\natexlab{b}}){Martins}, {Herv{\'e}},
  {Bouret}, {Marcolino}, {Wade}, {Neiner}, {Alecian}, {Grunhut}, \&
  {Petit}}]{martins_2015}
{Martins}, F., {Herv{\'e}}, A., {Bouret}, J.~C., {et~al.} 2015{\natexlab{b}},
  \aap, 575, A34, \dodoi{10.1051/0004-6361/201425173}

\bibitem[{{Mason} {et~al.}(1998){Mason}, {Gies}, {Hartkopf}, {Bagnuolo}, {ten
  Brummelaar}, \& {McAlister}}]{Mason_1998}
{Mason}, B.~D., {Gies}, D.~R., {Hartkopf}, W.~I., {et~al.} 1998, \aj, 115, 821,
  \dodoi{10.1086/300234}

\bibitem[{{Mason} {et~al.}(2009){Mason}, {Hartkopf}, {Gies}, {Henry}, \&
  {Helsel}}]{mason_2009}
{Mason}, B.~D., {Hartkopf}, W.~I., {Gies}, D.~R., {Henry}, T.~J., \& {Helsel},
  J.~W. 2009, \aj, 137, 3358, \dodoi{10.1088/0004-6256/137/2/3358}

\bibitem[{{Mayer} {et~al.}(2010){Mayer}, {Harmanec}, {Wolf}, {Bo{\v{z}}i{\'c}},
  \& {{\v{S}}lechta}}]{Mayer_2010}
{Mayer}, P., {Harmanec}, P., {Wolf}, M., {Bo{\v{z}}i{\'c}}, H., \&
  {{\v{S}}lechta}, M. 2010, \aap, 520, A89, \dodoi{10.1051/0004-6361/200913796}

\bibitem[{{Mel'nik} \& {Dambis}(2017)}]{Melnik_2017}
{Mel'nik}, A.~M., \& {Dambis}, A.~K. 2017, \mnras, 472, 3887,
  \dodoi{10.1093/mnras/stx2225}

\bibitem[{{Mihalas} \& {Conti}(1980)}]{Mihalas_1980}
{Mihalas}, D., \& {Conti}, P.~S. 1980, \apj, 235, 515, \dodoi{10.1086/157654}

\bibitem[{{Miller-Jones} {et~al.}(2021){Miller-Jones}, {Bahramian}, {Orosz},
  {Mandel}, {Gou}, {Maccarone}, {Neijssel}, {Zhao}, {Zi{\'o}{\l}kowski},
  {Reid}, {Uttley}, {Zheng}, {Byun}, {Dodson}, {Grinberg}, {Jung}, {Kim},
  {Marcote}, {Markoff}, {Rioja}, {Rushton}, {Russell}, {Sivakoff}, {Tetarenko},
  {Tudose}, \& {Wilms}}]{Miller_Jones_2021}
{Miller-Jones}, J. C.~A., {Bahramian}, A., {Orosz}, J.~A., {et~al.} 2021,
  Science, 371, 1046, \dodoi{10.1126/science.abb3363}

\bibitem[{{Moffat} \& {Michaud}(1981)}]{Moffat_1981}
{Moffat}, A.~F.~J., \& {Michaud}, G. 1981, \apj, 251, 133,
  \dodoi{10.1086/159447}

\bibitem[{{Munoz} {et~al.}(2022){Munoz}, {Wade}, {Faes}, {Carciofi}, \&
  {Labadie-Bartz}}]{Munoz_2022}
{Munoz}, M.~S., {Wade}, G.~A., {Faes}, D.~M., {Carciofi}, A.~C., \&
  {Labadie-Bartz}, J. 2022, \mnras, 511, 3228, \dodoi{10.1093/mnras/stab3767}

\bibitem[{{Munoz} {et~al.}(2020){Munoz}, {Wade}, {Naz{\'e}}, {Puls}, {Bagnulo},
  \& {Szyma{\'n}ski}}]{Munoz_2020}
{Munoz}, M.~S., {Wade}, G.~A., {Naz{\'e}}, Y., {et~al.} 2020, \mnras, 492,
  1199, \dodoi{10.1093/mnras/stz2904}

\bibitem[{{Opli{\v{s}}tilov{\'a}} {et~al.}(2023){Opli{\v{s}}tilov{\'a}},
  {Mayer}, {Harmanec}, {Bro{\v{z}}}, {Pigulski}, {Bo{\v{z}}i{\'c}}, {Zasche},
  {{\v{S}}lechta}, {Pablo}, {Ko{\l}aczek-Szyma{\'n}ski}, {Moffat}, {Lovekin},
  {Wade}, {Zwintz}, {Popowicz}, \& {Weiss}}]{Oplivstilova_2023}
{Opli{\v{s}}tilov{\'a}}, A., {Mayer}, P., {Harmanec}, P., {et~al.} 2023, \aap,
  672, A31, \dodoi{10.1051/0004-6361/202245272}

\bibitem[{{Owocki} {et~al.}(2016){Owocki}, {ud-Doula}, {Sundqvist}, {Petit},
  {Cohen}, \& {Townsend}}]{owocki_2016}
{Owocki}, S.~P., {ud-Doula}, A., {Sundqvist}, J.~O., {et~al.} 2016, \mnras,
  462, 3830, \dodoi{10.1093/mnras/stw1894}

\bibitem[{{Parker} {et~al.}(2005){Parker}, {Phillipps}, {Pierce}, {Hartley},
  {Hambly}, {Read}, {MacGillivray}, {Tritton}, {Cass}, {Cannon}, {Cohen},
  {Drew}, {Frew}, {Hopewell}, {Mader}, {Malin}, {Masheder}, {Morgan}, {Morris},
  {Russeil}, {Russell}, \& {Walker}}]{Parker_2005}
{Parker}, Q.~A., {Phillipps}, S., {Pierce}, M.~J., {et~al.} 2005, \mnras, 362,
  689, \dodoi{10.1111/j.1365-2966.2005.09350.x}

\bibitem[{{Paxton} {et~al.}(2011){Paxton}, {Bildsten}, {Dotter}, {Herwig},
  {Lesaffre}, \& {Timmes}}]{Paxton_2011}
{Paxton}, B., {Bildsten}, L., {Dotter}, A., {et~al.} 2011, \apjs, 192, 3,
  \dodoi{10.1088/0067-0049/192/1/3}

\bibitem[{{Paxton} {et~al.}(2019){Paxton}, {Smolec}, {Schwab}, {Gautschy},
  {Bildsten}, {Cantiello}, {Dotter}, {Farmer}, {Goldberg}, {Jermyn}, {Kanbur},
  {Marchant}, {Thoul}, {Townsend}, {Wolf}, {Zhang}, \& {Timmes}}]{Paxton_2019}
{Paxton}, B., {Smolec}, R., {Schwab}, J., {et~al.} 2019, \apjs, 243, 10,
  \dodoi{10.3847/1538-4365/ab2241}

\bibitem[{{Petit} {et~al.}(2013){Petit}, {Owocki}, {Wade}, {Cohen},
  {Sundqvist}, {Gagn{\'e}}, {Ma{\'\i}z Apell{\'a}niz}, {Oksala}, {Bohlender},
  {Rivinius}, {Henrichs}, {Alecian}, {Townsend}, {ud-Doula}, \& {MiMeS
  Collaboration}}]{petit_2013}
{Petit}, V., {Owocki}, S.~P., {Wade}, G.~A., {et~al.} 2013, \mnras, 429, 398,
  \dodoi{10.1093/mnras/sts344}

\bibitem[{{Petit} {et~al.}(2017){Petit}, {Keszthelyi}, {MacInnis}, {Cohen},
  {Townsend}, {Wade}, {Thomas}, {Owocki}, {Puls}, \& {ud-Doula}}]{Petit_2017}
{Petit}, V., {Keszthelyi}, Z., {MacInnis}, R., {et~al.} 2017, \mnras, 466,
  1052, \dodoi{10.1093/mnras/stw3126}

\bibitem[{{Petit} {et~al.}(2019){Petit}, {Wade}, {Schneider}, {Fossati},
  {Kamp}, {Neiner}, {David-Uraz}, {Alecian}, \& {MiMeS
  Collaboration}}]{Petit_2019}
{Petit}, V., {Wade}, G.~A., {Schneider}, F.~R.~N., {et~al.} 2019, \mnras, 489,
  5669, \dodoi{10.1093/mnras/stz2469}

\bibitem[{{Piskunov} {et~al.}(1995){Piskunov}, {Kupka}, {Ryabchikova}, {Weiss},
  \& {Jeffery}}]{piskunov_1995}
{Piskunov}, N.~E., {Kupka}, F., {Ryabchikova}, T.~A., {Weiss}, W.~W., \&
  {Jeffery}, C.~S. 1995, \aaps, 112, 525

\bibitem[{Pope(2019)}]{10.17909/t9-6wj4-eb32}
Pope, B. 2019, The K2 Bright Star Survey ("halo"),  STScI/MAST,
  \dodoi{10.17909/T9-6WJ4-EB32}

\bibitem[{{Pope} {et~al.}(2019){Pope}, {White}, {Farr}, {Yu}, {Greklek-McKeon},
  {Huber}, {Aerts}, {Aigrain}, {Bedding}, {Boyajian}, {Creevey}, \&
  {Hogg}}]{Pope_2019}
{Pope}, B. J.~S., {White}, T.~R., {Farr}, W.~M., {et~al.} 2019, \apjs, 245, 8,
  \dodoi{10.3847/1538-4365/ab3d29}

\bibitem[{{Preston}(1967)}]{preston_1967}
{Preston}, G.~W. 1967, \apj, 150, 547, \dodoi{10.1086/149358}

\bibitem[{{Pr{\v{s}}a} {et~al.}(2016){Pr{\v{s}}a}, {Harmanec}, {Torres},
  {Mamajek}, {Asplund}, {Capitaine}, {Christensen-Dalsgaard}, {Depagne},
  {Haberreiter}, {Hekker}, {Hilton}, {Kopp}, {Kostov}, {Kurtz}, {Laskar},
  {Mason}, {Milone}, {Montgomery}, {Richards}, {Schmutz}, {Schou}, \&
  {Stewart}}]{Prsa_2016}
{Pr{\v{s}}a}, A., {Harmanec}, P., {Torres}, G., {et~al.} 2016, \aj, 152, 41,
  \dodoi{10.3847/0004-6256/152/2/41}

\bibitem[{{Puls} {et~al.}(2005){Puls}, {Urbaneja}, {Venero}, {Repolust},
  {Springmann}, {Jokuthy}, \& {Mokiem}}]{puls_2005}
{Puls}, J., {Urbaneja}, M.~A., {Venero}, R., {et~al.} 2005, \aap, 435, 669,
  \dodoi{10.1051/0004-6361:20042365}

\bibitem[{{Ramachandran} {et~al.}(2025){Ramachandran}, {Sander}, {Oskinova},
  {Sch{\"o}sser}, {Pauli}, {Hamann}, {Mahy}, {Bernini-Peron}, {Brigitte}, \&
  {Kub{\'a}tov{\'a}}}]{Ramachandran_2025}
{Ramachandran}, V., {Sander}, A.~A.~C., {Oskinova}, L.~M., {et~al.} 2025, \aap,
  698, A37, \dodoi{10.1051/0004-6361/202554184}

\bibitem[{{Ramiaramanantsoa} {et~al.}(2014){Ramiaramanantsoa}, {Moffat},
  {Chen{\'e}}, {Richardson}, {Henrichs}, {Desforges}, {Antoci}, {Rowe},
  {Matthews}, {Kuschnig}, {Weiss}, {Sasselov}, {Rucinski}, \&
  {Guenther}}]{Ramiaramanantsoa_2014}
{Ramiaramanantsoa}, T., {Moffat}, A. F.~J., {Chen{\'e}}, A.-N., {et~al.} 2014,
  \mnras, 441, 910, \dodoi{10.1093/mnras/stu619}

\bibitem[{{Ramiaramanantsoa} {et~al.}(2018){Ramiaramanantsoa}, {Moffat},
  {Harmon}, {Ignace}, {St-Louis}, {Vanbeveren}, {Shenar}, {Pablo},
  {Richardson}, {Howarth}, {Stevens}, {Piaulet}, {St-Jean}, {Eversberg},
  {Pigulski}, {Popowicz}, {Kuschnig}, {Zoc{\l}o{\'n}ska}, {Buysschaert},
  {Handler}, {Weiss}, {Wade}, {Rucinski}, {Zwintz}, {Luckas}, {Heathcote},
  {Cacella}, {Powles}, {Locke}, {Bohlsen}, {Chen{\'e}}, {Miszalski}, {Waldron},
  {Kotze}, {Kotze}, \& {B{\"o}hm}}]{Ramiaramanantsoa_2018}
{Ramiaramanantsoa}, T., {Moffat}, A. F.~J., {Harmon}, R., {et~al.} 2018,
  \mnras, 473, 5532, \dodoi{10.1093/mnras/stx2671}

\bibitem[{{Raskin} {et~al.}(2011){Raskin}, {van Winckel}, {Hensberge},
  {Jorissen}, {Lehmann}, {Waelkens}, {Avila}, {de Cuyper}, {Degroote},
  {Dubosson}, {Dumortier}, {Fr{\'e}mat}, {Laux}, {Michaud}, {Morren}, {Perez
  Padilla}, {Pessemier}, {Prins}, {Smolders}, {van Eck}, \&
  {Winkler}}]{raskin_2011}
{Raskin}, G., {van Winckel}, H., {Hensberge}, H., {et~al.} 2011, \aap, 526,
  A69, \dodoi{10.1051/0004-6361/201015435}

\bibitem[{{Rauw} {et~al.}(2023){Rauw}, {Naz{\'e}}, {ud-Doula}, \&
  {Neiner}}]{GREGOR_2023}
{Rauw}, G., {Naz{\'e}}, Y., {ud-Doula}, A., \& {Neiner}, C. 2023, \mnras, 521,
  2874, \dodoi{10.1093/mnras/stad693}

\bibitem[{{Ricker} {et~al.}(2015){Ricker}, {Winn}, {Vanderspek}, {Latham},
  {Bakos}, {Bean}, {Berta-Thompson}, {Brown}, {Buchhave}, {Butler}, {Butler},
  {Chaplin}, {Charbonneau}, {Christensen-Dalsgaard}, {Clampin}, {Deming},
  {Doty}, {De Lee}, {Dressing}, {Dunham}, {Endl}, {Fressin}, {Ge}, {Henning},
  {Holman}, {Howard}, {Ida}, {Jenkins}, {Jernigan}, {Johnson}, {Kaltenegger},
  {Kawai}, {Kjeldsen}, {Laughlin}, {Levine}, {Lin}, {Lissauer}, {MacQueen},
  {Marcy}, {McCullough}, {Morton}, {Narita}, {Paegert}, {Palle}, {Pepe},
  {Pepper}, {Quirrenbach}, {Rinehart}, {Sasselov}, {Sato}, {Seager},
  {Sozzetti}, {Stassun}, {Sullivan}, {Szentgyorgyi}, {Torres}, {Udry}, \&
  {Villasenor}}]{ricker_2015}
{Ricker}, G.~R., {Winn}, J.~N., {Vanderspek}, R., {et~al.} 2015, Journal of
  Astronomical Telescopes, Instruments, and Systems, 1, 014003,
  \dodoi{10.1117/1.JATIS.1.1.014003}

\bibitem[{{Rodrigo} \& {Solano}(2020)}]{Rodrigo_2020}
{Rodrigo}, C., \& {Solano}, E. 2020, in XIV.0 Scientific Meeting (virtual) of
  the Spanish Astronomical Society, 182

\bibitem[{{Rogers} {et~al.}(2013){Rogers}, {Lin}, {McElwaine}, \&
  {Lau}}]{Rogers_2013}
{Rogers}, T.~M., {Lin}, D.~N.~C., {McElwaine}, J.~N., \& {Lau}, H.~H.~B. 2013,
  \apj, 772, 21, \dodoi{10.1088/0004-637X/772/1/21}

\bibitem[{Rolph {et~al.}(2017)Rolph, Stein, \& Stunder}]{ROLPH2017210}
Rolph, G., Stein, A., \& Stunder, B. 2017, Environmental Modelling \& Software,
  95, 210, \dodoi{https://doi.org/10.1016/j.envsoft.2017.06.025}

\bibitem[{{Royer} {et~al.}(2024){Royer}, {Merle}, {Dsilva}, {Sekaran}, {Van
  Winckel}, {Fr{\'e}mat}, {Van der Swaelmen}, {Gebruers}, {Tkachenko},
  {Laverick}, {Dirickx}, {Raskin}, {Hensberge}, {Abdul-Masih}, {Acke},
  {Alonso}, {Bandhu Mahato}, {Beck}, {Behara}, {Bloemen}, {Buysschaert}, {Cox},
  {Debosscher}, {De Cat}, {Degroote}, {De Nutte}, {De Smedt}, {de Vries},
  {Dumortier}, {Escorza}, {Exter}, {Goriely}, {Gorlova}, {Hillen}, {Homan},
  {Jorissen}, {Kamath}, {Karjalainen}, {Karjalainen}, {Lampens}, {Lobel},
  {Lombaert}, {Marcos-Arenal}, {Menu}, {Merges}, {Moravveji}, {Nemeth},
  {Neyskens}, {Ostensen}, {P{\'a}pics}, {Perez}, {Prins}, {Royer},
  {Samadi-Ghadim}, {Sana}, {Sans Fuentes}, {Scaringi}, {Schmid}, {Siess},
  {Siopis}, {Smolders}, {S{\'o}dor}, {Thoul}, {Triana}, {Vandenbussche}, {Van
  de Sande}, {Van De Steene}, {Van Eck}, {van Hoof}, {Van Marle}, {Van Reeth},
  {Vermeylen}, {Volpi}, {Vos}, \& {Waelkens}}]{Royer_2024}
{Royer}, P., {Merle}, T., {Dsilva}, K., {et~al.} 2024, \aap, 681, A107,
  \dodoi{10.1051/0004-6361/202346847}

\bibitem[{{Ryabchikova} {et~al.}(2015){Ryabchikova}, {Piskunov}, {Kurucz},
  {Stempels}, {Heiter}, {Pakhomov}, \& {Barklem}}]{ryabchikova_2015}
{Ryabchikova}, T., {Piskunov}, N., {Kurucz}, R.~L., {et~al.} 2015, \physscr,
  90, 054005, \dodoi{10.1088/0031-8949/90/5/054005}

\bibitem[{{Sana} {et~al.}(2013){Sana}, {de Koter}, {de Mink}, {Dunstall},
  {Evans}, {H{\'e}nault-Brunet}, {Ma{\'\i}z Apell{\'a}niz},
  {Ram{\'\i}rez-Agudelo}, {Taylor}, {Walborn}, {Clark}, {Crowther}, {Herrero},
  {Gieles}, {Langer}, {Lennon}, \& {Vink}}]{Sana_2013}
{Sana}, H., {de Koter}, A., {de Mink}, S.~E., {et~al.} 2013, \aap, 550, A107,
  \dodoi{10.1051/0004-6361/201219621}

\bibitem[{{Sana} {et~al.}(2014){Sana}, {Le Bouquin}, {Lacour}, {Berger},
  {Duvert}, {Gauchet}, {Norris}, {Olofsson}, {Pickel}, {Zins}, {Absil}, {de
  Koter}, {Kratter}, {Schnurr}, \& {Zinnecker}}]{sana_2014}
{Sana}, H., {Le Bouquin}, J.~B., {Lacour}, S., {et~al.} 2014, \apjs, 215, 15,
  \dodoi{10.1088/0067-0049/215/1/15}

\bibitem[{{Santolaya-Rey} {et~al.}(1997){Santolaya-Rey}, {Puls}, \&
  {Herrero}}]{Santolaya-Rey_1997}
{Santolaya-Rey}, A.~E., {Puls}, J., \& {Herrero}, A. 1997, \aap, 323, 488

\bibitem[{{Scargle}(1982)}]{Scargle_1982}
{Scargle}, J.~D. 1982, \apj, 263, 835, \dodoi{10.1086/160554}

\bibitem[{{Schneider} {et~al.}(2014){Schneider}, {Langer}, {de Koter}, {Brott},
  {Izzard}, \& {Lau}}]{Schneider_2014}
{Schneider}, F.~R.~N., {Langer}, N., {de Koter}, A., {et~al.} 2014, \aap, 570,
  A66, \dodoi{10.1051/0004-6361/201424286}

\bibitem[{{Schneider} {et~al.}(2019){Schneider}, {Ohlmann}, {Podsiadlowski},
  {R{\"o}pke}, {Balbus}, {Pakmor}, \& {Springel}}]{Schneider_2019}
{Schneider}, F. R.~N., {Ohlmann}, S.~T., {Podsiadlowski}, P., {et~al.} 2019,
  \nat, 574, 211, \dodoi{10.1038/s41586-019-1621-5}

\bibitem[{{Schnerr} {et~al.}(2007){Schnerr}, {Rygl}, {van der Horst},
  {Oosterloo}, {Miller-Jones}, {Henrichs}, {Spoelstra}, \&
  {Foley}}]{Schnerr_2007}
{Schnerr}, R.~S., {Rygl}, K.~L.~J., {van der Horst}, A.~J., {et~al.} 2007,
  \aap, 470, 1105, \dodoi{10.1051/0004-6361:20066299}

\bibitem[{{Sch{\"o}ller} {et~al.}(2017){Sch{\"o}ller}, {Hubrig}, {Fossati},
  {Carroll}, {Briquet}, {Oskinova}, {J{\"a}rvinen}, {Ilyin}, {Castro}, {Morel},
  {Langer}, {Przybilla}, {Nieva}, {Kholtygin}, {Sana}, {Herrero}, {Barb{\'a}},
  {de Koter}, \& {BOB Collaboration}}]{Scholler_2017}
{Sch{\"o}ller}, M., {Hubrig}, S., {Fossati}, L., {et~al.} 2017, \aap, 599, A66,
  \dodoi{10.1051/0004-6361/201628905}

\bibitem[{{Schultz} {et~al.}(2022){Schultz}, {Bildsten}, \&
  {Jiang}}]{Schultz_2022}
{Schultz}, W.~C., {Bildsten}, L., \& {Jiang}, Y.-F. 2022, \apjl, 924, L11,
  \dodoi{10.3847/2041-8213/ac441f}

\bibitem[{{Shappee} {et~al.}(2014){Shappee}, {Prieto}, {Stanek}, {Kochanek},
  {Holoien}, {Jencson}, {Basu}, {Beacom}, {Szczygiel}, {Pojmanski},
  {Brimacombe}, {Dubberley}, {Elphick}, {Foale}, {Hawkins}, {Mullins},
  {Rosing}, {Ross}, \& {Walker}}]{Shappee_2014}
{Shappee}, B., {Prieto}, J., {Stanek}, K.~Z., {et~al.} 2014, in American
  Astronomical Society Meeting Abstracts, Vol. 223, American Astronomical
  Society Meeting Abstracts \#223, 236.03

\bibitem[{{Sharma} {et~al.}(2024){Sharma}, {Ravi}, {Connor}, {Law}, {Ocker},
  {Sherman}, {Kosogorov}, {Faber}, {Hallinan}, {Harnach}, {Hellbourg}, {Hobbs},
  {Hodge}, {Hodges}, {Lamb}, {Rasmussen}, {Somalwar}, {Weinreb}, {Woody},
  {Leja}, {Anand}, {Das}, {Qin}, {Rose}, {Dong}, {Miller}, \&
  {Yao}}]{Kritti_2024}
{Sharma}, K., {Ravi}, V., {Connor}, L., {et~al.} 2024, \nat, 635, 61,
  \dodoi{10.1038/s41586-024-08074-9}

\bibitem[{{Shenar} {et~al.}(2015){Shenar}, {Oskinova}, {Hamann}, {Corcoran},
  {Moffat}, {Pablo}, {Richardson}, {Waldron}, {Huenemoerder}, {Ma{\'\i}z
  Apell{\'a}niz}, {Nichols}, {Todt}, {Naz{\'e}}, {Hoffman}, {Pollock}, \&
  {Negueruela}}]{Shenar_2015}
{Shenar}, T., {Oskinova}, L., {Hamann}, W.~R., {et~al.} 2015, \apj, 809, 135,
  \dodoi{10.1088/0004-637X/809/2/135}

\bibitem[{{Shenar} {et~al.}(2017){Shenar}, {Oskinova}, {J{\"a}rvinen},
  {Luckas}, {Hainich}, {Todt}, {Hubrig}, {Sander}, {Ilyin}, \&
  {Hamann}}]{shenar_2017}
{Shenar}, T., {Oskinova}, L.~M., {J{\"a}rvinen}, S.~P., {et~al.} 2017, \aap,
  606, A91, \dodoi{10.1051/0004-6361/201731291}

\bibitem[{{Shultz} \& {Wade}(2017)}]{Shultz_2017}
{Shultz}, M., \& {Wade}, G.~A. 2017, \mnras, 468, 3985,
  \dodoi{10.1093/mnras/stx759}

\bibitem[{{Sim{\'o}n-D{\'\i}az} {et~al.}(2024){Sim{\'o}n-D{\'\i}az},
  {Britavskiy}, {Castro}, {Holgado}, \& {de Burgos}}]{Simon_2024}
{Sim{\'o}n-D{\'\i}az}, S., {Britavskiy}, N., {Castro}, N., {Holgado}, G., \&
  {de Burgos}, A. 2024, arXiv e-prints, arXiv:2405.11209,
  \dodoi{10.48550/arXiv.2405.11209}

\bibitem[{{Sim{\'o}n-D{\'\i}az} {et~al.}(2011){Sim{\'o}n-D{\'\i}az}, {Castro},
  {Herrero}, {Puls}, {Garcia}, \& {Sab{\'\i}n-Sanjuli{\'a}n}}]{simon_diaz_2011}
{Sim{\'o}n-D{\'\i}az}, S., {Castro}, N., {Herrero}, A., {et~al.} 2011, in
  Journal of Physics Conference Series, Vol. 328, Journal of Physics Conference
  Series (IOP), 012021, \dodoi{10.1088/1742-6596/328/1/012021}

\bibitem[{{Sim{\'o}n-D{\'\i}az} {et~al.}(2017){Sim{\'o}n-D{\'\i}az}, {Godart},
  {Castro}, {Herrero}, {Aerts}, {Puls}, {Telting}, \&
  {Grassitelli}}]{Simon_2017}
{Sim{\'o}n-D{\'\i}az}, S., {Godart}, M., {Castro}, N., {et~al.} 2017, \aap,
  597, A22, \dodoi{10.1051/0004-6361/201628541}

\bibitem[{{Sim{\'o}n-D{\'\i}az} \& {Herrero}(2014)}]{Simon_2014}
{Sim{\'o}n-D{\'\i}az}, S., \& {Herrero}, A. 2014, \aap, 562, A135,
  \dodoi{10.1051/0004-6361/201322758}

\bibitem[{{Sim{\'o}n-D{\'\i}az} {et~al.}(2006){Sim{\'o}n-D{\'\i}az}, {Herrero},
  {Esteban}, \& {Najarro}}]{simon_2006}
{Sim{\'o}n-D{\'\i}az}, S., {Herrero}, A., {Esteban}, C., \& {Najarro}, F. 2006,
  \aap, 448, 351, \dodoi{10.1051/0004-6361:20053066}

\bibitem[{{Sim{\'o}n-D{\'\i}az} {et~al.}(2020){Sim{\'o}n-D{\'\i}az}, {P{\'e}rez
  Prieto}, {Holgado}, {de Burgos}, \& {Iacob Team}}]{simon-diaz_2020}
{Sim{\'o}n-D{\'\i}az}, S., {P{\'e}rez Prieto}, J.~A., {Holgado}, G., {de
  Burgos}, A., \& {Iacob Team}. 2020, in XIV.0 Scientific Meeting (virtual) of
  the Spanish Astronomical Society, 187

\bibitem[{{Sim{\'o}n-D{\'\i}az} {et~al.}(2015){Sim{\'o}n-D{\'\i}az},
  {Negueruela}, {Ma{\'\i}z Apell{\'a}niz}, {Castro}, {Herrero}, {Garcia},
  {P{\'e}rez-Prieto}, {Caon}, {Alacid}, {Camacho}, {Dorda}, {Godart},
  {Gonz{\'a}lez-Fern{\'a}ndez}, {Holgado}, \& {R{\"u}bke}}]{simon-diaz_2015}
{Sim{\'o}n-D{\'\i}az}, S., {Negueruela}, I., {Ma{\'\i}z Apell{\'a}niz}, J.,
  {et~al.} 2015, in Highlights of Spanish Astrophysics VIII, 576--581.
\newblock \doarXiv{1504.04257}

\bibitem[{{Smithsonian Astrophysical Observatory}(2000)}]{smithsonian_2000}
{Smithsonian Astrophysical Observatory}. 2000, {SAOImage DS9: A utility for
  displaying astronomical images in the X11 window environment}, Astrophysics
  Source Code Library, record ascl:0003.002

\bibitem[{{Sota} {et~al.}(2014){Sota}, {Ma{\'\i}z Apell{\'a}niz}, {Morrell},
  {Barb{\'a}}, {Walborn}, {Gamen}, {Arias}, \& {Alfaro}}]{sota_2014}
{Sota}, A., {Ma{\'\i}z Apell{\'a}niz}, J., {Morrell}, N.~I., {et~al.} 2014,
  \apjs, 211, 10, \dodoi{10.1088/0067-0049/211/1/10}

\bibitem[{{Sota} {et~al.}(2011){Sota}, {Ma{\'\i}z Apell{\'a}niz}, {Walborn},
  {Alfaro}, {Barb{\'a}}, {Morrell}, {Gamen}, \& {Arias}}]{sota_2011}
{Sota}, A., {Ma{\'\i}z Apell{\'a}niz}, J., {Walborn}, N.~R., {et~al.} 2011,
  \apjs, 193, 24, \dodoi{10.1088/0067-0049/193/2/24}

\bibitem[{{Stibbs}(1950)}]{Stibbs_1950}
{Stibbs}, D.~W.~N. 1950, \mnras, 110, 395, \dodoi{10.1093/mnras/110.4.395}

\bibitem[{{Subramanian} {et~al.}(2022){Subramanian}, {Balsara}, {ud-Doula}, \&
  {Gagn{\'e}}}]{Subramanian_2022}
{Subramanian}, S., {Balsara}, D.~S., {ud-Doula}, A., \& {Gagn{\'e}}, M. 2022,
  \mnras, 515, 237, \dodoi{10.1093/mnras/stac1778}

\bibitem[{{Sundqvist} {et~al.}(2013){Sundqvist}, {Sim{\'o}n-D{\'\i}az}, {Puls},
  \& {Markova}}]{Sundqvist_2013}
{Sundqvist}, J.~O., {Sim{\'o}n-D{\'\i}az}, S., {Puls}, J., \& {Markova}, N.
  2013, \aap, 559, L10, \dodoi{10.1051/0004-6361/201322761}

\bibitem[{{Sundqvist} {et~al.}(2012){Sundqvist}, {ud-Doula}, {Owocki},
  {Townsend}, {Howarth}, \& {Wade}}]{sundqvist_2012}
{Sundqvist}, J.~O., {ud-Doula}, A., {Owocki}, S.~P., {et~al.} 2012, \mnras,
  423, L21, \dodoi{10.1111/j.1745-3933.2012.01248.x}

\bibitem[{{Telting} {et~al.}(2014){Telting}, {Avila}, {Buchhave}, {Frandsen},
  {Gandolfi}, {Lindberg}, {Stempels}, {Prins}, \& {NOT staff}}]{telting_2014}
{Telting}, J.~H., {Avila}, G., {Buchhave}, L., {et~al.} 2014, Astronomische
  Nachrichten, 335, 41, \dodoi{10.1002/asna.201312007}

\bibitem[{{Thompson} {et~al.}(2024){Thompson}, {Herwig}, {Woodward}, {Mao},
  {Denissenkov}, {Bowman}, \& {Blouin}}]{Thompson_2024}
{Thompson}, W., {Herwig}, F., {Woodward}, P.~R., {et~al.} 2024, \mnras, 531,
  1316, \dodoi{10.1093/mnras/stae1162}

\bibitem[{{Tokovinin} {et~al.}(2020){Tokovinin}, {Mason}, {Mendez}, {Costa}, \&
  {Horch}}]{Tokovinin_2020}
{Tokovinin}, A., {Mason}, B.~D., {Mendez}, R.~A., {Costa}, E., \& {Horch},
  E.~P. 2020, \aj, 160, 7, \dodoi{10.3847/1538-3881/ab91c1}

\bibitem[{{Townsend} {et~al.}(2010){Townsend}, {Oksala}, {Cohen}, {Owocki}, \&
  {ud-Doula}}]{Townsend_2010}
{Townsend}, R.~H.~D., {Oksala}, M.~E., {Cohen}, D.~H., {Owocki}, S.~P., \&
  {ud-Doula}, A. 2010, \apjl, 714, L318, \dodoi{10.1088/2041-8205/714/2/L318}

\bibitem[{{Turner} {et~al.}(2008){Turner}, {ten Brummelaar}, {Roberts},
  {Mason}, {Hartkopf}, \& {Gies}}]{turner_2008}
{Turner}, N.~H., {ten Brummelaar}, T.~A., {Roberts}, L.~C., {et~al.} 2008, \aj,
  136, 554, \dodoi{10.1088/0004-6256/136/2/554}

\bibitem[{{ud-Doula} \& {Owocki}(2002)}]{ud-Doula_2002}
{ud-Doula}, A., \& {Owocki}, S.~P. 2002, \apj, 576, 413, \dodoi{10.1086/341543}

\bibitem[{{ud-Doula} {et~al.}(2008){ud-Doula}, {Owocki}, \&
  {Townsend}}]{udDoula_2008}
{ud-Doula}, A., {Owocki}, S.~P., \& {Townsend}, R. H.~D. 2008, \mnras, 385, 97,
  \dodoi{10.1111/j.1365-2966.2008.12840.x}

\bibitem[{{ud-Doula} {et~al.}(2009){ud-Doula}, {Owocki}, \&
  {Townsend}}]{ud-Doula_2009}
---. 2009, \mnras, 392, 1022, \dodoi{10.1111/j.1365-2966.2008.14134.x}

\bibitem[{{ud-Doula} {et~al.}(2013){ud-Doula}, {Sundqvist}, {Owocki}, {Petit},
  \& {Townsend}}]{ud-Doula_2013}
{ud-Doula}, A., {Sundqvist}, J.~O., {Owocki}, S.~P., {Petit}, V., \&
  {Townsend}, R.~H.~D. 2013, \mnras, 428, 2723, \dodoi{10.1093/mnras/sts246}

\bibitem[{{Van Cleve} {et~al.}(2016){Van Cleve}, {Howell}, {Smith}, {Clarke},
  {Thompson}, {Bryson}, {Lund}, {Handberg}, \& {Chaplin}}]{van_cleve_2016}
{Van Cleve}, J.~E., {Howell}, S.~B., {Smith}, J.~C., {et~al.} 2016, \pasp, 128,
  075002, \dodoi{10.1088/1538-3873/128/965/075002}

\bibitem[{{van Leeuwen}(1997)}]{vanLeeuwen_1997_hipparocs_mission}
{van Leeuwen}, F. 1997, \ssr, 81, 201, \dodoi{10.1023/A:1005081918325}

\bibitem[{{van Leeuwen}(2007)}]{van_Leeuwen_2007}
---. 2007, \aap, 474, 653, \dodoi{10.1051/0004-6361:20078357}

\bibitem[{{van Leeuwen} {et~al.}(1997){van Leeuwen}, {Evans}, {Grenon},
  {Grossmann}, {Mignard}, \& {Perryman}}]{vanLeeuwen_1997}
{van Leeuwen}, F., {Evans}, D.~W., {Grenon}, M., {et~al.} 1997, \aap, 323, L61

\bibitem[{{VanderPlas}(2018)}]{vanderplas_2018}
{VanderPlas}, J.~T. 2018, \apjs, 236, 16, \dodoi{10.3847/1538-4365/aab766}

\bibitem[{{Vidal} {et~al.}(2019){Vidal}, {C{\'e}bron}, {ud-Doula}, \&
  {Alecian}}]{Vidal_2019}
{Vidal}, J., {C{\'e}bron}, D., {ud-Doula}, A., \& {Alecian}, E. 2019, \aap,
  629, A142, \dodoi{10.1051/0004-6361/201935658}

\bibitem[{Virtanen {et~al.}(2020)Virtanen, Gommers, Oliphant, Haberland, Reddy,
  Cournapeau, Burovski, Peterson, Weckesser, Bright, {van der Walt}, Brett,
  Wilson, Millman, Mayorov, Nelson, Jones, Kern, Larson, Carey, Polat, Feng,
  Moore, {VanderPlas}, Laxalde, Perktold, Cimrman, Henriksen, Quintero, Harris,
  Archibald, Ribeiro, Pedregosa, {van Mulbregt}, \& {SciPy 1.0
  Contributors}}]{2020SciPy-NMeth}
Virtanen, P., Gommers, R., Oliphant, T.~E., {et~al.} 2020, Nature Methods, 17,
  261, \dodoi{10.1038/s41592-019-0686-2}

\bibitem[{{Wade} {et~al.}(2000){Wade}, {Donati}, {Landstreet}, \&
  {Shorlin}}]{wade_2000}
{Wade}, G.~A., {Donati}, J.~F., {Landstreet}, J.~D., \& {Shorlin}, S.~L.~S.
  2000, \mnras, 313, 851, \dodoi{10.1046/j.1365-8711.2000.03271.x}

\bibitem[{{Wade} \& {MiMeS Collaboration}(2015)}]{wade_2015}
{Wade}, G.~A., \& {MiMeS Collaboration}. 2015, in Astronomical Society of the
  Pacific Conference Series, Vol. 494, Physics and Evolution of Magnetic and
  Related Stars, ed. Y.~Y. {Balega}, I.~I. {Romanyuk}, \& D.~O. {Kudryavtsev},
  30, \dodoi{10.48550/arXiv.1411.3604}

\bibitem[{{Wade} {et~al.}(2011){Wade}, {Howarth}, {Townsend}, {Grunhut},
  {Shultz}, {Bouret}, {Fullerton}, {Marcolino}, {Martins}, {Naz{\'e}}, {Ud
  Doula}, {Walborn}, \& {Donati}}]{Wade_2011}
{Wade}, G.~A., {Howarth}, I.~D., {Townsend}, R.~H.~D., {et~al.} 2011, \mnras,
  416, 3160, \dodoi{10.1111/j.1365-2966.2011.19265.x}

\bibitem[{{Wade} {et~al.}(2012{\natexlab{a}}){Wade}, {Grunhut}, {Gr{\"a}fener},
  {Howarth}, {Martins}, {Petit}, {Vink}, {Bagnulo}, {Folsom}, {Naz{\'e}},
  {Walborn}, {Townsend}, \& {Evans}}]{wade_2012}
{Wade}, G.~A., {Grunhut}, J., {Gr{\"a}fener}, G., {et~al.} 2012{\natexlab{a}},
  \mnras, 419, 2459, \dodoi{10.1111/j.1365-2966.2011.19897.x}

\bibitem[{{Wade} {et~al.}(2012{\natexlab{b}}){Wade}, {Ma{\'\i}z Apell{\'a}niz},
  {Martins}, {Petit}, {Grunhut}, {Walborn}, {Barb{\'a}}, {Gagn{\'e}},
  {Garc{\'\i}a-Melendo}, {Jose}, {Moffat}, {Naz{\'e}}, {Neiner}, {Pellerin},
  {Penad{\'e}s Ordaz}, {Shultz}, {Sim{\'o}n-D{\'\i}az}, \&
  {Sota}}]{wade_ngc1624}
{Wade}, G.~A., {Ma{\'\i}z Apell{\'a}niz}, J., {Martins}, F., {et~al.}
  2012{\natexlab{b}}, \mnras, 425, 1278,
  \dodoi{10.1111/j.1365-2966.2012.21523.x}

\bibitem[{{Wade} {et~al.}(2015){Wade}, {Barb{\'a}}, {Grunhut}, {Martins},
  {Petit}, {Sundqvist}, {Townsend}, {Walborn}, {Alecian}, {Alfaro}, {Ma{\'\i}z
  Apell{\'a}niz}, {Arias}, {Gamen}, {Morrell}, {Naz{\'e}}, {Sota}, {ud-Doula},
  \& {MiMeS Collaboration}}]{Wade_2015_cpd}
{Wade}, G.~A., {Barb{\'a}}, R.~H., {Grunhut}, J., {et~al.} 2015, \mnras, 447,
  2551, \dodoi{10.1093/mnras/stu2548}

\bibitem[{{Wade} {et~al.}(2016){Wade}, {Neiner}, {Alecian}, {Grunhut}, {Petit},
  {Batz}, {Bohlender}, {Cohen}, {Henrichs}, {Kochukhov}, {Landstreet},
  {Manset}, {Martins}, {Mathis}, {Oksala}, {Owocki}, {Rivinius}, {Shultz},
  {Sundqvist}, {Townsend}, {ud-Doula}, {Bouret}, {Braithwaite}, {Briquet},
  {Carciofi}, {David-Uraz}, {Folsom}, {Fullerton}, {Leroy}, {Marcolino},
  {Moffat}, {Naz{\'e}}, {Louis}, {Auri{\`e}re}, {Bagnulo}, {Bailey},
  {Barb{\'a}}, {Blaz{\`e}re}, {B{\"o}hm}, {Catala}, {Donati}, {Ferrario},
  {Harrington}, {Howarth}, {Ignace}, {Kaper}, {L{\"u}ftinger}, {Prinja},
  {Vink}, {Weiss}, \& {Yakunin}}]{wade_2016}
{Wade}, G.~A., {Neiner}, C., {Alecian}, E., {et~al.} 2016, \mnras, 456, 2,
  \dodoi{10.1093/mnras/stv2568}

\bibitem[{{Walborn}(1972)}]{Walborn_1972}
{Walborn}, N.~R. 1972, \aj, 77, 312, \dodoi{10.1086/111285}

\bibitem[{{Webster} \& {Murdin}(1972)}]{Webster_1972}
{Webster}, B.~L., \& {Murdin}, P. 1972, \nat, 235, 37, \dodoi{10.1038/235037a0}

\bibitem[{{White} {et~al.}(2017){White}, {Pope}, {Antoci}, {P{\'a}pics},
  {Aerts}, {Gies}, {Gordon}, {Huber}, {Schaefer}, {Aigrain}, {Albrecht},
  {Barclay}, {Barentsen}, {Beck}, {Bedding}, {Fredslund Andersen}, {Grundahl},
  {Howell}, {Ireland}, {Murphy}, {Nielsen}, {Silva Aguirre}, \&
  {Tuthill}}]{White_2017}
{White}, T.~R., {Pope}, B.~J.~S., {Antoci}, V., {et~al.} 2017, \mnras, 471,
  2882, \dodoi{10.1093/mnras/stx1050}

\bibitem[{{Wilson}(1953)}]{wilson_1953}
{Wilson}, R.~E. 1953, Carnegie Institute Washington D.C. Publication, 0

\bibitem[{{Winecki} \& {Kochanek}(2024)}]{Winecki_2024}
{Winecki}, D., \& {Kochanek}, C.~S. 2024, \apj, 971, 61,
  \dodoi{10.3847/1538-4357/ad5a0b}

\bibitem[{{Zechmeister} \& {K{\"u}rster}(2009)}]{Zechmeister_2009}
{Zechmeister}, M., \& {K{\"u}rster}, M. 2009, \aap, 496, 577,
  \dodoi{10.1051/0004-6361:200811296}

\end{thebibliography}
\bibliographystyle{aasjournal}

\end{document}